\documentclass[lettersize,journal]{IEEEtran}
\usepackage{amsmath,amsfonts}
\usepackage{algorithmic}
\usepackage{algorithm}
\usepackage{array}
\usepackage[caption=false,font=normalsize,labelfont=sf,textfont=sf]{subfig}
\usepackage{textcomp}
\usepackage{stfloats}
\usepackage{url}
\usepackage{verbatim}
\usepackage{graphicx}
\usepackage{booktabs}
\usepackage{cite}
\usepackage{amssymb}
\usepackage{amsbsy}
\usepackage{bm}
\usepackage{caption}
\usepackage{tabularx}
\usepackage{nomencl}
\usepackage{cancel}
\usepackage{graphicx}
\usepackage{subcaption}
\usepackage{multirow}
\usepackage{tikz}
\usepackage{pgfplots}
\usepackage{arydshln}
\usepackage{soul}

\definecolor{darkergreen}{rgb}{0.0, 0.5, 0.0}
\definecolor{dgreen}{rgb}{0.3, 0.8, 0.0}

\definecolor{darkeryellow}{rgb}{0.8, 0.8, 0.0} 
\definecolor{darkerorange}{rgb}{1.0, 0.6, 0.0}
\definecolor{darkerred}{rgb}{2.0, 0.0, 0.0}

\usepackage{threeparttable}

\makenomenclature

\renewcommand{\nompreamble}{The below list presents several symbols that will be used within the body of the paper.}

\usepackage{etoolbox}
\renewcommand\nomgroup[1]{%
  \item[\bfseries
  \ifstrequal{#1}{M}{HDRM Parameters}{%
  \ifstrequal{#1}{O}{Robust Subsystem-Based Adaptive Control Parameters}{%
  \ifstrequal{#1}{E}{Electromechanical Linear Actuator Parameters}{%
  \ifstrequal{#1}{O}{Other symbols}{}}}}%
]}

\usepackage{siunitx}
\usepackage{amsmath,amssymb,amsfonts}
\usepackage{caption} 
\usepackage{array} 
\usepackage{xcolor} 
\usepackage{setspace} 
\begin{document}

\title{Robustness-Guaranteed Observer-Based Control Strategy with Modularity for Cleantech EMLA-Driven Heavy-Duty Robotic Manipulator}

\author{Mehdi Heydari Shahna, Mohammad Bahari, and Jouni Mattila
\thanks{Funding for this research was provided by the Business Finland partnership project Future All-Electric Rough Terrain Autonomous Mobile Manipulators (Grant No. 2334/31/2022).}
\thanks{M. H. Shahna, M. Bahari, and J. Mattila are with the Faculty of Engineering and Natural Sciences, Tampere University, Finland. \\(e-mail:
mehdi.heydarishahna@tuni.fi; mohammad.bahari@tuni.fi; and jouni.mattila@tuni.fi).
        }%
}

\maketitle

\begin{abstract}
This paper introduces an innovative observer-based modular control strategy in a class of $n_a$-degree-of-freedom (DoF) fully electrified heavy-duty robotic manipulators (HDRMs) to (1) guarantee robustness in the presence of uncertainties and disturbances, (2) address the complexities arising from several interacting mechanisms, (3) ensure uniformly exponential stability, and (4) enhance overall control performance. To begin, the dynamic model of HDRM actuation systems, which exploits the synergy between cleantech electromechanical linear actuators (EMLAs) and permanent magnet synchronous motors (PMSMs), is investigated. In addition, the reference trajectories of each joint are computed based on direct collocation with B-spline curves to extract the key kinematic and dynamic quantities of HDRMs. To guarantee robust tracking of the computed trajectories by the actual motion states, a novel control methodology, called robust subsystem-based adaptive (RSBA) control, is enhanced through an adaptive state observer. The RSBA control addresses inaccuracies inherent in motion, including modeling errors, non-triangular uncertainties, and both torque and voltage disturbances, to which the EMLA-driven HDRM is susceptible. Furthermore, this approach is presented in a unified generic equation format for all subsystems to mitigate the complexities of the overall control system. By applying the RSBA architecture, the uniformly exponential stability of the EMLA-driven HDRM is proven based on the Lyapunov stability theory. The proposed RSBA control performance is validated through simulations and experiments of the scrutinized PMSM-powered EMLA-actuated mechanisms.
\end{abstract}

\def\abstractname{Note to Practitioners}
\begin{abstract}
Following strict global regulations, such as the 2015 Paris Agreement, there has been significant attention paid to the electrification trend. In this regard, the advancement of zero-emission electromechanical linear actuator technology has played a substantial role in developing fully electrified HDRMs. However, these systems are highly nonlinear and complex, comprising several interacting components, such as electric motors, reduction gearboxes, screw mechanisms, and load-bearing structures. Each of these components is prone to adverse effects arising from inaccuracies in modeling equations, sensor readings, and torque or voltage disturbances. As a result, achieving high-performance control presents significant challenges for engineers and necessitates computationally intensive approaches in practice. This paper presents a subsystem-based approach, enhanced by a robust state observer, to (1) mitigate the impact of uncertainties and disturbances substantially, (2) alleviate the computational burden and complexity of the targeted system, (3) prove mathematical stability, and (4) offer highly accurate and fast tracking performance. The proposed approach employs the dynamic motion of the studied EMLA-actuated HDRM, decomposing it into distinct subsystems and introducing a unified generic equation control for all subsystems. This modularity feature paves the way for researchers to extend the proposed approach to address other intricate applications.
\end{abstract}


\nomenclature[M]{\(t_M\)}{Final time (\SI{}{\second}) if $t_0 = 0$}
\nomenclature[M]{\(n_a\)}{Manipulator number of degrees of freedom}
\nomenclature[M]{$\bf{q}(t,\bf{c})$}{Configuration vector}
\nomenclature[M]{$\bf{q}_{LB}$}{Lower bound configuration of joints (\SI{}{\radian})}
\nomenclature[M]{$\bf{q}_{UB}$}{Upper bound configuration of joints (\SI{}{\radian})}
\nomenclature[M]{$\bf{q}_{E}$}{Configuration vector of joints at ending point (\SI{}{\radian})}
\nomenclature[M]{$\bf{q}_{S}$}{Configuration vector of joints at starting point (\SI{}{\radian})}
\nomenclature[M]{$\bf{v}_{UB}$}{Upper bound velocity vector of joints (\SI{}{\meter\per\second})}
\nomenclature[M]{$\bf{v}_{LB}$}{Lower bound velocity vector
 of joints (\SI{}{\meter\per\second})}
\nomenclature[M]{$\bf{v}_{E}$}{Velocity vector of joints at ending point (\SI{}{\radian\per\second})}
\nomenclature[M]{$\bf{v}_{S}$}{Velocity vector of joints at starting point (\SI{}{\radian\per\second})}
\nomenclature[M]{$\bm{\dot{q}}$, $\bm{\ddot{q}}$}{First and second time derivatives of configuration vector}
\nomenclature[M]{\(\bf{B}(t)\)}{Basis function of the B-spline}
\nomenclature[M]{$\bm{\dot{B}}(t)$, $\bm{\ddot{B}}(t)$}{First and second time derivatives of basis function of the B-spline}
\nomenclature[M]{$\bf{f}_{LB}$}{Lower bound force vector of joints ($\SI{}{\kilo\newton}$)}
\nomenclature[M]{$\bf{f}_{UB}$}{Upper bound force vector of joints (\SI{}{\kilo\newton})}
\nomenclature[M]{\(t_{M_{m}}\)}{Maximum allowed time to accomplish the task ($\SI{}{\second}$)}
\nomenclature[M]{\(\bf{c}\)}{Control points}
\nomenclature[M]{\(\mathcal{X}\)}{Manipulator robot workspace}
\nomenclature[M]{\(\mathbf{p}_{0}^{\text{EE}}\)}{Manipulator end-effector position vector}
\nomenclature[M]{\({f}_{Lift}\)}{Force in lift joint of the manipulator (\SI{}{\kilo\newton})}
\nomenclature[M]{\({f}_{Tilt}\)}{Force in tilt joint of the manipulator (\SI{}{\kilo\newton})}
\nomenclature[M]{\({f}_{Tel}\)}{Force in telescope joint of the manipulator (\SI{}{\kilo\newton})}
\nomenclature[M]{\(\bf{v}_{Lift}\)}{Velocity in lift joint of the manipulator (\SI{}{\meter\per\second})}
\nomenclature[M]{\(\bf{v}_{Tilt}\)}{Velocity in tilt joint of the manipulator (\SI{}{\meter \per \second})}
\nomenclature[M]{\(\bf{v}_{Tel}\)}{Velocity in telescope joint of the manipulator (\SI{}{\meter\per\second})}
\nomenclature[M]{\(\bf{x}_{Lift}\)}{Position in lift joint of the manipulator (\SI{}{\meter\per\second})}
\nomenclature[M]{\(\bf{x}_{Tilt}\)}{Position in tilt joint of the manipulator (\SI{}{\meter\per\second})}
\nomenclature[M]{\(\bf{x}_{Tel}\)}{Position in telescope joint of the manipulator (\SI{}{\meter\per\second})}
\nomenclature[M]{\(\bf{a}_{Lift}\)}{Acceleration in lift joint of the manipulator (\SI{}{\meter\per\second^2})}
\nomenclature[M]{\(\bf{a}_{Tilt}\)}{Acceleration in tilt joint of the manipulator (\SI{}{\meter\per\second^2})}
\nomenclature[M]{\(\bf{a}_{Tel}\)}{Acceleration in telescope joint of the manipulator (\SI{}{\meter\per\second^2})}
\nomenclature[E]{ADC}{Analog-to-digital converter}
\nomenclature[E]{$i_N$}{Rated current of electric motor (\SI{}{\ampere})}
\nomenclature[E]{$u_N$}{Rated voltage of electric motor (\SI{}{\volt})}
\nomenclature[E]{$P_N$}{Rated power of electric motor (\SI{}{\kilo\watt})}
\nomenclature[E]{$F_{c0}$}{Continuous force at zero speed (\SI{}{\kilo\newton})}
\nomenclature[E]{$F_{c}$}{Continuous force at maximum speed (\SI{}{\kilo\newton})}
\nomenclature[E]{$F_{p0}$}{Peak force at zero speed (\SI{}{\kilo\newton})}
\nomenclature[E]{$F_{p}$}{Peak force at maximum speed (\SI{}{\kilo\newton})}
\nomenclature[E]{$v_{max}$}{Maximum linear velocity (\SI{}{\meter\per\second})}
\nomenclature[E]{$a_{max}$}{Maximum linear acceleration (\SI{}{\meter\per\second^2})}
\nomenclature[E]{$n_{N}$}{Rated speed of the electric motor (\SI{}{rpm})}
\nomenclature[E]{$i_{a,b,c}$}{Three-phase motor currents (\SI{}{\ampere})}
\nomenclature[E]{$u_{a,b,c}$}{Three-phase motor voltages (\SI{}{\volt})}
\nomenclature[E]{$s_1 \sim s_6$}{Switch commands of three-phase inverter}
\nomenclature[E]{$\omega_m$}{Electric motor angular velocity (\SI{}{\radian\per\second})}
\nomenclature[E]{\(\tau_m\)}{Electromagnetic torque of electric motor (\SI{}{\newton\cdot\meter})}
\nomenclature[E]{\(\tau_N\)}{Rated torque of electric motor (\SI{}{\newton\cdot\meter})}
\nomenclature[E]{\(\tau_{max}\)}{Maximum torque of electric motor (\SI{}{\newton\cdot\meter})}
\nomenclature[E]{\(L_N\)}{Rated inductance of electric motor (\SI{}{\milli\henry})}
\nomenclature[E]{$\omega_{GB}$}{Gearbox angular velocity (\SI{}{\radian\per\second})}
\nomenclature[E]{\(\tau_{GB}\)}{Gearbox torque (\SI{}{\newton\cdot\meter})}
\nomenclature[E]{\(F_L\)}{Load force (\SI{}{\newton})}
\nomenclature[E]{\({x}_{L}\)}{Linear position of screw (\SI{}{\meter})}
\nomenclature[E]{\(\dot{x}_{L}\)}{Linear velocity of screw (\SI{}{\meter\per\second})}
\nomenclature[E]{\(\ddot{x}_{L}\)}{Linear acceleration of screw (\SI{}{\meter\per\second\squared})}
\nomenclature[E]{\(\Phi_{PM}\)}{Permanent magnet flux linkage (\SI{}{\weber})}
\nomenclature[E]{\(R_{s}\)}{Electric motor stator resistance (\si{\ohm})}
\nomenclature[E]{\(N_{p}\)}{Number of motor pole pairs}
\nomenclature[E]{\(L_{d}\)}{Electric motor inductance in d-axis (\SI{}{\milli\henry})}
\nomenclature[E]{\(L_{q}\)}{Electric motor inductance in q-axis (\SI{}{\milli\henry})}
\nomenclature[E]{\(\rho\)}{Inverse gearbox ratio}
\nomenclature[E]{\(x(t)\)}{State vector of an EMLA mechanism}
\nomenclature[E]{\(u(t)\)}{Input vector of an EMLA mechanism}
\nomenclature[E]{\(l\)}{Screw lead (\SI{}{\meter})}
\nomenclature[E]{\(d\)}{Screw diameter (\SI{}{\meter})}
\nomenclature[E]{\(\eta_{GB}\)}{Gearbox efficiency}
\nomenclature[E]{\(F_{L}\)}{Applied load force of EMLA (\SI{}{\newton})}
\nomenclature[E]{\(k_{bearing}\)}{Thrust bearing stiffness (\SI{}{\newton\per\micro\meter})}
\nomenclature[E]{\(k_{screw}\)}{Screw (compression) stiffness (\SI{}{\newton\per\micro\meter})}
\nomenclature[E]{\(k_{nut}\)}{Ball nut stiffness (\SI{}{\newton\per\micro\meter})}
\nomenclature[E]{\(k_{tube}\)}{Thrust tube stiffness (\SI{}{\newton\per\micro\meter})}
\nomenclature[E]{\(k_{L}\)}{Stiffness of screw mechanism linear units (\SI{}{\newton\per\micro\meter})}
\nomenclature[E]{\(k_{\tau 1}\)}{Stiffness of motor shaft coupling (\SI{}{\newton\cdot\meter\per\radian})}
\nomenclature[E]{\(k_{\tau 2}\)}{Stiffness of gearbox connection (\SI{}{\newton\cdot\meter\per\radian})}
\nomenclature[E]{\(k_{\tau 3}\)}{Stiffness of screw mechanism (\SI{}{\newton\cdot\meter\per\radian})}

\nomenclature[E]{$J_{m}$}{Electric motor inertia (\SI{}{\kilogram\cdot\meter\squared})}
\nomenclature[E]{$J_{c}$}{Coupling inertia (\SI{}{\kilogram\cdot\meter\squared})}
\nomenclature[E]{$J_{GB}$}{Gearbox mechanism inertia (\SI{}{\kilogram\cdot\meter\squared})}
\nomenclature[E]{$M_{BS}$}{Screw mechanism mass (\SI{}{\kilogram})}
\nomenclature[E]{$B_{m}$}{Electric motor viscous
 friction (\SI{}{\newton\cdot\meter\cdot\second\per\radian})}
\nomenclature[E]{$B_{BS}$}{Screw mechanism viscous
 friction (\SI{}{\newton\cdot\second\per\meter})}
\nomenclature[E]{\(A_{eq}\)}{Equivalent mass at the load side (\SI{}{\kilo\gram})}
\nomenclature[E]{\(B_{eq}\)}{Equivalent damping of EMLA (\SI{}{\newton\cdot\second\per\meter})}
\nomenclature[E]{\(C_{eq}\)}{Equivalent stiffness of EMLA (\SI{}{\newton\per\meter})}
\nomenclature[E]{\(D_{eq}\)}{Load to torque conversion ratio}
\nomenclature[O]{${\hat{\eta}}$}{Observer adaptation law}
\nomenclature[O]{${{\eta^*}}$}{Unknown parameter for the observer adaptation law}
\nomenclature[O]{${\bar{\eta}}$}{Error of the observer adaptation law}
\nomenclature[O]{${m}$}{Positive and continuous function}
\nomenclature[O]{${\ell}$}{Positive constant}
\nomenclature[O]{${H}$}{Positive and continuous function}
\nomenclature[O]{${y}$}{System output of motion information}
\nomenclature[O]{${\hat{y}}$}{Estimated sensor output}
\nomenclature[O]{${\bar{y}}$}{Error of the estimated sensor output}
\nomenclature[O]{$u_i$}{Control signals}
\nomenclature[O]{$g_i$}{Known functions of the EMLA modeling system}
\nomenclature[O]{$F_i$}{Unknown non-triangular uncertainties}
\nomenclature[O]{$F^*_i$}{Derivative of the virtual position control $a_1$}
\nomenclature[O]{$\bar{F}_i$}{Tracking uncertainties comprising $F_i$ and $F^*_i$}
\nomenclature[O]{$d_i$}{Disturbances with uncertain magnitudes and
timings}
\nomenclature[O]{$i$}{Number of subsystems in each EMLA ranging from $1$ to $4$}
\nomenclature[O]{$k$}{EMLA-actuated joint number ranging from $1$ to $n_a$}
\nomenclature[O]{$\bm{A}$}{State matrix in the state-space representation}
\nomenclature[O]{$\bm{B}$}{Input matrix in the state-space representation}
\nomenclature[O]{$\bm{C}$}{Output matrix in the state-space representation}
\nomenclature[O]{$\bm{K}$}{Unknown vector specifying nonlinearities}
\nomenclature[O]{$\bm{g}$}{Known vector specifying modeling parameters}
\nomenclature[O]{$\bm{u}$}{Control input vector in the state representation}
\nomenclature[O]{$f$}{Finite, positive, and continuous function}
\nomenclature[O]{$\bm{x}$}{Vector of the EMLA states ($x_1$, $x_2$, $x_3$, and $x_4$)}
\nomenclature[O]{$x_1$}{Linear position state of the EMLA (m)}
\nomenclature[O]{$x_2$}{Linear velocity state of the EMLA (m/s)}
\nomenclature[O]{$x_3$}{q-axis current state of the PMSM (A)}
\nomenclature[O]{$x_4$}{d-axis current state of the PMSM (A)}
\nomenclature[O]{$\bm{\bar{x}}$}{Vector of motion states (position ${x}_1$ and velocity ${x}_2$)}
\nomenclature[O]{$\bm{x_d}$}{Reference vector of the system in tracking tasks}
\nomenclature[O]{$\bm{\hat{x}}$}{Vector of the estimated position and velocity states}
\nomenclature[O]{$\bm{{x}_{eo}}$}{Error vector of the estimated states}
\nomenclature[O]{${P}_{i}$}{Transformation of space-state into tracking form}
\nomenclature[O]{${a}_{1}$}{Virtual position control signal}
\nomenclature[O]{${A}_{i}$}{Coefficient of the control signal}
\nomenclature[O]{$\beta_i$}{Positive constant}
\nomenclature[O]{$\zeta_i$}{Positive constant}
\nomenclature[O]{$\delta_i$}{Positive constant}
\nomenclature[O]{$\sigma_i$}{Positive constant}
\nomenclature[O]{$\hat{\theta}_i$}{Control adaptation law}
\nomenclature[O]{$\tilde{{\theta}}_i$}{Error of the control adaptation law}
\nomenclature[O]{${{\theta}}^*_i$}{Unknown parameter for the control adaptation law}
\nomenclature[O]{$\bm{Q}$}{Positive definite matrix}
\nomenclature[O]{$\bm{p}$}{Positive definite matrix}
\nomenclature[O]{$\bm{\bar{A}}$}{Hurwitz matrix}
\nomenclature[O]{$\bm{\alpha}$}{Feedback gain matrix for the observer}
\nomenclature[O]{$d_{max(i)}$}{Magnitude bound of disturbance in $i$th subsystem}
\nomenclature[O]{$\Omega_i$}{Magnitude bound of $\dot{x}_{id}$}
\nomenclature[O]{$x_{id}$}{Reference trajectory of the state $x_{i}$}
\nomenclature[O]{$\mu_i$}{Unknown positive constant}
\nomenclature[O]{$\nu_i$}{Unknown positive constant}
\nomenclature[O]{$\psi_i$}{Unknown positive constant}
\nomenclature[O]{$p_{min}$}{Minimum eigenvalue of $\bm{p}$}
\nomenclature[O]{$\phi_{i}$}{Unknown positive parameter}
\nomenclature[O]{$\iota$}{Unknown positive parameter}
\printnomenclature

\section{Introduction}
\subsection{Background Context}
\label{intro}

\IEEEPARstart{D}{riven} by the urgent need to mitigate climate change, automation systems are undergoing a rapid evolution. The undeniable impact of greenhouse gas emissions has spurred international agreements, such as the 2015 Paris Agreement for CO\textsubscript{2} reduction, underscoring the importance of various industries transitioning to clean energy \cite{agreement2015paris}. The push for decarbonization, alongside battery and charging infrastructure advancements, profoundly impacts various sectors, exemplified by the surge in zero-emission battery electric vehicle (BEV) development \cite{bischoff2010strategic,daily2017self,badue2021self}. Extending the concept of BEVs into the working machinery, this industry is witnessing the rise of a new class of electric vehicles, called electrified mobile manipulators (MMs). Traditionally, electro-hydraulic actuators (EHA) have been commonly used as the actuation mechanism for MMs, yet they pose challenges, such as energy inefficiency and leakage \cite{cao2011overview}. However, the development of electromechanical linear actuators (EMLAs) has significantly contributed to the electrification of MMs, offering enhanced mobility, efficiency, and safety, while decreasing maintenance requirements \cite{li2016review}. These electrified MMs are mobile platforms on which electrified manipulators are mounted, expanding their application across diverse sectors, including manufacturing, logistics, agriculture, and even search and rescue operations \cite{habibnejad2021designing}.

\subsubsection{Introduction to the EMLA Mechanism}
\begin{figure}[h] 
    \centering
    \scalebox{0.8}
    {\includegraphics[trim={0.1cm 0.1cm 0.1cm
    0.1cm},clip,width=\columnwidth]{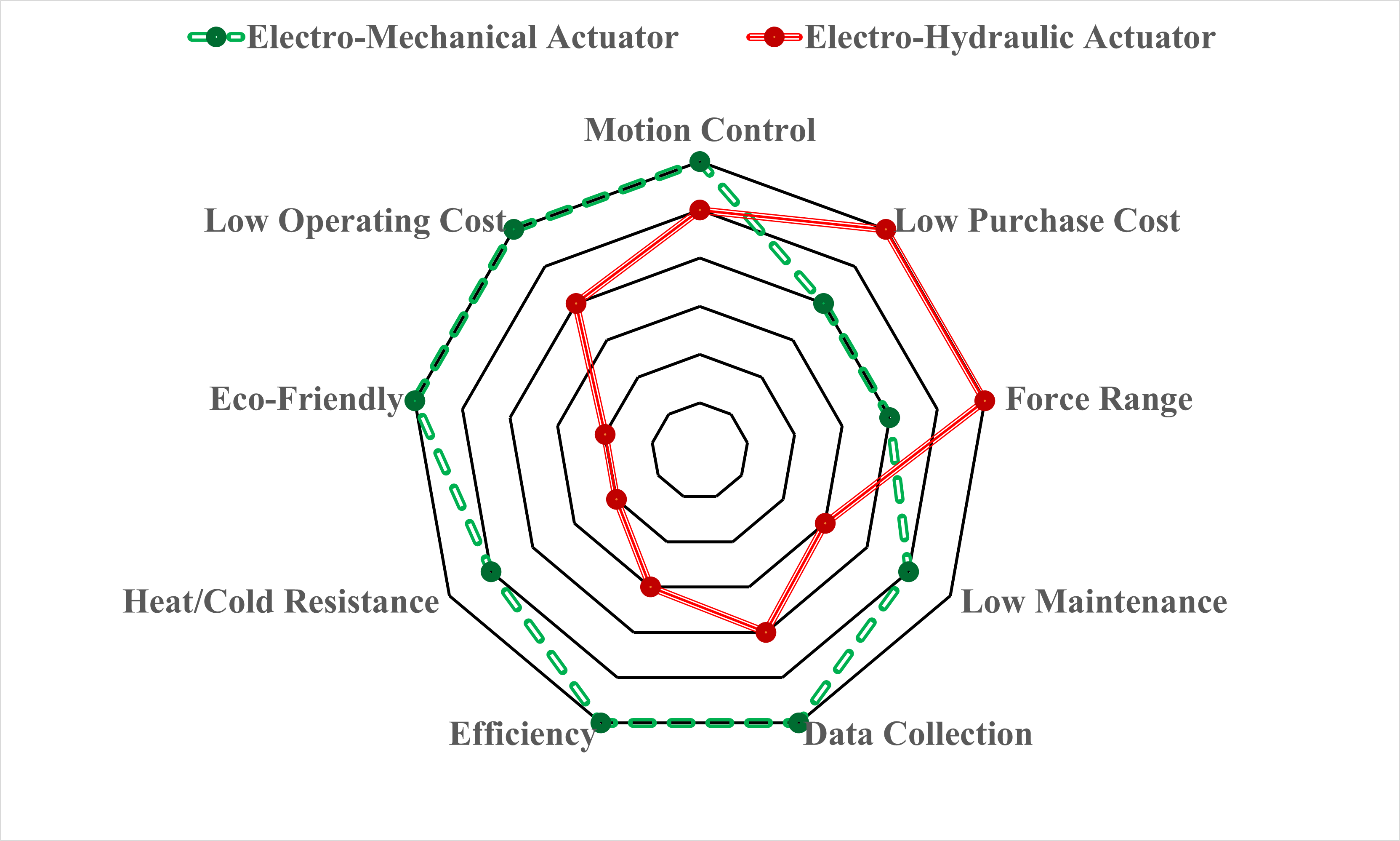}}
    \caption{Illustration of comparative key performance factors influencing EHAs and EMLAs}
    \label{fig:Comparing_spider}
\end{figure}

EMLAs are contributing to a growing trend of electrification in MMs, offering an alternative to traditional EHAs with several advantages. For a better understanding, the graphical representation depicted in Fig. \ref{fig:Comparing_spider} provides a comparative visualization of factors influencing the performance of both EHAs and EMLAs, encompassing such considerations as efficiency, force range, and motion controllability. By converting electrical energy into mechanical linear motion more efficiently, EMLAs significantly reduce overall energy consumption, a critical consideration for MMs \cite{10199841}. Moreover, their streamlined design, with fewer moving parts, results in reduced maintenance requirements compared to EHAs, mitigating issues such as oil leakages, which are commonly associated with traditional hydraulic systems \cite{fleming2021electrification}. The composition of EMLAs typically includes an electric motor, a reduction gearbox, a screw mechanism, and load-bearing components \cite{boldea1999linear,knabe2014design,redekar2022functionality}. Within this structure, the motor serves as the source of rotational power, while the lead/ball/roller screw converts this rotational movement into linear motion \cite{Väisänen2023}. One of the defining features of EMLAs lies in their integration with sensors and sophisticated electronic systems, enabling precise control of EMLAs that improves MMs' performance in tasks requiring accuracy.
\cite{nagel2017actuation,lequesne2015automotive}. What is more, permanent magnet synchronous motors (PMSMs)
 are favored for integration within EMLAs due to their exceptional efficiency, noteworthy torque density \cite{hassan2012efficiency}, and low cogging torque \cite{2016cogging}. Overall, the aforementioned features make EMLAs a well-suited choice for the energy-conscious design of MMs due to their limited battery storage \cite{li2003predictive,hagras2019nonlinear}.

\subsection{Literature Review and Control Challenges}
\label{motivations}
Based on our data collected from querying Google Scholar and other academic databases using the keywords ‘PMSM’ and ‘actuator,’ we noted with interest that over the past four years, the number of studies published on PMSM-powered actuators has been more than five times greater than that published over the 10 years from 2000 to 2010, $98\%$ of which emphasized ‘control.’
This high growth in academic and industrial interest invites further investigation into the factors driving this increased research focus, such as technological advancements, greater funding availability, and a heightened industrial and academic focus on electrification technologies. Basically, the PMSMS-driven EMLA control framework is divided into two main interconnected levels. The outer level is responsible for the motion dynamics of the EMLA and calculating the required torque/current of the motor to ensure the linear actuator's position or velocity aligns with the reference values. The inner level converts electrical energy to mechanical energy in the servomechanism and adjusts the sufficient motor voltage signal to generate the required torque/current motor, as determined by the outer level \cite{heydari2024robust, hang2022simplified}. In response to the increasing demand for high-performance control solutions to address myriad application-based tasks across diverse PMSM-driven actuator systems, various control strategies have been developed and implemented in various industries. Assuming the use of proportional-integral (PI) control at the inner level, researchers have commonly overlooked the nonlinearity between the motor voltage and current at the inner level and focused solely on outer-level control. For instance, Ref. \cite{wang2023pid} employed a genetic algorithm to optimize the coefficients of the traditional proportional–integral–derivative (PID) control solely for the outer level of the studied PMSM-powered actuator. Similarly, Ref. \cite{mahdi2020fuzzy} addressed the robustness of the sole outer-level control of the application by developing an adaptive PID-type sliding mode control. Following the discussion on robustness, Ref. \cite{wang2022nonlinear} proposed a robust control method that addresses the compensation for active nonlinear uncertainties in motion dynamic control, while similarly utilizing traditional PI control at the inner level. In addition, it assumed that the angular velocity of the PMSM is exactly equal to the derivative of the angular position, without considering sensor inaccuracies. Ref. \cite{zhao2024online} advanced the field by addressing both inner and outer level control for electric motors. It introduced a neural network voltage compensator to address aperiodic harmonic disturbances while guaranteeing Lyapunov stability. However, PMSM-powered EMLAs are coupled using complex multi-stage gearboxes and ball screw systems, introducing more intense nonlinearity at both the inner and outer system levels. Accordingly, sensor inaccuracies, along with disturbances in torque and voltage in EMLAs, are further complicated by the conversion of rotational motion to linear motion, necessitating advances in robustness and stability. In addition, the PMSM-powered EMLA under study is intended to actuate each joint of a $n_a$-degree-of-freedom (DoF) fully electrified heavy-duty robotic manipulator (HDRM), which further intensifies the uncertainties and nonlinearities. In other words, the electrified HDRM is equipped with multiple PMSM-powered EMLAs, each responsible for driving a high-load-bearing joint or segment of the manipulator. These operate in environments characterized by uncertainties and unknown forces from diverse sources, which affect performance in Cartesian space. Consequently, designing such a control solution for these mechanisms poses significant challenges, the primary of which are as follows:
\subsubsection{Sensor Inaccuracies}
Position sensors are integral components of motion control systems and are commonly employed in inverter-driven PMSMs \cite{bahari2019new}. However, these sensors are vulnerable to environmental factors, such as temperature, workplace contamination, external magnetic fields, mechanical shocks, and humidity, which impact their accuracy and cause noises on the received signal \cite{9802805}. Hence, deriving the position signal to obtain the actual velocity value for the control feedback is impractical, as this intensifies the noise effects. Although the velocity can be estimated through the Euler approximation method, this technique exhibits limited resolution, particularly when applied in scenarios involving low-speed operations \cite{shin2011position}. In addition, the provision of velocity sensors for the multiple EMLA-actuated joints of the HDRM mechanism can significantly increase the total cost, as well as the vulnerability to environmental factors, while also reducing reliability.
\subsubsection{Existence of Non-Triangular Uncertainties}
In contrast to triangular uncertainty structures, as explored in \cite{xing2023dynamic,jiang2020cooperative,wang2021output,bernard2020adaptive,koivumaki2022subsystem}, non-triangular uncertainty structures suggest uncertainties that could potentially be influenced by all states within the control system. These structures require a thorough exploration of a broader range of dependencies and interactions among system states \cite{cai2016adaptive, cai2022decentralized}. In this context, modeling equations from both levels governing PMSM-powered EMLAs incorporates interacting terms dependent on all system states \cite{heydari2024robust}. Therefore, modeling the inaccuracies of such mechanisms could lead to uncertainties being structured into non-triangular forms. To avoid this limitation, numerous PMSM control studies rely on simplifying assumptions that may not be valid in all industrial applications, particularly HDRMs, which are actuated by several EMLAs. For instance, certain investigations, as exemplified by \cite{li2009adaptive} and \cite{zhang2023adaptive}, assume that the d-axis current of a PMSM precisely equals zero ($i_d=0$). This assumption allows them to achieve a triangular structure in dynamic equations by eliminating terms associated with $i_d$. However, if $i_d$ deviates from zero, these terms bring non-triangular uncertainties that impact control performance.
\subsubsection{Influence of Time-Varying Disturbances}
The control system of the inner level system, inverter-driven PMSM, is vulnerable to external disturbances, consequently degrading the overall control efficacy of EMLA-actuated joints in the task execution of the HDRM. Distinguished disturbances are itemized as follows:
 
\begin{itemize}
    \item Torque disturbance: This type of disturbance originates from flux harmonics within the PMSM, resulting from the non-sinusoidal flux density distribution in the airgap \cite{yang2016disturbance}.
    \item Voltage disturbance: This disturbance arises from fluctuations or anomalies in the electrical supply and the switching action of the inverter, employed to convert DC to AC power for the PMSM \cite{de2017robust}.
\end{itemize}
\subsubsection{Management of System Interactions}
Formulating a control design to manage the complex dynamics of a PMSM-powered, EMLA-driven $n_a$-DoF HDRM mechanism constitutes a formidable task, as it consists of multiple coupling mechanical components. Concurrently, any alterations to the dynamics of the mechanism often necessitate a comprehensive redesign of the control system. Subsequently, effectively managing the interactions between various components to ensure the stability and high performance of the overall system, while mitigating the burden of control system redesigns in response to dynamic changes, poses a significant challenge.

\subsection{Paper Contributions and Organization}
In Section \ref{motivations}, we highlight the control challenges of HDRMs due to the incorporation of multiple joints actuated by EMLAs. To address these intricate issues effectively, this paper proposes a novel strategy termed robust subsystem-based adaptive (RSBA) control. 
The key findings of this research are as follows: 
\begin{itemize}
    \item This study establishes a comprehensive, dynamic model of a PMSM-powered EMLA to capture the intricacies of the mechanism's motion. Afterward, reference trajectories for individual joints of the HDRM are determined through direct collocation with B-spline curves to define a control task. This modeling framework ultimately lays the groundwork for designing effective control.
    \item Contrary to the motion state observers for PMSM-powered applications presented in Refs. \cite{zuo2023review}, \cite{lascu2020pll}, and \cite{li2021generalized}, which estimated the true values of the angular position and velocity of the motor, this paper proposes a robust adaptive state observer to estimate accurately the linear position and velocity states at the load side of EMLAs, which are coupled via a complex multi-stage gearbox and ball screw systems with the motor, thus compensating for all possible measurement inaccuracies in different stages. In addition, the parameter settings of the proposed observer are investigated, focusing on balancing the trade-off between robustness and responsiveness.
    \item The assured robustness of the proposed control guarantees its effectiveness in addressing non-triangular uncertainties and both torque and voltage disturbances in the EMLA-actuated HDRM.
    \item The proposed approach demonstrates modularity, enabling the design of control for HDRM joints using a single generic equation form. This ensures that joint dynamics' modifications, additions, or removals do not affect the control laws governing the remaining system. This modular characteristic offers the opportunity to extend the proposed approach further to address diverse and complex applications beyond its initial scope.
    \item The term ``stability connector," which was introduced in \cite{koivumaki2022subsystem}, is designed to capture triangular dynamic interactions between subsystems that effectively offset the instability of the system to achieve asymptotic stability. Expanding upon this concept, we advance its application to EMLA-actuated HDRMs, accommodating non-triangular uncertainties and leading to exponentially stable analysis of the entire system.
    \item The proposed control strategy offers accurate and fast convergence of states toward the reference trajectories while reducing the torque effort. For the sake of comparison, we analyzed the proposed control performance with recent studies \cite{liu2023command,zhang2023adaptive} through simulations and experiments. 
\end{itemize}

The remainder of the paper is structured as follows:
\subsubsection{Section \ref{system_modeling}}
This section provides an in-depth exploration of modeling an EMLA-driven HDRM, establishing the foundation for subsequent analysis and control: 
\begin{itemize}
    \item Part \ref{EMLA_motion} outlines the modeling of motion dynamics for an EMLA to convert electrical into mechanical power.
    \item Part \ref{manipulator_motion} investigates a trajectory generation approach for the HDRM, defining a control task through the utilization of the direct collocation with the B-spline curves method.
\end{itemize}
\subsubsection{Section \ref{paper}}
This section elaborates on the design of the RSBA control. This strategy decomposes the control architecture into distinct subsystems for both inner and outer levels and introduces a unified generic equation control applicable to all subsystems.
\begin{itemize}
    \item Part \ref{observer} introduces an adaptive state observer to estimate accurately the true linear motion states of each PMSM-driven EMLA at the load side. In addition, a step-by-step summary of the proposed observer is provided in \textbf{Algorithm 1}.
    \item Part \ref{control} outlines the design of the RSBA control, which is compatible with the observer estimator. It addresses both non-triangular uncertainties and torque and voltage disturbances to track the defined control task effectively. As well, \textbf{Algorithm 2} is included to provide a comprehensive overview of the design.
    \item Part \ref{stability} illustrates how incorporating the entire control strategy into the modeled PMSM-powered EMLA-driven $n_a$-DoF HDRM mechanism ensures exponential stability and robustness.
\end{itemize}

\subsubsection{Section \ref{simulation}}
In this section, simulation results and comprehensive analysis are provided:
\begin{itemize}
\item Part \ref{manipulator_simulation} conducts a desired task performed on a 3-DoF EMLA-actuated HDRM with a $470$-kg payload at the end-effector in the Cartesian space. Then, reference trajectories of each PMSM-powered EMLA-actuated joint are computed based on direct collocation with B-spline curves.
\item Part \ref{c-simulation} presents control simulations of each PMSM-powered EMLA-actuated joint to track reference trajectories defined in \ref{manipulator_simulation}.
\end{itemize}

\subsubsection{Section \ref{experiment}}
In this section, two experimental scenarios involving a prototype PMSM-powered EMLA mechanism are presented to evaluate the RSBA's robustness and responsiveness in practice. This PMSM-powered EMLA is intended to actuate one of the joints in an upcoming electrified HDRM.
\begin{itemize}
\item Part \ref{exp1} presents the EMLA prototype performance to track a desired linear position at an upper-moderate velocity of $0.026$ m/s under a gradually increasing load ranging from $7$ kN to $70$ kN.
\item Part \ref{exp2} presents the EMLA prototype performance to track the same desired linear position but at a nominally high velocity of $0.03$ m/s under the heaviest load of $70$ kN. 
\end{itemize}

\section{System Modeling of EMLA-Equipped HDRM}
\label{system_modeling}
\subsection{EMLA Motion Dynamic Modeling}
\label{EMLA_motion}
\begin{figure}[h] 
    \centering
    \scalebox{0.8}
    {\includegraphics[trim={0cm 0.0cm 0.0cm
    0cm},clip,width=\columnwidth]{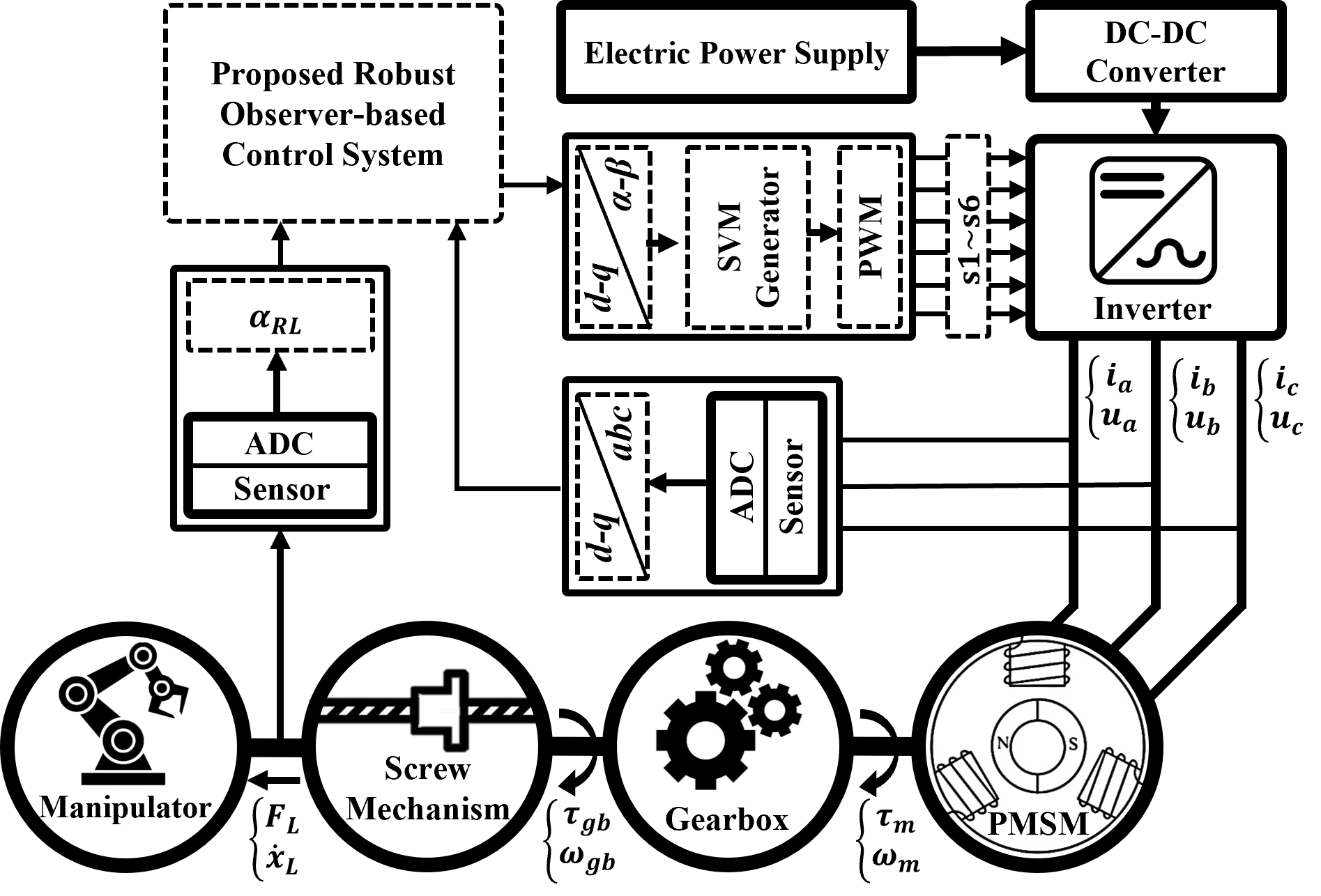}}
    \caption{1-DoF EMLA mechanism and controller schematic}
    \label{fig:1DOF}
\end{figure}

In this section, we investigate an EMLA model equipped with a PMSM with nonlinearities, aiming to transform rotational motion into linear motion. The fundamental elements of an EMLA include a cylinder housing, attachments, a thrust tube, a ball screw or roller screw, a gearbox, and a motor. The power transmission sequence of the EMLA commences with the motor, which generates torque and rotational speed through electrical power. The gearbox is responsible for decreasing the rotational speed while elevating the torque to the desired magnitude, whereas the screw shaft and nut assembly effectively convert the rotational motion into linear motion and can actuate the manipulator joints (Fig. \ref{fig:1DOF}). The velocity and force of this linear motion maintain a proportional relationship with the lead of the screw mechanism. 

The control procedure is executed within direct and quadrature axes in the \textit{d-q} axis, followed by the implementation of Park transformations ($\bf{T}$) to transform a three-phase (\textit{3-Ph}) signal, as described in (\ref{equation:Park_Transformation}) and (\ref{equation:abc_to_dq}):
\begin{equation}
\hspace{-0.1cm} \bf{T} = \left[\begin{array}{ccc}
cos(\omega t)  & cos(\omega t - \frac{2 \pi}{3})  & cos(\omega t + \frac{2 \pi}{3}) \\
-sin(\omega t) & -sin(\omega t - \frac{2 \pi}{3}) & -sin(\omega t + \frac{2 \pi}{3}) \\
\frac{1}{2}    & \frac{1}{2}                      & \frac{1}{2}
\end{array}\right]
\label{equation:Park_Transformation}
\end{equation}
\begin{equation}
\left[\begin{array}{l}
u_d \\
u_q \\
u_0
\end{array}\right]
=\frac{2 \bf{T}}{3} 
\left[\begin{array}{l}
u_{\mathrm{a}} \\
u_{\mathrm{b}} \\
u_{\mathrm{c}}
\end{array}\right]
\label{equation:abc_to_dq}
\end{equation}

The electrical characterization of the PMSM in the \textit{d-q} reference frame is outlined in \eqref{equation:PMSM} \cite{krishnan2017permanent}:
\begin{equation}
\hspace{-0.3cm}
\left\{
\begin{alignedat}{3}
&\frac{d i_q}{d t}&&=\frac{1}{L_q}\left(u_q-R_s i_q- N_p \omega_m i_d L_d - N_p \omega_m \Phi_{PM}\right) \\
&\frac{d i_d}{d t}&&=\frac{1}{L_d}\left(u_d-R_s i_d+ N_p \omega_m i_q L_q\right) \\
&\tau_m&&=\frac{3}{2} N_p \left[i_q\left(i_d L_d+\Phi_{PM}\right)-i_d i_q L_q\right]
\end{alignedat}
\right.
\label{equation:PMSM}
\end{equation}

Regarding the mechanical components depicted in Fig. \ref{fig:1DOF}, the torque generated by the PMSM is transmitted to the gearbox and screw mechanism, respectively, to power the manipulator's joints. To simplify the rotary to linear motion conversion ratio, $\alpha_{RL}$ can be defined as in \eqref{equation:alpha}:
\begin{equation}
\alpha_{RL} = \frac{2\pi}{\rho l}
\label{equation:alpha}
\end{equation}
Applying the principles of Newton's law of motion, we can deduce the torque balance equations for the PMSM ($\tau_m$) and thereby achieve a comprehensive performance of an EMLA mechanism, as achieved in \eqref{equation:motor_torque}-\eqref{equation:MotionLaw}:
\begin{equation}
\tau_m = A_{eq} \ddot{x}_L + B_{eq} \dot{x}_L + C_{eq} {x}_L + D_{eq} F_{L}
\label{equation:motor_torque}
\end{equation}
\begin{equation}
\hspace{-0.5cm}\text{where}
\left\{
\begin{alignedat}{4}
&A_{eq}      &&= \alpha_{RL} \left(J_m+J_c+J_{G B}+\frac{1}{\alpha_{RL}^2} M_{BS}\right) \\
&B_{eq}      &&= \alpha_{RL} \left(B_m +\frac{1}{\alpha_{RL}^2} B_{BS}\right)\\
&C_{eq}      &&= \alpha_{RL}^2 \bigg(\frac{1}{k_{\tau 1}} + \frac{1}{k_{\tau 2}} + \frac{1}{{\rho}^2 k_{\tau 3}} + \frac{\alpha_{RL}^2}  {k_{L}}\bigg)^{-1}\\
&D_{eq}      &&= \frac{1}{\alpha_{RL} \eta_{G B}}
\end{alignedat}
\right.
\label{equation:MotionLaw}
\end{equation}
%

Meanwhile, the parameter $k_{L}$ denotes the collective linear stiffness encompassing the thrust bearing, ball screw, ball nut, and thrust tube components, and it can be calculated using (\ref{equation:linear_components_stiffness}):
\begin{equation}
k_L=\left(\frac{1}{k_{\text {bearing }}}+\frac{1}{k_{\text {screw }}}+\frac{1}{k_{\text {nut }}}+\frac{1}{k_{\text {tube }}}\right)^{-1}
\label{equation:linear_components_stiffness}
\end{equation}

The primary goal of control is to attain the targeted speed set points while ensuring $i_d$ is equal to zero for energy efficiency, while $u_d$ and $u_q$ stay within predefined operational boundaries. In the context of the studied EMLA, the state vector ${X(t)}\in\mathbb{R}^4$ and input vector ${U(t)}\in\mathbb{R}^3$ for the state-space model of the studied EMLA can be defined as in \eqref{equation:StateSpace}:
\begin{equation}
\left\{
\begin{alignedat}{2}
{x(t)} &= {[{x}_L (t)\;\;\dot{x}_L (t)\;\;i_q(t)\;\;i_d(t)}]^T\\
{u(t)} &= {[i^*_q (t)\;\;u_q (t)\;\;u_d (t)]^T}
\end{alignedat}
\right.
\label{equation:StateSpace}
\end{equation}

Ultimately, we can define the state space vector for the given case by considering $x_1 = {x}_L (t)$, $x_2 = \dot{x}_L (t)$, $x_3 = i_q(t)$, and $x_4 = i_d(t)$, as expressed in \eqref{equation:StateSpace3}:
\begin{equation}
\hspace{-0.3cm}\left\{\begin{alignedat}{5}
\dot{x}_1= &\hspace{0.3em} x_2\\
\dot{x}_2= &\hspace{0.3em} \frac{1}{A_{eq}}\left[\frac{3}{2} N_{p} \left(x_3 x_4 L_d+i^*_q \Phi_{PM}-x_3 x_4 L_q\right) \right. \\
&\quad-\left.B_{eq} x_2 - C_{eq} x_1 - D_{eq} F_L\right]\\
\dot{x}_3= &\hspace{0.3em} \frac{1}{L_q}\left[u_q - R_s x_3- N_{p} \alpha_{RL} x_2 \left( x_4 L_d + \Phi_{PM} \right) \right] \\
\dot{x}_4= &\hspace{0.3em} \frac{1}{L_d}\left[u_d - R_s x_4+ N_{p} \alpha_{RL} x_2 x_3 L_q\right]
\end{alignedat}\right.
\label{equation:StateSpace3}
\end{equation}

Thus, \eqref{equation:StateSpace3} implies that each PMSM-driven EMLA-actuated joint can have four states, consequently leading to four subsystems. By considering $x_{id}$ as the reference trajectory for the $i$th subsystem, the control signal of the second subsystem ($i_q^*$) is assigned to force the true linear velocity state ($x_{2}$) to track the velocity reference $x_{2d}$. Then, the $i_q^*$ signal will be the reference value ($x_{3d}=i_q^*$) for $x_3=i_q$ in the third subsystem. Eventually, the q-voltage ($u_q$) and d-voltage ($u_d$) signals are assigned to force real $x_3$ and $x_4=i_d$ states to track $i_q^*$ and $i_d^*=0$, respectively. Converging $i_d$ to zero removes the magnetic field along the $d$-axis, ensuring that the torque produced by the electric machine is unaffected by the rotor's position \cite{heydari2024robust}. That is, to maximize torque per ampere, $u_d$ the d-axis reference current is set to track zero. Based on these details, we can establish the relationship between the desired torque ($\tau_m^*$) and currents, derived from \eqref{equation:PMSM}, as follows:
\begin{equation}
\tau^*_m=\frac{3}{2} N_p i^*_q\Phi_{PM}, \hspace{0.2cm} i_d^*=0
\label{desired}
\end{equation}

Finally, the pulse width modulation block generated the corresponding rectangular waveforms used to control the power switches $s_1 {\sim} s_6$ of the inverter.\\
\indent \textbf{Remark 1} 
In contrast to \cite{li2009adaptive, zhang2023adaptive}, we do not assume that $i_d$ is exactly zero ($i_d=i^*_d=0$) but consider $i_d$ as the fourth state intended to track the reference $i^*_d=0$. As demonstrated in \eqref{equation:StateSpace3}, the second subsystem ($\dot{x}_2$) equation encompasses the preceding ($x_1$), present ($x_2$), and subsequent ($x_3$ and $x_4$) states. This pattern is mirrored in the third subsystem ($\dot{x}_3$) equation as well. Consequently, any inaccuracies in modeling each joint actuated by EMLA or sensor information may result in uncertainties with a non-triangular structure, relying on all states, in the whole dynamic of the manipulator. To achieve a high-performance HDRM with robust control whose joints are actuated by PMSM-powered EMLAs, addressing non-triangular complexities is necessary.

\subsection{Manipulator's Optimized Motion with B-spline Trajectory Generation}
\label{manipulator_motion}
To analyze the motion profile of the EMLAs, we employ a trajectory generation method based on direct collocation with B-spline curves, as described by \cite{lee2005newton}. This approach involves transcribing the problem into a finite-dimensional nonlinear programming
problem, which can then be solved efficiently using numerical optimization techniques.
We divide time into discrete intervals, defined by the set of points described in \eqref{equation:time}:
\begin{equation}
\mathcal{T} \triangleq\left\{t_0 \cdots t_k \cdots t_M\right\}
\label{equation:time} 
\end{equation}

The configuration vector of joints $\bf{q}$ belonging to $\mathbb{R}^{n_a}$, along with its first two time derivatives $\dot{\bf{q}}$ and $\ddot{\bf{q}}$, is parameterized by B-spline curves, as defined in {(\ref{equation:matrix_basis})}, where $M$ denotes the number of partitions and $t$ represents the collocation points. In this context, $n_a$ represents the number of DoFs, specifically denoting how many actuated joints as passive joints can be formulated as functions of the actuated ones. In addition, $\bf{c}$ signifies the control points, and $\bf{B}(t)$ corresponds to the basis functions of the B-spline \cite{paz2019practical}, along with its time derivatives, as described in \eqref{equation:matrix_basis}: 
\begin{equation}
\bf{q}(t,\bf{c}) = \bf{B}(t)\bf{c} 
\quad \,\, \dot{\bf{q}}(t,\bf{c}) = \dot{\bf{B}}(t)\bf{c} 
\quad \,\, \ddot{\bf{q}}(t,\bf{c}) = \ddot{\bf{B}}(t)\bf{c}
\label{equation:matrix_basis}
\end{equation}

The basis function $\bf{B}(t)$, mapped from the set of time values $\mathcal{T}$ to the interval $[0,1]$, is defined, and its first and second time derivatives, denoted as $\dot{\bf{B}}(t)$ and $\ddot{\bf{B}}(t)$, respectively, are computed. Applying the inverse dynamics procedure outlined in \cite{petrovic2022mathematical}, we can derive velocities ${\bf{v}_x}$ and forces ${\bf{f}_x}$ within the linear actuators. This involves employing a recursive Newton-Euler algorithm designed to handle closed kinematic chains {(\ref{equation:rnea})}:
\begin{equation}
    (\bf{v}_{\!L},\bf{f}_{\!L}) \ = \ \bf{RNEA}(\bf{q},\dot{\bf{q}},\ddot{\bf{q}})
    \label{equation:rnea}
\end{equation}

We frame the problem as minimizing the integrated sum of delivered power in the joints at time $t_k$ and computed using the output of \eqref{equation:rnea}. This optimizes the power consumption for the linear actuators, and it is formulated by the following quadratic multiplication form of velocity and force in \eqref{equation:costf2}, as well as the constraints of the problem, as defined in \eqref{equation:const2}:
\begin{eqnarray}
\underset{\bf{c}, \ t_M}{\operatorname{minimize}} & \hspace*{-0.8cm} & J(\bf{c}) \ = \ \tfrac{1}{2} \sum_{t=t_0}^{t_M} \Delta_t (\bf{v}_{\!L}^{\top} \bf{f}_{\!L})^{2}
\label{equation:costf2} \\
\nonumber \\
\mbox{subject to} & \hspace*{-0.8cm} &
\left\{\begin{array}{lcl}
\bf{q}(t_0,\bf{c}) & \!\!\! = & \!\!\! \bf{q}_I \\
\bf{q}(t_M,\bf{c}) & \!\!\! = & \!\!\! \bf{q}_F \\
\dot{\bf{q}}(t_0,\bf{c}) & \!\!\! = & \!\!\! {\bf{v}_S} \\
\dot{\bf{q}}(t_M,\bf{c}) & \!\!\! = & \!\!\! {\bf{v}_F} \\
{\bf{q}_{L\!B}}  & \hspace*{-1.6cm} \leq & \hspace*{-0.8cm} \bf{q}(t,\bf{c}) \ \leq \ {\bf{q}_{U\!B}} \\
{\bf{f}_{L\!B}}  & \hspace*{-1.6cm} \leq & \hspace*{-0.8cm} \bf{f}_{\!L}(t,\bf{c}) \leq \ {\bf{f}_{U\!B}} \\
{\bf{v}_{L\!B}}  & \hspace*{-1.6cm} \leq & \hspace*{-0.65cm} \dot{\bf{q}}(t,\bf{c}) \leq \ {\bf{v}_{U\!B}} \\
t_M & \hspace*{-1.6cm} \leq  & \hspace*{-0.8cm} t_{M_{m}}\\
\end{array} \right.
\label{equation:const2}
\end{eqnarray}
where $\Delta_t = t_k-t_{k-1}$ and $J(\bf{c})$ is the scalar cost function. 

 The robot's workspace, described in \eqref{eq:setX}, is sampled while adhering to joint limits, and the resulting sampled workspace representation in $\mathbb{R}^{2}$ is denoted as $\mathcal{\bar{X}}$.
To follow the outlined poses and paths, the manipulator's end-effector traverses and is sampled into n points, accordingly, and the trajectory line is considered within $\mathcal{\bar{X}}$:
\begin{equation}
\mathcal{\bar{X}} = \{ \mathbf{x} \mid \mathbf{x} = \mathbf{p}_{0}^{\text{EE}}(\mathbf{q}), \ \bf{q}_{LB} \leq \bf{q} \leq \ \bf{q}_{UB} \}
\label{eq:setX}
\end{equation}
\section{Robust Observer-Based Modular Control with Exponential Stability Connector}
\label{paper}
\subsection{Robust Adaptive State Observer for Linear Position and Velocity of EMLAs}
\label{observer}
Position sensors commonly used in electric motors include resolvers, encoders, Hall effect sensors, synchros, potentiometers, and linear variable differential transformers. They can generate signals proportional to the position status, serving as the output of systems. To estimate the true linear position and velocity on each EMLA in the presence of uncertainties, by considering the system output $y(t):\mathbb{R}^{1\times2} \times \mathbb{R}^{2\times1} \rightarrow \mathbb{R}$, we can further illustrate the motion dynamic system (two first subsystems of \eqref{equation:StateSpace3}), as follows:
\begin{equation}
\begin{aligned}
\label{17}
&\bm{\dot{\bar{x}}}(t)=\bm{A\bar{x}}(t)+\bm{Bu}(t)+\bm{g}(\bm{\bar{x}},t)+\bm{K}(\bm{\bar{x}},t)\\
&y(t)=\bm{C\bar{x}}(t) 
\end{aligned}
\end{equation}
where $\bm{\bar{x}}=[{x}_1,{x}_2]^{\top}$ is the actual value of the state vector (linear position and velocity of each EMLA), and $\bm{A}\in \mathbb{R}^{2 \times 2}$, $\bm{B}\in \mathbb{R}^2$, and $\bm{C}\in \mathbb{R}^{1 \times 2}$ are constant coefficients, assuming that $\bm{g}(.):\mathbb{R} \rightarrow \mathbb{R}^2$ comprises known modeling nonlinearities. We define $\bm{K}(.)=[K_{1},K_{2}]^{\top}$ as representing uncertainties, sensor noises, and external disturbances. In this context, $\bm{u}(t)$ is equal to $[0, i_q^*]^{\top}$ in \eqref{equation:StateSpace3}, although we will show that the modeling term $\bm{g}(.)$ and control input $\bm{u}(t)$ are mathematically ineffectual in the observer estimation.\\
\indent{\textbf{Assumption 1}} We can assume that matrices $\bm{A}$ and $\bm{C}$, as provided in \eqref{17}, are observable. Then, a feedback gain matrix $\bm{\alpha}$ belonging to $\mathbb{R}^2$ can be found, such that $\bm{\bar{A}} = \bm{A} - \bm{\alpha} \bm{C}$ is a Hurwitz matrix \cite{chen1990adaptive}.\\
\indent{\textbf{Assumption 2}} Following Assumption 1, we can assume that $K_{j}(.)$ is bounded. For all $(y(t), t) \in \mathbb{R} \times \mathbb{R}$, we can find two positive definite matrices $\bm{Q} \in \mathbb{R}^{2 \times 2}$ and $\bm{p} \in \mathbb{R}^{2 \times 2}$ in the following equation \cite{chen1990adaptive}:
\begin{equation}
\begin{aligned}
\label{18}
-\bm{Q} = \bm{p} \bm{\bar{A}}+\bm{\bar{A}^{\top}} \bm{p}
\end{aligned}
\end{equation}
such that an unknown positive constant $\eta^* \in \mathbb{R}^+$ and a continuous positive function $H(\cdot):\mathbb{R} \times \mathbb{R} \rightarrow \mathbb{R}^{+}$ can be assumed to meet the subsequent condition:
\begin{equation}
\begin{aligned}
\label{19}
||\hspace{0.1cm} \bm{K}(\bm{\bar{x}}, t)|| \leq \bm{p^{-1}} ||\bm{C^{\top}}|| \hspace{0.1cm} \eta^* \hspace{0.1cm} H(y(t), t)
\end{aligned}
\end{equation}
where $\|\cdot\|$ denotes the squared Euclidean norm.\\
\indent After defining the error of state observation $\bm{{x}_{eo}}=\bm{\bar{x}}-\bm{\hat{x}}$, we define $\bm{\hat{x}}: \mathbb{R} \rightarrow \mathbb{R}^2$ as the estimation vector of system states, as follows:
\begin{equation}
\begin{aligned}
\label{20}
\bm{\dot{\hat{x}}}(t)=&\bm{A \hat{x}}(t)+\bm{B u}(t)+\bm{g}(\bm{\bar{x}},t)+\bm{\alpha}(y-\hat{y})\\
&+\bm{p^{-1}} \bm{C^{\top}} f\\
 \hat{y}(t)=&\bm{C\hat{x}}(t), \hspace{0.3cm} \bar{y}=y-\hat{y}, \hspace{0.3cm} \bar{y}(t)=\bm{C{x}_{eo}}(t)\\
\bm{\dot{x}_{eo}}=& \bm{\bar{A} {x}_{eo}}+\bm{K}(\bm{\bar{x}}(t),t)-\bm{p^{-1}} \bm{C^{\top}} f
\end{aligned}
\end{equation}
where a finite and continuous function $f(\cdot):\mathbb{R} \times \mathbb{R} \rightarrow \mathbb{R}$ can be proposed as follows:
\begin{equation}
\begin{aligned}
\label{21}
f(\bar{y}, \hat{\eta}(t), t)=\frac{\hat{\eta}^2 \hspace{0.1cm} . \hspace{0.1cm} H(y, t)^2 \hspace{0.1cm}.\hspace{0.1cm} \bar{y}}{\hat{\eta} \hspace{0.1cm}H(y, t)\hspace{0.1cm}\| \bar{y}\|\hspace{0.1cm}+\hspace{0.1cm}m(t)}
\end{aligned}
\end{equation}
{}{where $m(t):\mathbb{R}\rightarrow \mathbb{R}^+$ is a positive and continuous function constrained by the following condition:}
\begin{equation}
\begin{aligned}
\label{22}
\lim _{t \rightarrow \infty} \int_{t_0}^t m(\tau) d \tau\leq \bar{m}<\infty
\end{aligned}
\end{equation}
and {$H(y,t)$ follows} \eqref{19}. The function $\hat{\eta}:\mathbb{R} \rightarrow \mathbb{R}$ represents the observer adaptation law, as follows:
\begin{equation}
\begin{aligned}
\label{23}
\dot{\hat{\eta}}=-m \hspace{0.1cm} \ell \hspace{0.1cm} \hat{\eta}+\ell \hspace{0.1cm} H(y, t)\hspace{0.1cm}\| \bar{y}\|
\end{aligned}
\end{equation}
where $\ell$ is a positive constant. From \eqref{19}, we defined an unknown positive parameter, denoted by $\eta^*$, and now we can define the observer adaptation error system $\bar{\eta}=\hat{\eta}-\eta^*$, such that:

\begin{equation}
\begin{aligned}
\label{24}
\dot{\bar{\eta}}= & -m \hspace{0.1cm} \ell \hspace{0.1cm} \bar{\eta}+\ell \hspace{0.1cm} H(y, t) \hspace{0.1cm}\| \bar{y}\| \hspace{0.1cm} -m \hspace{0.1cm} \ell \hspace{0.1cm}\eta^*
\end{aligned}
\end{equation}
\indent \textbf{Remark 2} According to the general solution of the given linear first-order ordinary differential equation in \eqref{23}, and assuming $\hat{\eta}\left(t_0\right) > 0$, we can say $\hat{\eta}(t)>0$ \cite{shahna2023exponential}. In addition, as with \eqref{21}, we can say:
\begin{equation}
\begin{aligned}
\label{25}
\|f(\bar{y}, \hat{\eta}(t), t)\| \leq \hat{\eta} H(y(t), t)
\end{aligned}
\end{equation}

\indent The deployment procedures for the proposed robust state observer to estimate the linear motion states of each EMLA are described in \textbf{Algorithm \ref{observer.a}}, which provides a summary and step-by-step guide for implementing the algorithm into each PMSM-driven EMLA-actuated joint system to calculate $\hat{x}$ based on the system output.

\begin{algorithm}[H]
\small
\caption{Robust state observer for each PMSM-driven EMLA-actuated joint}\label{observer.a}
\begin{algorithmic}
\STATE 
\STATE {\textsc{\textbf{IF} $\bm{\bar{A}}$ \textbf{is not Hurwitz}}}
    \STATE \hspace{0.7cm} $\bm{\alpha}=randn(2,1)$;
    \STATE \hspace{0.7cm} $\bm{\bar{A}}=\bm{A}-\bm{\alpha}*\bm{C}$;
\STATE {\textsc{\textbf{else}}}
\STATE\hspace{0.9cm}{\textsc{\textbf{If} $\bm{p}$\hspace{0.1cm}\textbf{is not positive definite matrices}}}
    \STATE \hspace{1.3cm} $\bm{Q} = randn(2,2)$;
    \STATE \hspace{1.3cm} $\bm{Q} = \bm{Q}*\bm{Q}^{\top}$;
    \STATE \hspace{1.3cm} $\bm{p} = \text{lyap}(\bm{\bar{A}}^{\top}, \bm{Q})$;
\STATE\hspace{0.9cm}\textsc{\textbf{ELSE}}
    \STATE \hspace{1.3cm} $\dot{\hat{\eta}}=-m \ell \hat{\eta}+\ell H\| y-\bm{C\hat{x}}\|$;
    \STATE \hspace{1.3cm} $f=(\hat{\eta}^2 H^2 \bar{y})/(\hat{\eta} H\| \bar{y}\|+m(t))$;
    \STATE \hspace{1.3cm} $\bm{\dot{\hat{x}}}= \bm{A} \bm{\hat{x}}+\bm{B} \bm{u}+\bm{g}+\bm{\alpha}\bar{y}+\bm{p}^{-1} \bm{C}^{\top} f$;
\STATE\hspace{0.9cm}\textsc{\textbf{end}}
\STATE \textsc{\textbf{end}}
\end{algorithmic}
\label{alg2}
\end{algorithm}

\vspace{-0.75cm}
\begin{table}[H]
\scriptsize
\begin{tablenotes} 
\item[-]- \text{lyap(.)} solves the Lyapunov equation.
\item[-]- \text{randn(.)} generates random numbers (0,1).
\item[-]- \text{eig(.)} computes the eigenvalues of a matrix.
\end{tablenotes}
\end{table}

\subsection{Robust Subsystem-Based Adaptive Control}
\label{control}
\subsubsection{Establishing Fundamental Concepts}
To propose the control methodology, after receiving the true linear position and velocity values from Section \ref{observer}, and by considering system states $\bm{x}=[{x}_1,{x}_2, x_3, x_4]^{\top}$, we can alter \eqref{equation:StateSpace3} into the following equations:
\begin{equation}
\left\{\begin{aligned}
\label{26}
\dot{x}_{1}(t) & =A_{1} x_{2}(t) +g_{1}(x)+F_{1}(x)+ d_{1}(t) \\
\dot{x}_{2}(t) & =A_{2} i_q^*(t)+g_{2}(x) +F_{2}\left(x\right)+ d_{2} (t)\\
\dot{x}_{3}(t) & =A_{3} u_{q}(t)+g_{3}(x) +F_{3}\left(x\right)+ d_{3} (t)\\
\dot{x}_{4}(t) & =A_{4} u_{d}(t)+g_{4}(x) +F_{4}\left(x\right)+ d_{4} (t)\\
\end{aligned}\right.
\end{equation}
For $i=1,\ldots,4$, $A_{i} \in \mathbb{R}$ is any non-zero coefficient, $g_{i}(x)$ is a known functional term originating from a model of the system, and $F_{i}(x)$ represents non-triangular uncertainties relying on all state variables and resulting from incomplete knowledge of system parameters or modeling inaccuracy. Meanwhile, $d_{i}:\mathbb{R} \rightarrow \mathbb{R}$ is a time-variant disturbance with uncertain magnitudes and timings. To clarify, Table \ref{strict} presents the definitions of each parameter used in \eqref{26} derived from the EMLA dynamic in \eqref{equation:StateSpace3}.
Note that \eqref{26} is valid for all $n_a$ EMLA-actuated joints. Consequently, we will have $4 \times n_a$ subsystems for the whole EMLA-actuated HDRM system. Our next step is to propose a controller with a modular structure that is compatible with the aforementioned observer. We can define the tracking error $x_{e_j}$ in the control system as follows:
\begin{equation}
\begin{aligned}
\label{27}
 x_{e_i}=&\hspace{+0.1cm}{x}_i-{x}_{id}, \hspace{0.3cm} i=1,...,4
\end{aligned}
\end{equation}
where $x_{1d}$ and $x_{2d}$ are derived from the control task outlined in Section \ref{manipulator_motion}. In addition, $x_{3d}$ is defined as $i_q^*$, while $x_{4d}$ is set to $i_d^*=0$.

\begin{table}[h]
\renewcommand{\arraystretch}{1.5}
    \small
    \centering
    \caption{Actual physical meaning of parameters in \eqref{26}}
\begin{tabular}{c||c}
        \hline
        \hline
          \multicolumn{1}{c}{\textbf{Term}} & \multicolumn{1}{c}{\textbf{EMLA parameters in \eqref{equation:StateSpace3}}}\\
        \hline
        \hline
            $A_1$& $1$\\
        \hline
            $A_2$& $\frac{3}{2A_{eq}}N_{p} \Phi_{PM}$\\
        \hline
            $A_3$& $\frac{1}{L_q}$\\
        \hline
            $A_4$& $\frac{1}{L_d}$ \\
        \hline
            $g_1(x)$& $0$ \\
        \hline
            $g_2(x)$& $\frac{1.5N_{p}(x_3 x_4 L_d-x_3 x_4 L_q)-B_{eq}x_2-C_{eq}x_1}{A_{eq}}$\\
        \hline
            $g_3(x)$& $\frac{- R_s x_3- N_{p} \alpha_{RL} x_2 ( x_4 L_d + \Phi_{PM})}{L_q}$ \\
        \hline
            $g_4(x)$& $\frac{- R_s x_4+ N_{p} \alpha_{RL} x_2 x_3 L_q}{L_d}$\\
        \hline
            $d_1(t)$& External disturbances\\
        \hline
            $d_2(t)$& Torque disturbance \hspace{0.1cm}$-\frac{D_{eq} F_L}{A_{eq}}$ \\
        \hline
            $d_3(t),d_4(t)$& Voltage disturbances in q- and d-axis\\
        \hline
            $F_{1,...,4}(x)$&  Non-triangular uncertainties \\
        \hline
        \hline
    \end{tabular}
    \label{strict}
\end{table}

\indent \textbf{Definition 1 }\cite{corless1993bounded, heydari2024robust} Consider that $\bm{x_d}$ and $\bm{x}$ are the reference and real state vectors of the manipulator system, respectively. For $t \geq t_0$, the HDRM system tracking error $\bm{{x}_e}=\bm{x}-\bm{x_d}$ is uniformly exponentially stable if the following condition is satisfied:
\begin{equation}
\begin{aligned}
\label{28}
&\|\bm{x_e}\|=\|\bm{x}-\bm{x_d}\| \leq \bar{c} e^{-O (t-t_0)} \|\bm{x_e}(t_0)\| + \tilde{\mu}
\end{aligned}
\end{equation}

where $\bar{c}$,$\tilde{\mu}$, and  $O$ are positive constants, and $\bm{x}_{e}(t_0)$ is any initial condition. More precisely, $\bm{x_e}$ is uniformly exponentially stabilized within a defined region $g\left(\tau\right)$, depending on the disturbance and non-triangular uncertainty bounds, as follows:
\begin{equation}
\begin{aligned}
\label{29}
&g\left(\tau\right):=\left\{\bm{x_e} \mid\|\bm{x_e} \| \leq \tau := {\tilde{\mu}}\right\}
\end{aligned}
\end{equation}

\begin{figure*}[h] 
  \centering
\scalebox{1.5}
    {\includegraphics[trim={0cm 0.0cm 0.0cm
    0cm},clip,width=\columnwidth]{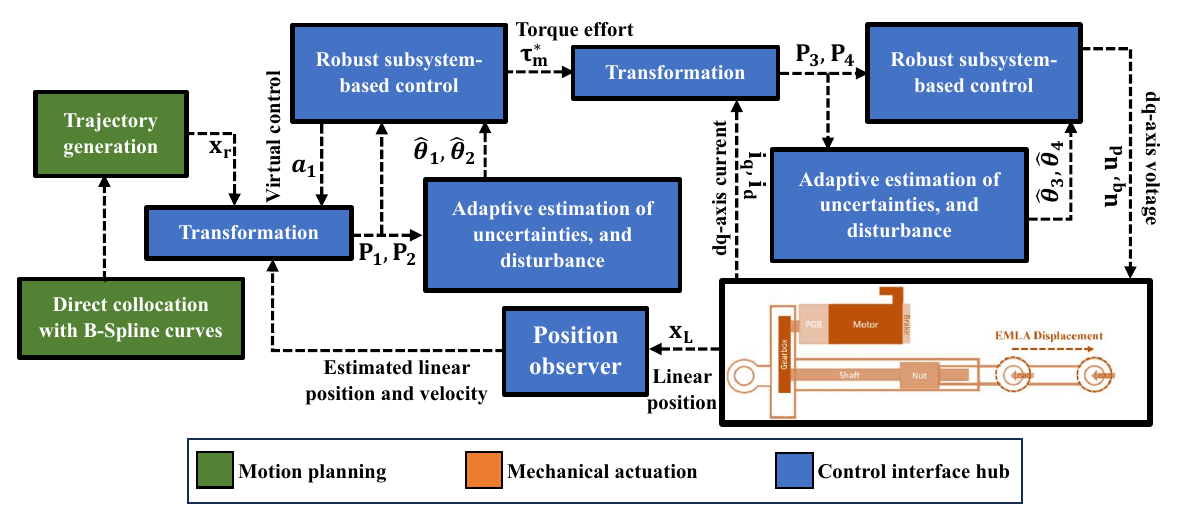}}
  \caption{Schematic of the RSBA control system for an EMLA-actuated joint of the HDRM}
  \label{fig:bigphoto}
\end{figure*}

\subsubsection{Transitioning System State into Tracking Form}
Now, we employ a transformation approach that alters the system states into the modular form of tracking:
\begin{equation}
\begin{aligned}
\label{30}
P_i = \begin{cases} x_{e_i} & \hspace{0.2cm} \textbf{if} \hspace{0.2cm}i=1,3,4 \\
x_{e_i} - a_1 & \hspace{0.2cm} \textbf{if} \hspace{0.2cm}i=2 \end{cases}
\end{aligned}
\end{equation}
$a_{1}:\mathbb{R} \rightarrow \mathbb{R}$ serves as a virtual position control input, and we can define it as shown:
\begin{equation}
\begin{aligned}
\label{31}
 {a_{1}}=&-\frac{1}{2 A_1}(\beta_{1}+\zeta_{1}\hat{\theta}_{1}){P_{1}}-x_{2d}-\frac{1}{A_1} g_1
\end{aligned}
\end{equation}
and the adaptation parameter $\hat{\theta}_{i}:\mathbb{R} \rightarrow \mathbb{R}$ is defined as follows:
\begin{equation}
\small
\begin{aligned}
\label{32}
&\dot{\hat{\theta}}_{i} = -{\delta_i}{\sigma_i}\hat{\theta}_{i}+\frac{1}{2}\zeta_{i}{\delta_i}|{P_i}|^2 , \hspace{0.3cm} i=1,...,4
\end{aligned}
\end{equation}
where $\zeta_i$, $\delta_i$, $\sigma_i$, and $\beta_{i}$ are positive constants.
The tracking equation structure in \eqref{30} and control adaptation law in \eqref{32} exhibit built-in modularity for all EMLAs, preventing undue complexity escalation as the system order $n_a$ increases. Each subsystem's equation can be designed with a generic form, ensuring the tracking equation's modularity.
Let $\hat{\theta}_i(0) \geq 0$ be an initial condition for the adaptive system. Following Remark 2, we can generate $\hat{\theta}_i(t) > 0$. Next, we can define the subsequent modular functions, assigning non-triangular uncertainties, assuming they may be unknown, as follows:
\begin{equation}
\begin{aligned}
\label{33}
\overline{F}_{i} = \begin{cases} {F_{i}}({x_1},...,{x_{4}}) & \hspace{0.2cm} \textbf{if} \hspace{0.2cm}i=1,3,4 \\
{F_{i}}({x_1},...,{x_{4}})
- {F}^* & \hspace{0.2cm} \textbf{if} \hspace{0.2cm}i=2 \end{cases}
\end{aligned}
\end{equation}
where:
\begin{equation}
\begin{aligned}
\label{34}
{F}^* =
 \sum_{i=1}^{4}\frac{\partial a_{1}}{\partial x_{i}} \frac{\mathrm{d} x_{i}}{\mathrm{~d} t}+\frac{\partial a_{1}}{\partial \hat{\theta}_{1}} \frac{\mathrm{d} \hat{\theta}_{1}}{\mathrm{~d} t}+\frac{\partial a_{1}}{\partial {x_{2d}}}\frac{\mathrm{d} x_{2d}}{\mathrm{~d} t}
\end{aligned}
\end{equation}
\indent {\textbf{Assumption 3}} According to \eqref{34}, we can practically assume that the linear position and velocity of the EMLA are differentiable and bounded; see the constraints provided in \eqref{equation:const2} and Figs. (\ref{fig:force_velocity}) and (\ref{fig:position_acceleration}). Furthermore, we can also assume that uncertainties and disturbances are bounded. Then, we can define a positive smooth function $r_i: \mathbb{R} \rightarrow \mathbb{R}^+$, along with positive constants $\Lambda_i$, $d_{max(i)}$, and $\Omega_i \in \mathbb{R}^+$ for each subsystem, which may all be unknown, such that:
\begin{equation}
\begin{aligned}
\label{35}
&\mid \overline{F}_i \mid \hspace{0.1cm} \leq \Lambda_i r_i \hspace{0.1cm}, \hspace{0.1cm} \mid d_i \mid \hspace{0.1cm} \leq  d_{max(i)}, \hspace{0.1cm} \mid \dot{x}_{id} \mid \leq \Omega_i
\end{aligned}
\end{equation}
Now, by differentiating \eqref{30}, inserting \eqref{26} and \eqref{27} into it, and considering \eqref{31} and \eqref{33}, we will have:
\begin{equation}
\begin{aligned}
\label{36}
\dot{P_1}=& A_1 {u_1}+A_1 x_{2d}+g_1+\overline{F}_1+A_1 a_1+d_1-\dot{x}_{1d} \\
\dot{P_2}=& A_2 u_2+g_2+\overline{F}_2 - \overset{.}{x}_{2d}+d_2\\
\dot{P_3}=& A_3 u_3+g_3+\overline{F}_3 - \overset{.}{x}_{3d}+d_3\\
\dot{P_4}=& A_4 u_4+g_4+\overline{F}_4 - \overset{.}{x}_{4d}+d_4
\end{aligned}
\end{equation}
where $u_1=P_{2}$. $u_2=i_q^*$, $u_3=u_q$, and $u_4=u_d$ are control signals. If we define:
\begin{equation}
\begin{aligned}
\label{38}
&G_i = A_i u_{i}+g_i+\overline{F}_i+d_i-\dot{x}_{id}
\end{aligned}
\end{equation}
For modularity, we have \eqref{36}:

\begin{equation}
\begin{aligned}
\label{37}
\dot{P_i} = \begin{cases}G_i+A_i x_{(i+1)d}+A_i a_i & \hspace{0.2cm} \textbf{if} \hspace{0.2cm}i=1 \\
G_i & \hspace{0.2cm} \textbf{if} \hspace{0.2cm}i=2,3,4 \end{cases}
\end{aligned}
\end{equation}

\subsubsection{Subsystem-Based Control Signal Design}
By considering the details mentioned, we propose the actual control inputs as follows:
\begin{equation}
\begin{aligned}
\label{39}
u_1=& \hspace{0.1cm} P_2\\
u_2=&-\frac{1}{2 A_{2}}(\beta_{2}+\zeta_{2}\hat{\theta}_{2}){P_{2}}-\frac{A_{1}}{A_{2}}P_{1}-\frac{1}{A_{2}} g_{2}\\
u_3=&-\frac{1}{2 A_{3}}(\beta_{3}+\zeta_{3}\hat{\theta}_{3}){P_{3}}-\frac{1}{A_{3}} g_{3}\\
u_4=&-\frac{1}{2 A_{4}}(\beta_{4}+\zeta_{4}\hat{\theta}_{4}){P_{4}}-\frac{1}{A_{4}} g_{4}\\
\end{aligned}
\end{equation}
\indent \textbf{Remark 3} Like the tracking equation dynamics of each EMLA-actuated joint in \eqref{37}, the control signals proposed in \eqref{39} can use a modular-structured equation as follows:
\begin{equation}
\begin{aligned}
\label{40}
u_i = \begin{cases}P_{i+1} & \hspace{0.2cm} \textbf{if} \hspace{0.2cm}i=1 \\
W_i-\frac{A_{i-1}}{A_{i}}P_{i-1} & \hspace{0.2cm} \textbf{if} \hspace{0.2cm}i=2 \\
W_i & \hspace{0.2cm} \textbf{if} \hspace{0.2cm}i=3,4 \end{cases}
\end{aligned}
\end{equation}
where:
\begin{equation}
\begin{aligned}
\label{41}
W_i = -\frac{1}{2 A_{i}}(\beta_{i}+\zeta_{i}\hat{\theta}_{i}){P_{i}}-\frac{1}{A_{i}} g_{i}
\end{aligned}
\end{equation}
\indent By assuming $\theta^*_i \in \mathbb{R}^+$ is an unknown positive constant to tune the adaptation law, we define the adaptation error ${\tilde{\theta}}_i=\hat{\theta}_i-{\theta^*_i}$.
Thus, we can obtain from \eqref{32}:
\begin{equation}
\begin{aligned}
\label{42}
&\dot{\tilde{\theta}}_{i}=-{\delta_i}{\sigma_i}{\tilde{\theta}}_{i}+\frac{1}{2}\zeta_{i}{\delta_i}|{P_i}|^2-{\delta_i}{\sigma_i}{\theta}^*_{i}
\end{aligned}
\end{equation} 
$\theta^*_i$ can be defined as follows:
\begin{equation}
\begin{aligned}
\label{43}
 \theta_i^* =\zeta_i^{-1}[\mu_i {}\Lambda_i^2+ \nu_i {}d_{max(i)}^2 +\psi_{{i}}\left(\Omega_i\right)^2]
\end{aligned}
\end{equation}
where $\mu_i$, $\nu_i$, $\psi_{{i}}$, and $\Omega_i$ are unknown positive constants, and $\zeta_i$ was introduced previously in \eqref{32}.
To clarify, Fig. \ref{fig:bigphoto} demonstrates how RSBA control operates for one EMLA-actuated joint. The deployment procedures for the RSBA control are described in \textbf{Algorithm \ref{controller}}, which provides a summary and step-by-step guide for incorporating the RSBA control algorithm into each EMLA of the HDRM system. Note that the inputs of the mentioned algorithm are the outputs of \textbf{Algorithm \ref{observer.a}}, current sensor information, and control parameters, while the output of \textbf{Algorithm \ref{alg2}} is control signals $u_2$, $u_3$, and $u_4$.

\begin{algorithm}[H]
\small
\caption{Implementation of RSBA control for each EMLA}\label{controller}
\begin{algorithmic}
\STATE 
\STATE {\textsc{\textbf{\textbf{For} $i=1:4$ \textbf{do}}}}
\STATE \hspace{0.3cm}{\textsc{\textbf{\textbf{if} $i=1$ \textbf{do}}}}
    \STATE \hspace{0.7cm} $x_{e_i}={x}_i-x_{id}$;\\
    \STATE \hspace{0.7cm} $P_i=x_{e_i}$;\\
    \STATE \hspace{0.7cm} $\dot{\hat{\theta}}_{i} = -{\delta_i}{\sigma_i}\hat{\theta}_{i}+\frac{1}{2}\zeta_{i}{\delta_i}|{P_i}|^2$;\\
    \STATE \hspace{0.7cm} ${a_{i}}=-\frac{1}{2 A_i}(\beta_{i}+\zeta_{i}\hat{\theta}_{i}){P_{i}}-x_{({i+1})d}-\frac{1}{A_i} g_i$;\\

    \STATE \hspace{0.3cm}
    {\textsc{\textbf{\textbf{if} $i=2$ \textbf{do}}}}
    \STATE \hspace{0.7cm} $x_{e_i}={x}_i-x_{id}$;\\
    \STATE \hspace{0.7cm} $P_i=x_{e_i}-a_{i-1}$;\\
    \STATE \hspace{0.7cm} $\dot{\hat{\theta}}_{i} = -{\delta_i}{\sigma_i}\hat{\theta}_{i}+\frac{1}{2}\zeta_{i}{\delta_i}|{P_i}|^2$;\\
   \STATE \hspace{0.7cm} $u_i=-\frac{1}{2 A_{i}}(\beta_{i}+\zeta_{i}\hat{\theta}_{i}){P_{i}}-\frac{A_{i-1}}{A_{i}}P_{i-1}-\frac{1}{A_{i}} g_{i}$\\
    
    \STATE \hspace{0.3cm} {\textsc{\textbf{\textbf{else} \textbf{do}}}}
        \STATE \hspace{0.7cm} $x_{e_i}={x}_i-x_{id}$;\\
    \STATE \hspace{0.7cm} $P_i=x_{e_i}$;\\
    \STATE \hspace{0.7cm} $\dot{\hat{\theta}}_{i} = -{\delta_i}{\sigma_i}\hat{\theta}_{i}+\frac{1}{2}\zeta_{i}{\delta_i}|{P_i}|^2$;\\
\STATE \hspace{0.7cm} $u_i=-\frac{1}{2 A_{i}}(\beta_{i}+\zeta_{i}\hat{\theta}_{i}){P_{i}}-\frac{1}{A_{i}} g_{i}$
\STATE \textsc{\textbf{end}}
\end{algorithmic}
\label{alg2}
\end{algorithm}

\begin{figure}[h] 
  \centering
\scalebox{1.0}
    {\includegraphics[trim={0cm 0.0cm 0.0cm
    0cm},clip,width=\columnwidth]{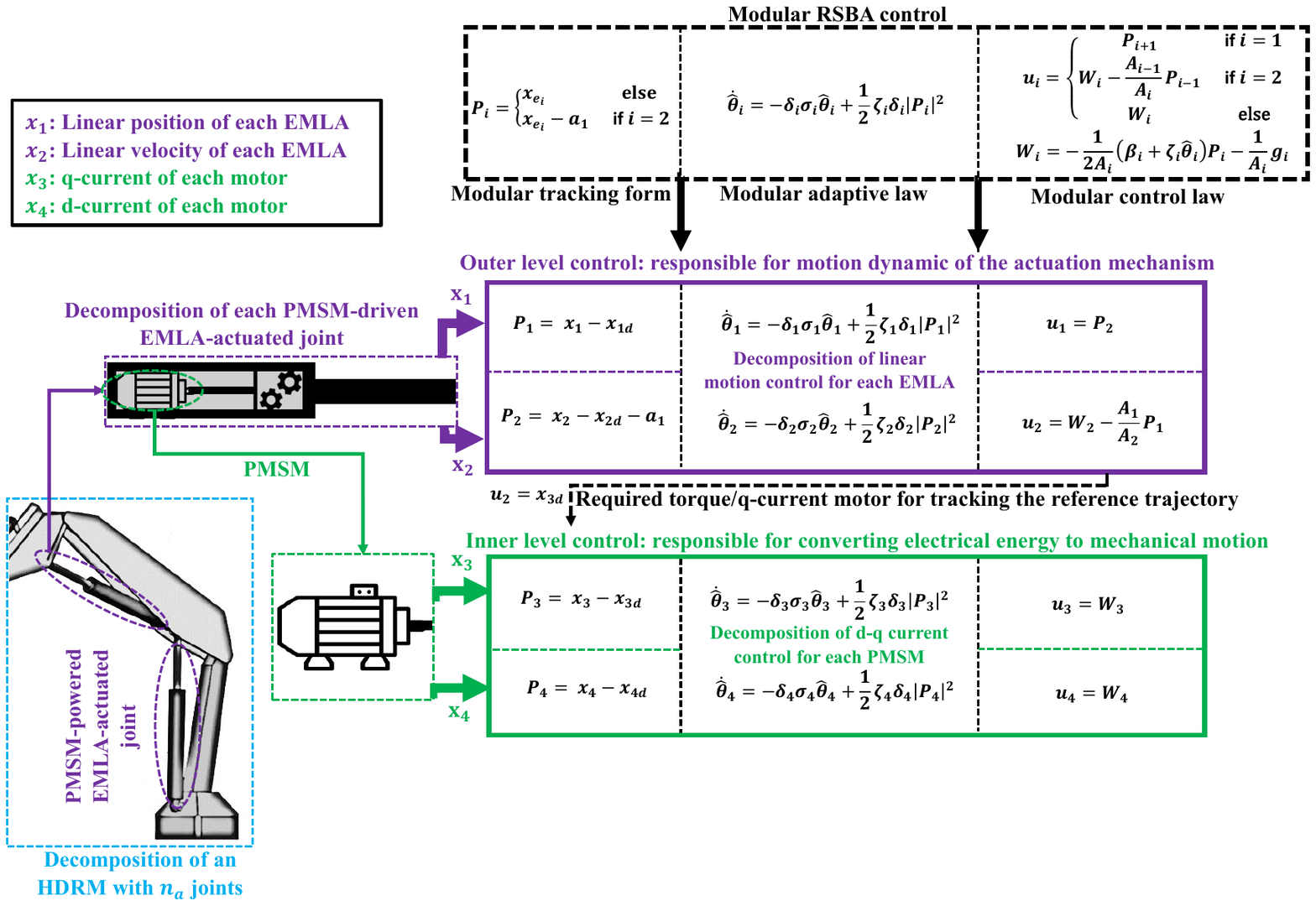}}
  \caption{Decomposition of each PMSM-powered EMLA-actuated joint of a $n_a$-DoF HDRM into two main subsystems: the motion dynamics of EMLA and energy conversion formulation of PMSM, along with the modularity feature of the RSBA control framework.}
  \label{modular}
\end{figure}

Fig. \ref{modular} illustrates the modularity feature of the RSBA control, which provides the unified control formulation applicable for all $n_a$ PMSM-powered EMLA-actuated joints of the HDRM. Each joint is decomposed into two interconnected subsystem-based levels: the outer level (which is further decomposed into two subsystems: the linear position and velocity subsystems) is responsible for the motion dynamics of the EMLA mechanism and calculating the required q-current of the motor ($i^*_q = u_2$) to ensure the linear actuator's position or velocity aligns with the reference values. The resulting q-current then becomes the desired q-current for the inner level subsystem control (which is further decomposed into subsystems for d- and q-current systems and converts electrical energy to mechanical one) to adjust the sufficient motor voltage signals (see Eqs. \eqref{equation:StateSpace3} and \eqref{desired}). The decomposition of PMSM-powered EMLA-actuated HDRMs and the proposal of modular RSBA control enable modifications to the dynamics of the HDRM, such as altering motors or other dynamic components or adding/removing joints, without affecting the unified control formulation. This modular characteristic of the RSBA control framework and the decomposition of the multi-component HDRM into smaller subsystems provide the opportunity for further subsystem-level analysis, such as optimization strategies. It can also be extended to address other servo-driven actuator applications, such as EHA mechanisms, regardless of the type of motor or actuator mechanism involved.

\subsection{Stability Analysis through Subsystem Connectivity}
\label{stability}
\begin{figure}[h!] 
    \centering
    \scalebox{0.8}
    {\includegraphics[trim={0.1cm 0.1cm 0.1cm
    0.1cm},clip,width=\columnwidth]{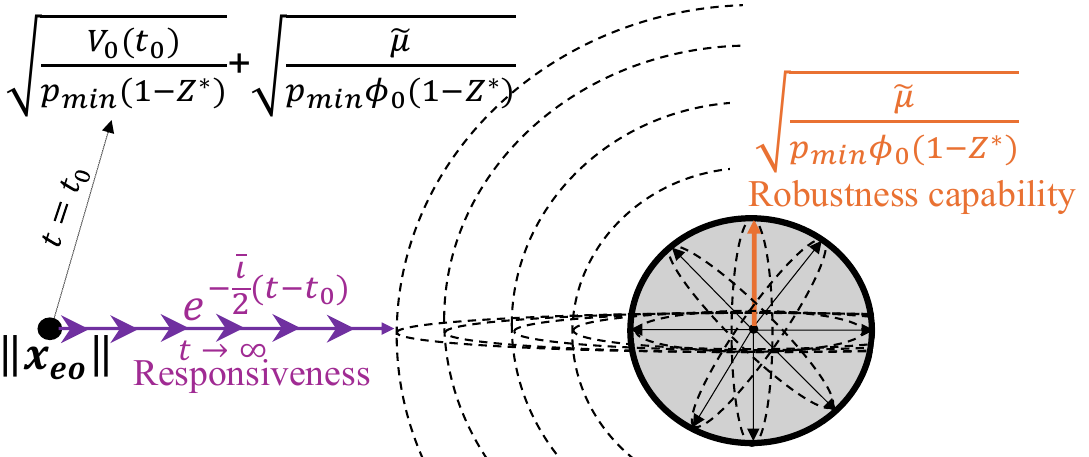}}
    \caption{Convergence of the proposed adaptive observer, signifying the trade-off between noise filtering and real-time responsiveness.}
    \label{circle}
\end{figure}

\textbf{Theorem 1} Consider a PMSM-driven EMLA-actuated joint system whose modular tracking equations are provided in \eqref{37}. By employing the modular RSBA control specified in \eqref{40}, along with the robust observer method given in Section \ref{observer}, in conjunction with adaptive laws presented in Eqs. \eqref{23} and \eqref{32} for the system, the tracking error of the reference trajectories provided in Section \ref{manipulator_motion} uniformly and exponentially converges to the bounded value $\bar{\tau}_0$, even in the presence of non-triangular uncertainties and time-variant disturbances.\\
\indent \textbf{Proof} We introduce a Lyapunov function for the proposed observer in the following manner:
\begin{equation}
\label{44}
V_0 = \bm{{x}_{eo}^\top p {x}_{eo}} + \frac{1}{\ell} \bar{\eta}^2
\end{equation}
After taking the derivative of the Lyapunov function and inserting \eqref{20}, we have:
\begin{equation}
\begin{aligned}
\label{45}
\dot{V}_0 =& 2 \bm{{x}_{eo}^{\top}p \bar{A} {x}_{eo}} +2 \bm{{x}_{eo}^{\top} p}\left[\bm{K}-\bm{p^{-1} C^{\top}} f\right]\\
&+2 \ell^{-1} \bar{\eta} \dot{\bar{\eta}}
\end{aligned}
\end{equation} 
Furthermore, \eqref{18}, \eqref{19}, and \eqref{20} imply that for all $t \geq t_0$:
\begin{equation}
\begin{aligned}
\label{46}
\dot{V}_0  \leq &  -\bm{{x}_{eo}^{\top} Q {x}_{eo}}+2 \eta^* \| \bar{y}\| H -2 \bar{y} f+2 \ell^{-1} \bar{\eta} \dot{\bar{\eta}}
\end{aligned}
\end{equation} 
Substituting Eqs. \eqref{21} and \eqref{24} into \eqref{46}, we obtain:
\begin{equation}
\begin{aligned}
\label{47}
 \dot{V}_0 \leq &  -\bm{{x}_{eo}^{\top} Q {x}_{eo}}+2 \eta^* H\| \bar{y}\| - \frac{2 \hat{\eta}^2 H^2 \bar{y}^2}{\hat{\eta} H\| \bar{y}\|+m} \\
&+2 \bar{\eta} H\| \bar{y}\|-2 m\bar{\eta}^2-2 m \bar{\eta} \eta^*
\end{aligned}
\end{equation} 
By considering $\tilde{\eta}=\hat{\eta}-\eta^*$, we have:
\begin{equation}
\begin{aligned}
\label{48}
\dot{V}_0 \leq & -\bm{{x}_{eo}^{\top} Q {x}_{eo}}+ \frac{2 \hat{\eta} H\| \bar{y}\| \cdot m}{\hat{\eta} H\| \bar{y}\|+m} -2 m \bar{\eta}^2-2 m \bar{\eta} \eta^*\\
\leq & -\bm{{x}_{eo}^{\top} Q {x}_{eo}}+ \frac{2m (\hat{\eta} H\| \bar{y}\| + m) - 2m^2}{\hat{\eta} H\| \bar{y}\|+m}\\
&-2 m \bar{\eta}^2-2 m \bar{\eta} \eta^*
\end{aligned}
\end{equation} 
By eliminating the negative part from the right-hand side of \eqref{48}, we have the option to deduce that:
\begin{equation}
\begin{aligned}
\label{49}
 \dot{V}_0 & \leq -\bm{{x}_{eo}^{\top} Q {x}_{eo}}+2 m-2 m \bar{\eta}^2-2 m \bar{\eta} \eta^* 
\end{aligned}
\end{equation} 
Then, considering $-2\bar{\eta}\eta^* \leq \bar{\eta}^2+ {\eta^*}^2$, we obtain:
\begin{equation}
\begin{aligned}
\label{50}
 \dot{V}_0  \leq&-\bm{{x}_{eo}^{\top} Q {x}_{eo}} +2 m-2 m \bar{\eta}^2+ m (\bar{\eta}^2+ {\eta^*}^2)\\
& =-\bm{{x}_{eo}^{\top} Q {x}_{eo}} - m \bar{\eta}^2+ m (2+ {\eta^*}^2)
\end{aligned}
\end{equation} 
Hence, by knowing that $\bm{p}$ and $\bm{Q}$ are positive definite matrices, we can assume there is a positive constant $L\in \mathbb{R^+}$ in which:
\begin{equation}
\begin{aligned}
\label{51}
 -\bm{{x}_{eo}^{\top} Q {x}_{eo}} \leq -\bm{{x}_{eo}^{\top} p {x}_{eo}} + L
\end{aligned}
\end{equation}
Then:
\begin{equation}
\begin{aligned}
\label{52}
 \dot{V}_0 & \leq-\bm{{x}_{eo}^{\top} p {x}_{eo}} - m \bar{\eta}^2+ m (2+ {\eta^*}^2)+L
\end{aligned}
\end{equation}
By considering $\tilde{\mu}_o=(2+ {\eta^*}^2)$ and according to Young's inequality, we can reach:
\begin{equation}
\begin{aligned}
\label{53}
\dot{V}_0 \leq-\bm{{x}_{eo}^{\top} p {x}_{eo}} - m \bar{\eta}^2+ L +\frac{1}{2} \tilde{\mu}_o ^2+ \frac{1}{2} m^2
\end{aligned}
\end{equation}
By defining $L_{max}$ as the supremum of $L$ and the positive function $\tilde{\mu}:\mathbb{R} \rightarrow \mathbb{R}^+$:
\begin{equation}
\begin{aligned}
\label{54}
\tilde{\mu} =
\frac{1}{2} \tilde{\mu}_o ^2+L_{max}
\end{aligned}
\end{equation}
From \eqref{44} and \eqref{54}, we have \eqref{53}, as follows:
\begin{equation}
\begin{aligned}
\label{55}
\dot{V}_0 \leq-\phi_0 V_0(t_0) + \frac{1}{2} m^2 + \tilde{\mu}
\end{aligned}
\end{equation}
where $\phi_0=\min[1, \underline{m}\hspace{0.1cm}\ell]$ and $\underline{m}$ is the infimum of $m$. Based on Definition 1 and Appendix A, and according to \cite{heydari2024robust} and \cite{corless1993bounded}, the error of the observer $\|\bm{x_{eo}}\|$ is uniformly exponentially bounded within a specific ball, as shown in Fig. \ref{circle}, and:
\begin{equation}
\begin{aligned}
\label{00554}
\|\bm{x_{eo}}\| \leq \sqrt{\frac{V_0\left(t_0\right)}{p_{min} ({1-\overset{*}{Z}})} } e^{-\frac{\bar{\iota}}{2}(t-t_0)} + \sqrt{\frac{\frac{1}{p_{\min} \phi_{0}} \tilde{\mu}}{1-\overset{*}{Z}}}
\end{aligned}
\end{equation}
\indent \textbf{Remark 4:}  As discussed in \cite{gessing2011robustness} and \cite{solsona2003disturbance}, a state observer with high-response convergence tends to exhibit poorer filtering capabilities. See Fig. \ref{circle} and Eq. \eqref{00554}. The noise-filtering capability of the proposed observer (the radius of the sphere illustrated in Fig. \ref{circle}) relies on the time constant term: $\sqrt{\frac{\frac{1}{p_{\min} \phi_{0}} \tilde{\mu}}{1-\overset{*}{Z}}}$. Thus, decreasing the positive parameter $\phi_0$ and eigenvalues of the positive definite matrix $p$ increases the noise-filtering capability and robustness of the observer. On the other side, as the positive parameter $\bar{\iota}$ must be $\bar{\iota} < \phi_0 - \frac{1}{2 p_{min}}$ based on Eq. \eqref{804}, the small values of the parameter $\phi_0$ and eigenvalues of the positive definite matrix $p$ decrease real-time responsiveness. Thus, depending on the intensity of noises and disturbances, a trade-off between robustness and real-time responsiveness in setting the proposed observer parameters should be balanced.

We will use \eqref{55} later. Now, a Lyapunov function for the first subsystem of the EMLA equations is proposed, as shown:
\begin{equation}
\begin{aligned}
\label{56}
&V_1 =\frac{1}{2} \hspace{0.1cm} [{P_1}^2+{\delta^{-1}_1}\tilde{\theta}_1^2]
\end{aligned}
\end{equation}
After differentiating $V_1$ and inserting $u_1$ from \eqref{40} and $G_1$ from \eqref{38} into \eqref{37}, we have \eqref{56}, as follows:
\begin{equation}
\begin{aligned}
\label{57}
\dot{V_1} =& A_1 P_{1} P_2+A_1 P_1 x_{2d}+P_1 g_1+ P_{1}\overline{F}_1+ P_{1} d_1\\
&+ A_1 P_{1}a_1-P_1 \dot{x}_{1d}+\delta_1^{-1} {}\tilde{\theta}_1\dot{\tilde{\theta}}_1
\end{aligned}
\end{equation} 

\indent \textbf{Remark 5} Koivumaki et al. in \cite{koivumaki2022subsystem} proposed a control method ensuring asymptotic stability for the systems with triangular uncertainties and the absence of time-varying disturbances. Expanding upon their stability analysis, we utilize the concept of a stability connector for the PMSM-driven EMLA-actuated $n_a$-DoF HDRM, encompassing non-triangular uncertainties and time-varying disturbances. The stability connector $S_1$ is a destabilizing dynamic interaction among the first and second subsystems (position and velocity) of each EMLA-actuated joint system. Its aim is to eliminate instability among subsystems and result in an exponentially stable analysis of the entire system. This term is defined as follows:
\begin{equation}
\begin{aligned}
\label{58}
&S_1=A_1 P_1 \hspace{0.05cm} P_{2}
\end{aligned}
\end{equation}

By considering \eqref{35} and \eqref{57} and using \eqref{58}, we have:
\begin{equation}
\begin{aligned}
\label{59}
\dot{V_1}\leq& \hspace{0.1cm} S_1+A_1 P_1 x_{2d}+P_1 g_1+{}\mid P_{1} \mid \Lambda_1 r_1+ \mid P_1 \mid \Omega_1\\
&+{}\mid P_{1} \mid d_{max(1)}+A_1 P_{1}a_1+\delta_1^{-1} {}\tilde{\theta}_1\dot{\tilde{\theta}}_1
\end{aligned}
\end{equation} 
By considering positive constants $\psi_1$, $\nu_1$, and $\mu_1$ and following Young's inequality, we have: 
\begin{equation}
\begin{aligned}
\label{60}
\dot{V_1}\leq& S_1+A_1 P_1 x_{2d}+P_1 g_1+ \frac{1}{2} \mid P_1 \mid^2 \mu_1 \Lambda_1 ^2 
\\&+ \frac{1}{2} \mu_1^{-1} r_1^2 +A_1 P_{1}a_1+ \frac{1}{2} \mid P_1 \mid ^2 \nu_1 d_{max(1)}^2\\
&+ \frac{1}{2} \nu_1^{-1}+\frac{1}{2}\psi^{-1}_1+\frac{1}{2}\psi_1 \Omega^2_1 P^2_1+\delta^{-1} {}\tilde{\theta}_1\dot{\tilde{\theta}}_1
\end{aligned}
\end{equation} 
By considering the description  of $\theta^*_1$ in \eqref{43}, as well as the descriptions $\tilde{\theta}_1$ in \eqref{42}, we obtain:
\begin{equation}
\begin{aligned}
\label{61}
\dot{V_1}\leq& S_1 + \frac{1}{2} {}\zeta_1 \theta_1^* \mid P_1 \mid^2 + \frac{1}{2} \mu_1^{-1} r_1^2+ \frac{1}{2} \nu_1^{-1}\\
&-\frac{1}{2}{}\beta_{1}{P_{1}^2}+\frac{1}{2}\psi^{-1}_1-\frac{1}{2}{}\zeta_{1}\hat{\theta}_{1}{P_{1}^2}\\
&-{}{\sigma_1}{\tilde{\theta}}_{1}^2+\frac{1}{2}{}\zeta_{1}|{P_1}|^2\tilde{\theta}_1-{}{\sigma_1}{\theta}^*_{1}\tilde{\theta}_1
\end{aligned}
\end{equation} 
Because $\tilde{\theta}_1=\hat{\theta}_1-\theta_1^*$,
\begin{equation}
\begin{aligned}
\label{62}
\dot{V_1}\leq& S_1 + \frac{1}{2} \mu_1^{-1} r_1^2 + \frac{1}{2} \nu_1^{-1}-\frac{1}{2}{}\beta_{1}{P_{1}^2}-{}{\sigma_1}{\tilde{\theta}}_{1}^2+\frac{1}{2}\psi^{-1}_1\\
&-{}{\sigma_1}{\theta}^*_{1}\tilde{\theta}_1
\end{aligned}
\end{equation} 
After dividing ${}{\sigma_1}{\tilde{\theta}}_{1}^2$ into $\frac{1}{2}{}{\sigma_1}{\tilde{\theta}}_{1}^2+\frac{1}{2}{}{\sigma_1}{\tilde{\theta}}_{1}^2$ and considering \eqref{56}, we can arrive at:
\begin{equation}
\begin{aligned}
\label{63}
\dot{V_1}\leq& -\phi_1 V_1 + S_1 + \frac{1}{2} \mu_1^{-1} r_1^2 + \frac{1}{2} \nu_1^{-1}+\frac{1}{2}{}{\sigma_1}{\theta_1^*}^2\\
&+\frac{1}{2}\psi^{-1}_1
\end{aligned}
\end{equation} 
where:
\begin{equation}
\begin{aligned}
\label{64}
\phi_1 = \min [{\beta_1},\hspace{0.3cm}{\delta_1}\sigma_1]
\end{aligned}
\end{equation}
We will use \eqref{63} later. Just as in \eqref{56}, we present the same scenario for the $2$-th subsystem by defining the Lyapunov function as follows:
\begin{equation}
\begin{aligned}
\label{65}
&V_2 = \frac{1}{2}\hspace{0.1cm}{}[P_2^2+\delta_2^{-1} \tilde{\theta}_2^2]
\end{aligned}
\end{equation} 
By differentiating $V_2$ and inserting $\dot{P}_2$ in \eqref{37} and $G_2$ in \eqref{38}, we have:
\begin{equation}
\begin{aligned}
\label{66}
\dot{V_2}=& P_{2} {}[A_2 u_{2}+g_2+\overline{F}_2-\dot{x}_{2d}+d_2]+\delta_2^{-1} {}\tilde{\theta}_2\dot{\tilde{\theta}}_2
\end{aligned}
\end{equation}
Likewise, we continue by considering $u_2$ in \eqref{40} and $W_2$ in \eqref{41}, and the stability connector $S_{1}$ from \eqref{58}, inserted into \eqref{66}. Similar to the first subsystem, we will obtain:
\begin{equation}
\begin{aligned}
\label{67}
 \dot{V_2}\leq&-\phi_2 V_2 - S_{1} + \frac{1}{2} \nu_2^{-1}+\frac{1}{2} \mu_2^{-1} r_2^2+\frac{1}{2} {\sigma_2}{\theta_2^*}^2\\
&+\frac{1}{2}\psi^{-1}_2
\end{aligned}
\end{equation}
where:
\begin{equation}
\begin{aligned}
\label{68}
\phi_2 = \min [\beta_2,\hspace{0.3cm}{\delta_2}\sigma_2]
\end{aligned}
\end{equation}
Likewise, we can establish an analogous Lyapunov function for the $3$th and $4$th subsystems, as shown:
\begin{equation}
\begin{aligned}
\label{650}
&V_3 = \frac{1}{2}\hspace{0.1cm}{}[P_3^2+\delta_3^{-1} \tilde{\theta}_3^2], \hspace{0.2cm} V_4 = \frac{1}{2}\hspace{0.1cm}{}[P_4^2+\delta_4^{-1} \tilde{\theta}_4^2]
\end{aligned}
\end{equation} 
Similarly, we can obtain:
\begin{equation}
\begin{aligned}
\label{69}
 \dot{V_3}\leq&-\phi_3 V_3 + \frac{1}{2} \mu_3^{-1} r_3^2 + \frac{1}{2} \nu_3^{-1}+\frac{1}{2} {\sigma_3}{\theta_3^*}^2\\
 &+\frac{1}{2}\psi^{-1}_3
\end{aligned}
\end{equation}
and:
\begin{equation}
\begin{aligned}
\label{70}
\dot{V_4}\leq&-\phi_4 V_4 + \frac{1}{2} \mu_4^{-1} r_4^2 + \frac{1}{2} \nu_4^{-1}+\frac{1}{2} {\sigma_4}{\theta_4^*}^2\\
&+\frac{1}{2}\psi^{-1}_4
\end{aligned}
\end{equation}
where:
\begin{equation}
\begin{aligned}
\label{71}
\phi_3 = \min [\beta_3,\hspace{0.1cm}{\delta_3}\sigma_3], \hspace{0.1cm} \phi_4 = \min [\beta_4,\hspace{0.1cm}{\delta_4}\sigma_4]
\end{aligned}
\end{equation}
Now, we introduce the Lyapunov function ${V}$ for the entire subsystem of one EMLA-actuated joint, including the observer section, as follows:
\begin{equation}
\begin{aligned}
\label{72}
&V =V_0+V_1+V_2+V_3+V_4
\end{aligned}
\end{equation}
After the derivative of \eqref{72}, and calling on \eqref{55}, \eqref{63}, \eqref{67}, \eqref{69}, and \eqref{70}, we obtain:
\begin{equation}
\begin{aligned}
\label{74}
\dot{V} \leq& -\sum_{i=0}^4 \phi_i V_i + [\ \overset{0}{\cancel{S_{1} - S_{1}}} ]\ + \frac{1}{2} \sum_{i=1}^4 \mu_i^{-1} r_i^2+ \tilde{\mu}\\
&+ \frac{1}{2} \sum_{i=1}^{4} \nu_i^{-1}+ \frac{1}{2} \sum_{i=1}^{4} \psi_i^{-1}+\frac{1}{2}\sum_{i=1}^4{\sigma_i}{\theta_i^*}^2+ \frac{1}{2} m^2 
\end{aligned}
\end{equation}
As we can observe in \eqref{74}, based on the concept defined in Remark 4, the unstable term associated with dynamic interactions between subsystems has been effectively offset in this step. Generally speaking, we can transform the equations provided in \eqref{72} by considering \eqref{44}, \eqref{56}, \eqref{65}, and \eqref{69}:

\begin{equation}
\begin{aligned}
\label{75}
V=& \frac{1}{2} \mathbf{P}^{\top} \bm{\lambda} \mathbf{P} + \frac{1}{2}\bm{\tilde{\theta}^{\top}} \mathbf{\Delta^{-1}} \bm{\tilde{\theta}}
\end{aligned}
\end{equation}
where:
\begin{equation}
\begin{aligned}
\label{76}
\mathbf{P} = \begin{bmatrix}
    x_{eo} \\
    P_1 \\
    \vdots \\
    P_4 \\
\end{bmatrix}, \hspace{0.2cm}
\bm{\lambda}=\begin{bmatrix}
    2\bm{p} & 0 & 0 & \ldots  & 0 \\
    0 & 1 & 0 & \ldots & 0 \\
\vdots & \vdots & \vdots & \vdots & \vdots \\
    0 & \ldots & 0 & 0 & 1 \\
    \end{bmatrix}, \hspace{0.2cm}\\
    \bm{\tilde{\theta}} = \begin{bmatrix}
    \bar{\eta} \\
    \tilde{\theta}_1 \\
    \vdots\\
    \tilde{\theta}_4 \\
\end{bmatrix}, \hspace{0.2cm}
\bm{\Delta^{-1}} = \begin{bmatrix}
    2 \ell_1^{-1} & 0 & 0 & \ldots  & 0 \\
    0 & \delta_1^{-1} & 0 & \ldots & 0 \\
    \vdots & \vdots & \vdots & \vdots & \vdots \\
    0 & \ldots & 0 & 0 & \delta_4^{-1} \\
\end{bmatrix}
\end{aligned}
\end{equation}
Then, from \eqref{74},
\begin{equation}
\begin{aligned}
\label{78}
\dot{V} \leq& -\phi_{total} V + \frac{1}{2} \sum_{i=1}^4 \bar{\mu}^{-1} M_i^2+\bar{\mu}_{total}
\end{aligned}
\end{equation}
where $\mathbf{P}:\mathbb{R}^{2} \times \mathbb{R} \times \mathbb{R} \times \mathbb{R} \times \mathbb{R} \rightarrow \mathbb{R}^{6}$, $\bm{\lambda}:\mathbb{R}^{2\times 2}\times \mathbb{R} \times \mathbb{R} \times \mathbb{R} \times \mathbb{R} \rightarrow \mathbb{R}^{6 \times 6}$, $\bm{\tilde\theta}:\mathbb{R} \times \mathbb{R} \times \mathbb{R} \times \mathbb{R} \times \mathbb{R} \rightarrow \mathbb{R}^{5}$, and $\bm{\Delta}:\mathbb{R} \times \times \mathbb{R} \times \mathbb{R} \times \mathbb{R} \times \mathbb{R} \rightarrow \mathbb{R}^{5 \times 5}$.
From \eqref{74}:
\begin{equation}
\begin{aligned}
\label{770}
&\frac{1}{2}m^2+\frac{1}{2}\sum_{i=1}^{4}\mu_i^{-1}r^2_i=\frac{1}{2}\sum_{i=1}^{4}[\mu_i^{-1}r^2_i+\frac{1}{4}m^2]
\end{aligned}
\end{equation}
we can define:
\begin{equation}
\begin{aligned}
\label{77}
&\bar{\mu}^{-1}=\max(\frac{1}{4} ,  \tilde{\mu}^{-1}_1,..., \tilde{\mu}^{-1}_4), \hspace{0.2cm} M_i^2=m^2 + r_i^2 \\
&\bar{\mu}_{total}=\Tilde{\mu}+ \frac{1}{2} \sum_{i=1}^{4} \nu_i^{-1}+ \frac{1}{2} \sum_{i=1}^{4} \psi_i^{-1}+\frac{1}{2}\sum_{i=1}^4{\sigma_i}{\theta_i^*}^2\\
&\phi_{total}=\min[\phi_0, \phi_1,..., \phi_4]
\end{aligned}
\end{equation}
As per the definitions in Equation \eqref{78}, $\bar{\mu}_{total}$, $\bar{\mu}$, and $\phi_{total}$ are known to be positive constants, while $M_{i}$ is a function yielding solely positive values. Therefore, we can conclude the demonstration of Theorem 1; see Appendix B for more details. Similar to Remark 4, based on Definition 1, Appendix A, and according to \cite{heydari2024robust} and \cite{corless1993bounded}, each RSBA-applied EMLA is uniformly exponentially bounded within a specific ball, as:
\begin{equation}
\small
\begin{aligned}
\label{0055}
\|\bm{P}\| \leq \sqrt{\frac{2 V\left(t_0\right)}{\lambda_{min} ({1-\overset{*}{Z}})} } e^{-\frac{\bar{\iota}}{2}(t-t_0)} + \sqrt{\frac{2 \bar{\mu}_{total}}{\lambda_{\min} \phi_{total}(1-\overset{*}{Z})}}
\end{aligned}
\end{equation}

\indent \textbf{Theorem 2} Consider a PMSM-driven EMLA-actuated $n_a$-DoF HDRM system, and its EMLA equations can be provided by \eqref{26}. By employing the modular RSBA control specified in \eqref{40} for the entire manipulator system, along with the robust observer methodology given in Section \ref{observer}, in conjunction with the adaptive laws presented in Equations. \eqref{23} and \eqref{32}, the tracking error of the reference trajectories of all EMLAs uniformly and exponentially converges to the bounded value $\bar{\tau}_0$, even in the presence of non-triangular uncertainties and external disturbances.\\
\indent \textbf{Proof}: As observed, the exponential stability of each EMLA-actuated joint has been achieved by employing modular RSBA control in Theorem 1. By considering a Lyapunov function as the sum of the Lyapunov functions of all joints in the form of \eqref{72}, continuing the same steps, and increasing the dimension of the matrix $\bm{P}$ from $6$ to $6n_a$, exponential stability for the entire $n_a$-DoF EMLA-actuated manipulator is straightforwardly obtained; see Appendix C for more details. Similar to Remark 4, based on Definition 1 and Appendix A, and according to \cite{heydari2024robust} and \cite{corless1993bounded}, each RSBA-applied EMLA is uniformly exponentially bounded within a specific ball, the radius of which depends on the noise and disturbance intensity.
\\
\indent \textbf{Remark 6} This paper suggests a departure from the conventional approach of designing a control strategy for HDRMs to stabilize the entire system as a single entity. Instead, we introduce modular tools (as outlined in Remark 3, Theorem 1, and Theorem 2) that achieve the following objectives:
\begin{itemize}
\item Automatically stabilize adjacent subsystems of the EMLA-actuated HDRM to ensure uniformly exponential stability.
\item Prevent an undue increase in the complexity of the control design for high-order DoFs.
\item Allow independent control modifications for each subsystem (without affecting control laws in other subsystems), while still guaranteeing the stability of the whole EMLA-actuated HDRM system, as demonstrated in the proof of Theorem 2.
\end{itemize}
\section{Simulation Results}
\label{simulation}
\subsection{3-DoF HDRM Motion Simulation}
\label{manipulator_simulation}
\begin{figure}[h!] 
    \centering
    \scalebox{0.6}
    {\includegraphics[trim={0cm 0.0cm 0.0cm
    0cm},clip,width=\columnwidth]{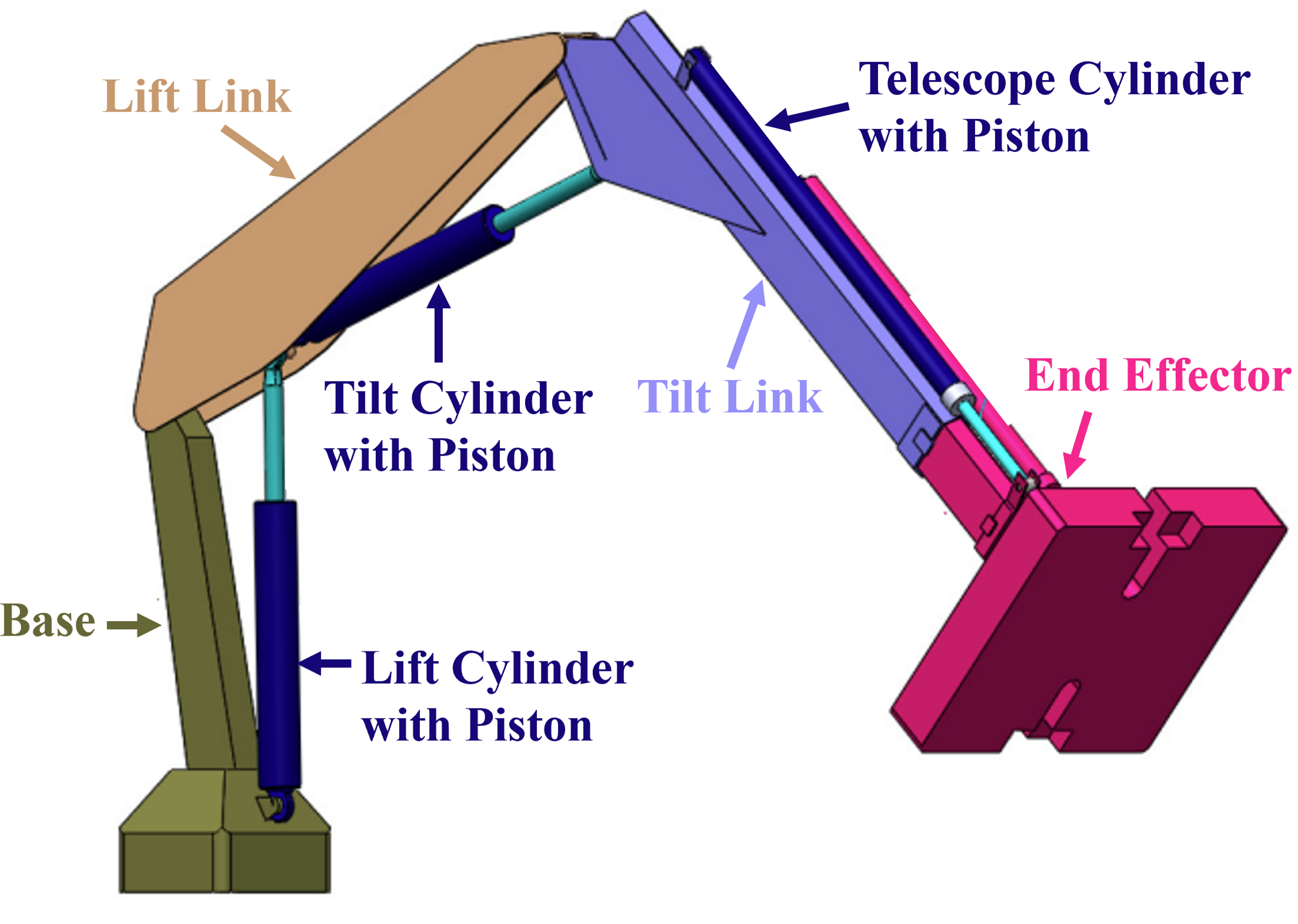}}
    \caption{Assembly of the studied 3-DoF HDRM}
    \label{fig:manipulator}
\end{figure}

In this part, we considered a 3-DoF parallel-serial manipulator actuated by PMSM-powered EMLAs for the case study, as shown in Fig. \ref{fig:manipulator}. For this case study, the values of the constraints of the motion generation optimization algorithm, according to \eqref{equation:costf2}-\eqref{equation:const2} in Section \ref{manipulator_motion}, are listed in Table \ref{manipulator_constraints}.
\begin{table}[h]
    \small
    \centering
    \caption{Parameters of the motion generation optimization algorithm for the 3-DoF HDRM}
    \begin{threeparttable}
        \begin{tabular}{c||c}
            \hline
            \hline
            \textbf{Term} & \textbf{Value}\\
            \hline
            \hline
            $\bf{q}_{S}$ & $\left[0.244, 0.389, 0.499\right]$ \SI{}{\meter}\\
            \hline
            $\bf{q}_{E}$ & $\left[0.270, 0.442, 0.348\right]$ \SI{}{\meter}\\
            \hline
            $\bf{v}_{S}$ & $\left[0.0016, 0, -0.0024\right]$ \SI{}{\meter\per\second}\\
            \hline
            $\bf{v}_{E}$ & $\left[0.0607, -0.0808, -0.5477\right]$ \SI{}{\meter\per\second} \\
            \hline
            $\bf{q}_{LB}$ & $\left[0, 0, 0\right]$ \SI{}{\meter}\\
            \hline
            $\bf{q}_{UB}$ & $\left[0.522, 0.611, 1\right]$ \SI{}{\meter}\\
            \hline
            $\bf{v}_{LB}$ & $\left[-0.136, -0.100, -0.210\right]$ \SI{}{\meter\per\second}\\
            \hline
            $\bf{v}_{UB}$ & $\left[0.136, 0.100, 0.210\right]$ \SI{}{\meter\per\second}\\
            \hline
            $\bf{f}_{LB}$ & $\left[-89.8, -68.6, -6.1\right]$ \SI{}{\kilo\newton}\\
            \hline
            $\bf{f}_{UB}$ & $\left[89.8, 68.6, 6.1\right]$ \SI{}{\kilo\newton} \\
            \hline
            $t_{M_{m}}$ & \SI{37.7}{\second} \\
            \hline
            \hline
        \end{tabular}
        \begin{tablenotes}  
            \small
            \item[- Lowercase bold symbols represent vectors.] 
        \end{tablenotes}
    \end{threeparttable}
    \label{manipulator_constraints}
\end{table}

\begin{figure}[h!] 
    \centering
    \scalebox{1}
    {\includegraphics[trim={0.1cm 0.1cm 0.1cm
    0.1cm},clip,width=\columnwidth]{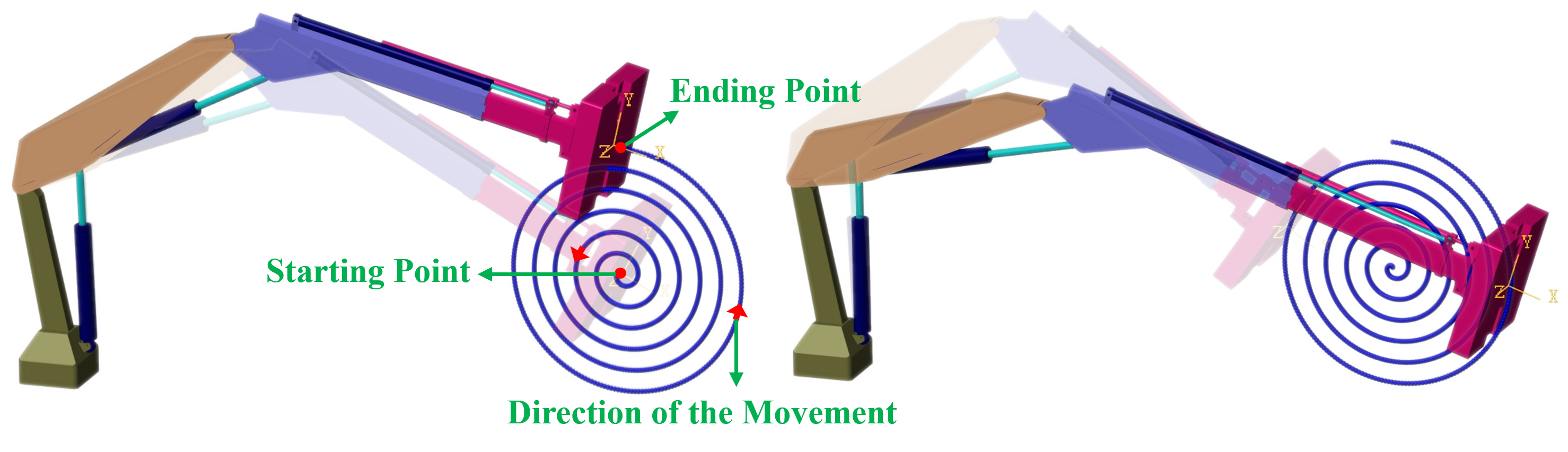}}
    \caption{Visualized movement of the studied 3-DoF HDRM in the Cartesian space}
    \label{fig:manipulator_motion}
\end{figure}

\begin{figure}[h!] 
    \centering
    \scalebox{0.9}
    {\includegraphics[trim={0.1cm 0.1cm 0.1cm
    0.1cm},clip,width=\columnwidth]{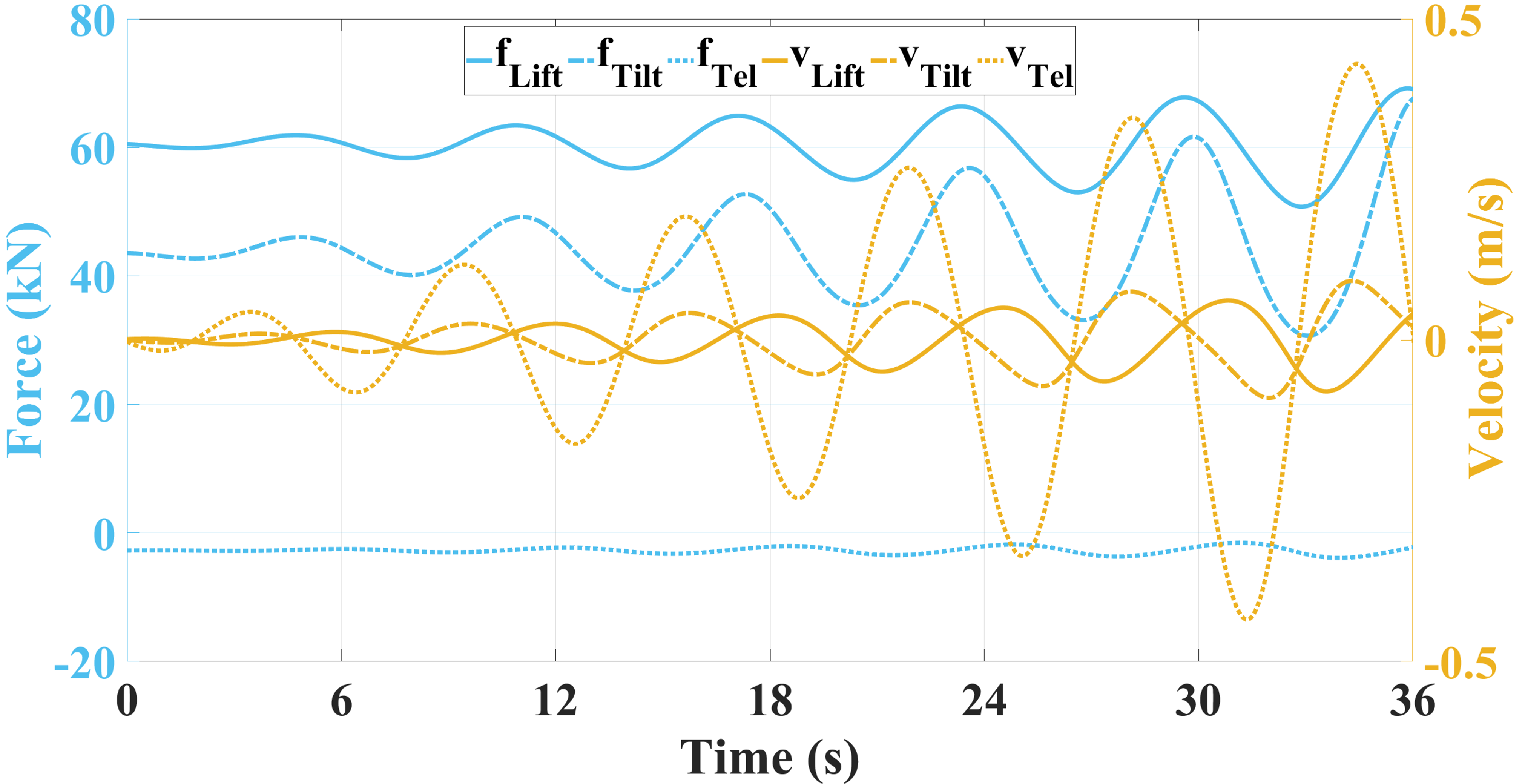}}
    \caption{Forces and velocities in the lift, tilt, and telescope pistons of the studied 3-DoF HDRM}
    \label{fig:force_velocity}
\end{figure}

\begin{figure}[h!] 
    \centering
    \scalebox{0.9}
    {\includegraphics[trim={0.1cm 0.1cm 0.1cm
    0.1cm},clip,width=\columnwidth]{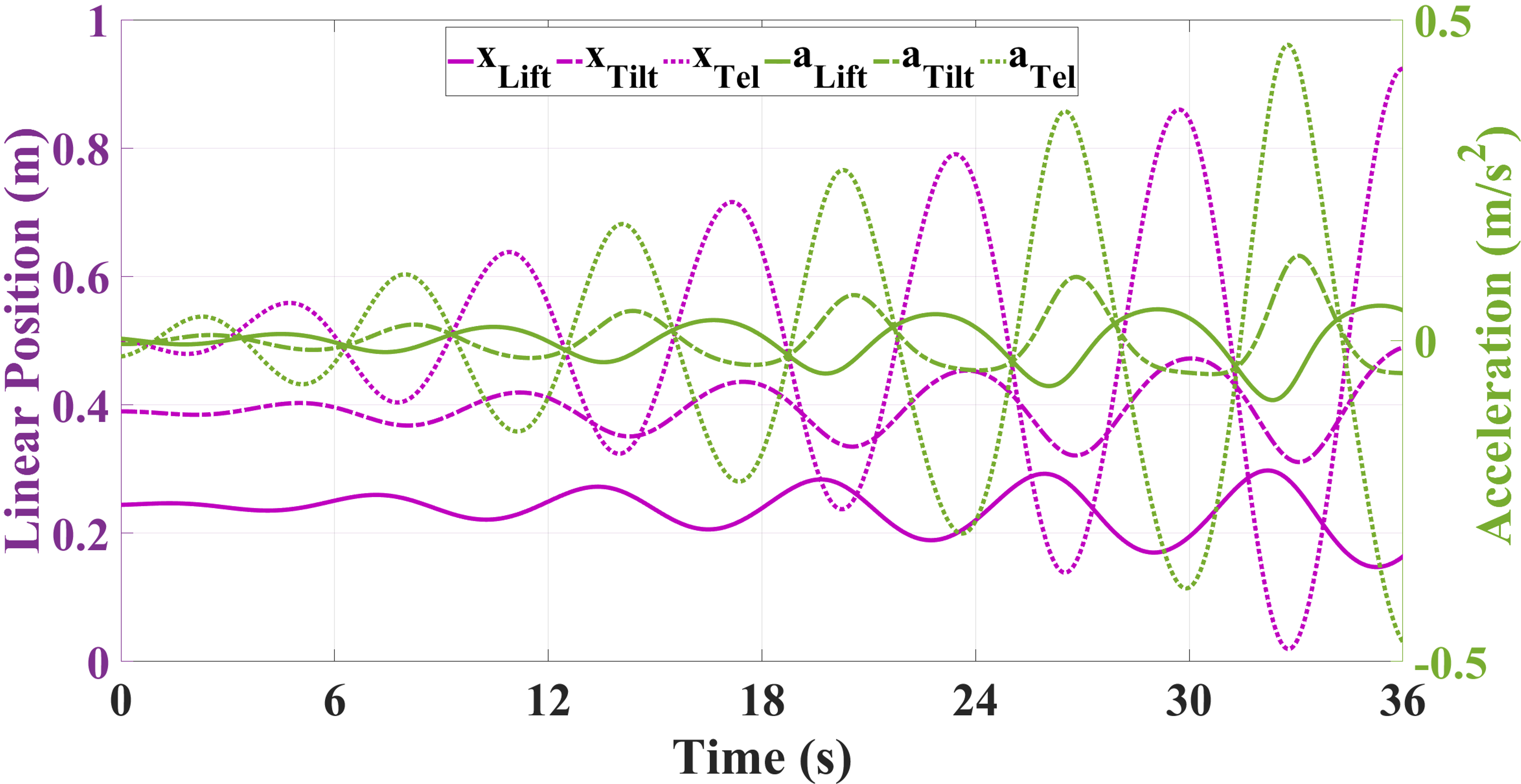}}
    \caption{Positions and accelerations in the lift, tilt, and telescope pistons of the studied 3-DoF HDRM}
    \label{fig:position_acceleration}
\end{figure}

The implemented predefined task for the HDRM is visualized in Fig. \ref{fig:manipulator_motion}. In addition, we incorporate a 470-kg payload at the manipulator's end-effector. Linear forces and linear velocities of EMAL-actuated joints of the HDRM, including lift, tilt, and telescope, are obtained as shown in Fig. \ref{fig:force_velocity}. Further, the linear position and acceleration of the mentioned joints of the HDRM are illustrated in Fig. \ref{fig:position_acceleration}.
The generated position and velocity trajectories of the three studied EMLAs will serve as reference trajectories for the motion dynamics, outlined in Eq. \eqref{equation:StateSpace3} in Section \ref{EMLA_motion}, to meet the control task at the joint level.
The three EMLA models that were chosen for actuating the HDRM in the simulation section of this paper are listed in Table \ref{EMLA_Table}.


\begin{table}[h!]
    \scriptsize
    \centering
    \caption{The list of selected EMLAs to actuate the joint pistons of the studied manipulator}
    \begin{tabular}{c|c|c}
        \hline
        \hline
        \textbf{Joint} & \textbf{EMLA Model} & \textbf{Motor/Manufacturer}\\
        \hline
        \hline
        \textbf{Lift} & SRSA-S-7520 & MCS19P29/Lenze\\
        \hline
        \textbf{Tilt} & SRSA-S-6010 & MCS19J30/Lenze \\
        \hline
        \textbf{Telescope} & CASM-100-BB & 1FK7064/Siemens\\
        \hline
        \hline
    \end{tabular}
    \label{EMLA_Table}
\end{table}
Meanwhile, Table \ref{EMLA_details} provides an overview of the essential features of the components within the EMLA mechanism, including linear units, the gearbox, and the electric motor.

\subsection{Implementation of RSBA Control for the Joint Level of the Simulated 3-DoF HDRM}
\label{c-simulation}
 To investigate the effectiveness of the proposed control, we apply the RSBA control to each PMSM-powered EMLA-actuated joint of the simulated HDRM to track the generated motion trajectories visualized in Figs. \ref{fig:force_velocity} and \ref{fig:position_acceleration} as the motion reference. This setup enables the studied manipulator to execute the visualized movement in Cartesian space, as illustrated in Fig. \ref{fig:manipulator_motion}, while supporting a payload of 470 kg at the end-effector. Except for the load force effects provided in Fig. \ref{fig:force_velocity}, we consider external disturbances and sensor noise ($d_2$ with $N$ unit) as follows:
\begin{equation}
\begin{aligned}
\label{98}
&\text{Lift:} \hspace{0.1cm} 0.8 \sin(8t) + 0.7 \sin(20t)+1.2\text{arctan}(x_2)e^{-3\text{t}} \\
&\text{Tilt:} \hspace{0.1cm} 0.8 \sin(7t) + 0.95 \sin(18t)+\text{rand}(0,2)\\
&\text{Telescope:} \hspace{0.1cm} 0.9 \sin(65t) -0.3 \sin(50t+ \pi/{4})
\end{aligned}
\end{equation}
As voltage disturbances ($d_3$ and $d_4$), we consider the following functions (with $v$ unit) for all three EMLAs: 
\begin{equation}
\begin{aligned}
\label{99}
&d_3=0.42 \cos(8t+ \frac{\pi}{3}) + 0.032 \sin(5t)\\
&d_4=-0.15 \cos(12t+ \frac{\pi}{6})
\end{aligned}
\end{equation}
In addition, we assume that the following non-triangular uncertainties occur for all three EMLAs:
\begin{equation}
\begin{aligned}
\label{100}
&F_2= \hspace{0.1cm} \frac{1.5N_{p}(x_3 x_4 L_d-x_3 x_4 L_q)-B_{eq}x_2-C_{eq}x_1}{100A_{eq}}\\
&F_3= \hspace{0.1cm} \frac{- R_s x_3- N_{p} \alpha_{RL} x_2 ( x_4 L_d + \Phi_{PM})}{500L_q}\\
&F_4= \hspace{0.1cm} \frac{- R_s x_4+ N_{p} \alpha_{RL} x_2 x_3 L_q}{1000L_d}
\end{aligned}
\end{equation}
\eqref{99} implies that the uncertainties imposed on the systems are $1\%$, $0.2\%$, and $0.1\%$ errors in the modeling of velocity, $q$-axis, and $d$-axis current equations. According to the nature of our work, we assume that all uncertainties and disturbances are unknown. 
Based on our understanding, we have identified two impressive pieces of research focusing on PMSM control with an adaptive approach.
The first one, cited in \cite{liu2023command}, presents a modified adaptive control scheme based on command-filtered backstepping. This command-filter-approximator-based adaptive
control (CAC) utilizes a stabilizing function through the hyperbolic tangent function. The second study, cited in \cite{zhang2023adaptive}, introduces an adaptive neural asymptotic tracking control (ANATC) scheme for PMSM systems, accounting for current constraints and unknown dynamics, while assuming that $i_d$ equals zero to avoid non-triangular uncertainties. Based on our findings, both papers demonstrate a strong control performance compatible with the studied PMSM-powered EMLAs compared with others. Therefore, we have chosen them as two reliable references to provide detailed comparisons, showcasing the effectiveness of our work. After ensuring consistent conditions across our application, including linear positions and velocities (see Section \ref{manipulator_simulation}), modeling parameters of the three EMLAs (see Table \ref{EMLA_details}), external disturbances (see Eqs. \eqref{98} and \eqref{99}), and uncertainties (see Eq. \eqref{100}), we selected the initial condition and the values of the control parameters corresponding to the aforementioned control strategies, as outlined in Table \ref{gains}.
\small
\begin{table}[h!]
    \small
    \centering
    \caption{Technical data and characteristics of the selected EMLAs for implementation in the manipulator's joints}
    \begin{tabular}{c||c||c||c}
        \hline
        \hline
          \multicolumn{1}{c}{\textbf{Term}} & \multicolumn{1}{c}{\textbf{Lift EMLA}} & \multicolumn{1}{c}{\textbf{\textcolor{black}{Tilt EMLA}}} & \multicolumn{1}{c}{\textbf{Tel. EMLA}}\\
        \hline
        \hline
          \multicolumn{4}{c}{\textit{Linear Unit and Gearbox}}\\
        \hline
        \hline
            $F_{c0}$& \SI{145.7}{\kilo\newton} & \SI{120.6}{\kilo\newton}&\SI{6.4}{\kilo\newton}\\
        \hline
            $F_{c}$&\SI{89.8}{\kilo\newton}&\SI{68.6}{\kilo\newton}&\SI{6.1}{\kilo\newton}\\
        \hline
            $F_{p0}$&\SI{261.1}{\kilo\newton}&\SI{199.7}{\kilo\newton}&\SI{17.1}{\kilo\newton}\\
        \hline
            $F_{p}$&\SI{147.4}{\kilo\newton}&\SI{145.5}{\kilo\newton}&\SI{17.1}{\kilo\newton}\\
        \hline
            $a_{max}$&\SI{3.4}{\meter\per\second^2}&\SI{2}{\meter\per\second^2}&\SI{6}{\meter\per\second^2}\\
        \hline
            $v_{max}$&\SI{136}{\milli\meter\per\second}&\SI{100}{\milli\meter\per\second}&\SI{210}{\milli\meter\per\second}\\
        \hline
            $M_{BS}$&\SI{156.5}{\kilo\gram}&\SI{83.6}{\kilo\gram}&\SI{30.4}{\kilo\gram}\\
        \hline
            $l$&\SI{0.02}{\meter}&\SI{0.01}{\meter}&\SI{0.01}{\meter}\\
        \hline
            $d$&\SI{0.075}{\meter}&\SI{0.06}{\meter}&\SI{0.04}{\meter}\\
        \hline
            $1/{\rho}$&\SI{7}{}&\SI{5}{}&\SI{1}{}\\
        \hline
        \hline
          \multicolumn{4}{c}{\textit{Electric Motor (PMSM)}}\\
        \hline
        \hline
            $J_m$&\SI{0.016}{\kilogram\meter^2}&\SI{0.0105}{\kilogram\meter^2}&\SI{0.00085} {\kilogram\meter^2}\\
        \hline
            $n_N$&\SI{2850}{rpm}&\SI{3000}{rpm}&\SI{3000}{rpm}\\
        \hline
            $\tau_N$&\SI{53}{\newton\cdot\meter}&\SI{29}{\newton\cdot\meter}&\SI{8}{\newton\cdot\meter}\\
        \hline
            $\tau_{max}$&\SI{190}{\newton\cdot\meter}&\SI{129}{\newton\cdot\meter}&\SI{32}{\newton\cdot\meter}\\
        \hline
            $i_N$&\SI{29.5}{\ampere} & \SI{18.5}{\ampere} & \SI{7.6}{\ampere}\\
        \hline
            $u_N$&\SI{315}{\volt} & \SI{300}{\volt} & \SI{600}{\volt}\\
        \hline
            $P_N$&\SI{15.8}{\kilo\watt} & \SI{9.1}{\kilo\watt} & \SI{2.5}{\kilo\watt}\\
        \hline
            $R_s$&\SI{0.14}{\ohm}&\SI{0.16}{\ohm}&\SI{0.35}{\ohm}\\
        \hline
            $L_N$&\SI{2.4}{\milli\henry}&\SI{3.2}{\milli\henry}&\SI{12}{\milli\henry}\\
        \hline
            $\Phi_{P\!M}$&\SI{0.15}{\weber}&\SI{0.13}{\weber}&\SI{0.12}{\weber}\\
        \hline
            $N_p$&\SI{8}{}&\SI{8}{}&\SI{6}{}\\
        \hline
        \hline
    \end{tabular}
    \label{EMLA_details}
\end{table}
\normalsize

Furthermore, to implement the RSBA control algorithm in the studied application, we utilized the proposed observer in Section \ref{observer} for all three EMLAs to estimate true linear positions and velocities. For all, we assumed $C=[1,0]$. The value of the vector $C$ implies that the observer utilizes the linear position information from the position sensor as the output of the system. We used the following values for the three EMLAs' observers:
\begin{equation}
\begin{aligned}
\label{101}
&\bm{\alpha} = 
\begin{bmatrix}
    0.3192 \\
    0.3129
\end{bmatrix}, \bm{p} = 
\begin{bmatrix}
    1.4078 & -0.1975 \\
   -0.1975 & 4.4535
\end{bmatrix},\\
&
\bm{\bar{A}}=
\begin{bmatrix}
   -0.3192 & 1 \\
   -0.3129 & 0
\end{bmatrix},
\bm{Q} = 
\begin{bmatrix}
    0.7752 & -0.0775 \\
   -0.0775 & 0.3949
\end{bmatrix}
\end{aligned}
\end{equation}
It is noteworthy that the matrices $\bm{p}$ and $\bm{Q}$ were calculated by the command $\bm{p}=\operatorname{lyap}\left(\bm{\bar{A}^{\top}}, \bm{Q}\right)$ in MATLAB. The function $H(y(t), t)$ was defined as follows:
\begin{equation}
\small
\begin{aligned}
\label{equaation: 102}
H(y(t), t)=20 \cos(y) ^4 +  20\sin(y) ^4
\end{aligned}
\end{equation}
Then, we selected: $\ell=1, \quad m = 200 e^{-0.001t}$. In addition, we assigned the initial conditions for the estimated observer parameters as follows:
\begin{equation}
\small
\begin{aligned}
\label{103}
&\text{Lift:}&& \hspace{0.1cm} \bm{\hat{x}_0}=[0.24, 0]^{\top}, \hat{\eta}(0)=1\\
&\text{Tilt:}&& \hspace{0.1cm} \bm{\hat{x}_0}=[0.358, 0]^{\top}, \hat{\eta}(0)=0.5\\
&\text{Telescope:}&& \hspace{0.1cm} \bm{\hat{x}_0}=[0.5, 0]^{\top}, \hat{\eta}(0)=0.2
\end{aligned}
\end{equation}

\begin{table}[h]
    \centering
    \caption{Initial condition of the system states and the parameters of control strategies}
      \small
    \begin{tabular}{c||c||c||c}
        \hline
        \hline
          \multicolumn{1}{c}{\textbf{Term}} & \multicolumn{1}{c}{\textbf{Lift}} & \multicolumn{1}{c}{\textbf{Tilt}} & \multicolumn{1}{c}{\textbf{Tel.}}\\
        \hline
        \hline
          \multicolumn{4}{c}{\textit{Initial values of the system states}}\\
        \hline
        \hline
            $x_L(0)$&\SI{0.24}{\meter}&$0.385$ m&$0.5$ m\\
        \hline
            $\dot{x}_L(0)$&$0$ m/s&$0$ m/s&$0$ m/s\\
        \hline
            $i_q(0)$&\SI{0}{\ampere}&\SI{0}{\ampere}&\SI{0}{\ampere}\\
        \hline
            $i_d(0)$&\SI{0}{\ampere}&\SI{0}{\ampere} &\SI{0}{\ampere}\\
        \hline
        \hline
        \multicolumn{4}{c}{\textit{RSBA control approach parameters}}\\
        \hline
        \hline
            $\beta_{1,2}$&$3000$&$2000$&$800$\\
        \hline
            $\zeta_{1,2}$&$100$&$100$&$0.001$\\
        \hline
            $\delta_{1,2}$&$100$&$100$&$1$\\
        \hline
            $\sigma_{1,2}$&$0.001$&$0.001$&$1$\\
        \hline
            $\beta_{3,4}$&$1000$&$750$&$420$\\
        \hline
            $\zeta_{3,4}$&$100$&$80$&$1$\\
        \hline
            $\delta_{3,4}$&$110$&$80$&$0.5$\\
        \hline
            $\sigma_{3,4}$&$0.01$&$0.01$&$1$\\
        \hline
        \hline
          \multicolumn{4}{c}{\textit{CAC approach parameters \cite{liu2023command}}}\\
        \hline
        \hline
            $k_{1,2}$& $230$ & $210$ & $175$\\
        \hline
            $k_{3,5}$&$0.05$&$0.03$&$0.03$\\
        \hline
            $k_4$&$1$&$1$&$0.5$\\
        \hline
            $\varpi_{1,2,4}$&$3$&$2$&$1$\\
        \hline
            $\varpi_{3,5}$&$0.03$&$0.02$&$0.03$\\
        \hline
            $m_{1,3}$&$1$&$1$&$0.9$\\
        \hline
            $m_2$&$12$&$10.5$&$8$\\
        \hline
        \hline
                  \multicolumn{4}{c}{\textit{ANATC approach parameters \cite{zhang2023adaptive}}}\\
        \hline
        \hline
            $k_{1}$& $0.14$ & $0.24$ & $0.1$\\
        \hline
            $k_{2}$&$0.7$&$0.75$&$0.7$\\
        \hline
            $g$&$0.075$&$0.075$&$0.075$\\
        \hline
            $\lambda_{1}$&$1.22$&$1.08$&$1.15$\\
        \hline
            $\lambda_{2}$&$0.66$&$0.85$&$0.78$\\
        \hline
            $\ell_{1,2}$&$0.03$&$0.03$&$0.01$\\
        \hline
            $b$&$2$&$2$&$2$\\
        \hline
        \hline
    \end{tabular}
    \label{gains}
\end{table}

\begin{figure}[h!] 
    \centering
    \scalebox{0.9}
    {\includegraphics[trim={0.1cm 0.1cm 0.1cm
    0.1cm},clip,width=\columnwidth]{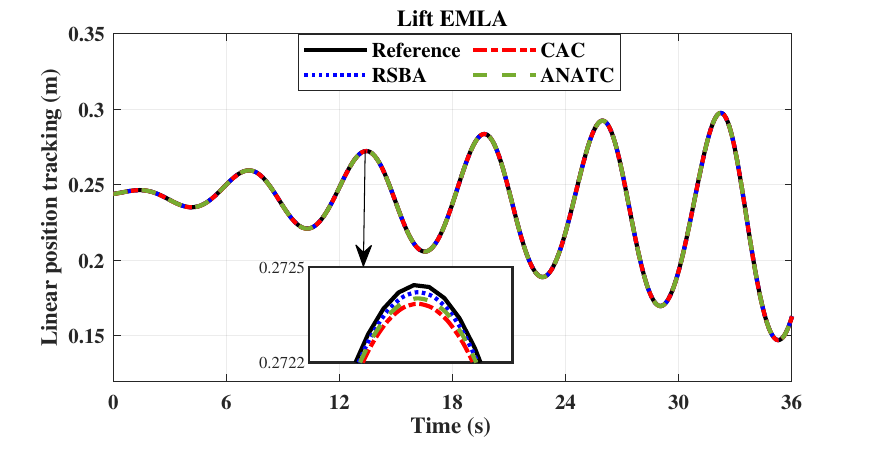}}
    \caption{Position tracking in the lift piston of the studied 3-DoF HDRM}
    \label{p_t1}
\end{figure}

\begin{figure}[h!] 
    \centering
    \scalebox{0.9}
    {\includegraphics[trim={0.1cm 0.1cm 0.1cm
    0.1cm},clip,width=\columnwidth]{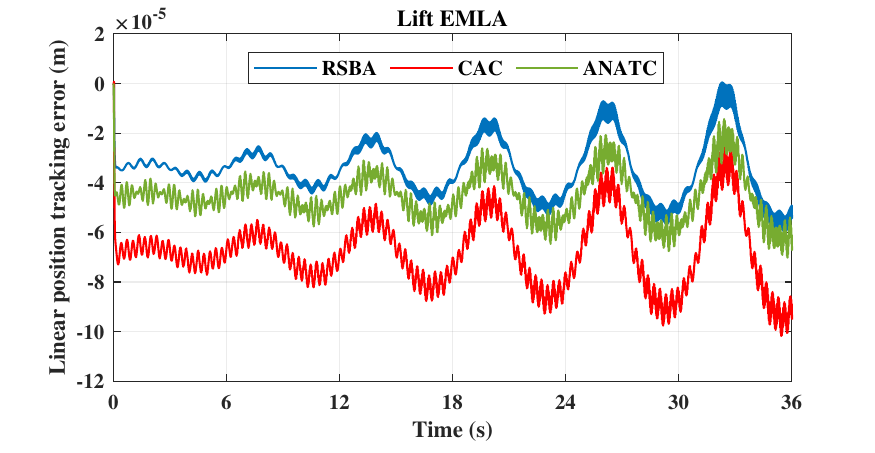}}
    \caption{Position tracking error in the lift piston of the studied 3-DoF HDRM}
    \label{p_e1}
\end{figure}

\begin{figure}[h!] 
    \centering
    \scalebox{0.9}
    {\includegraphics[trim={0.1cm 0.1cm 0.1cm
    0.1cm},clip,width=\columnwidth]{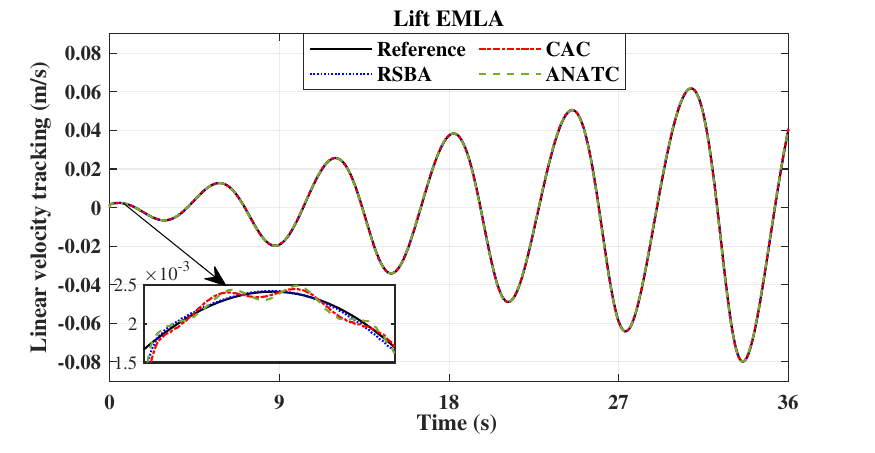}}
    \caption{Velocity tracking in the lift piston of the studied 3-DoF HDRM}
    \label{v_t1}
\end{figure}

\begin{figure}[h!] 
    \centering
    \scalebox{0.9}
    {\includegraphics[trim={0.1cm 0.1cm 0.1cm
    0.1cm},clip,width=\columnwidth]{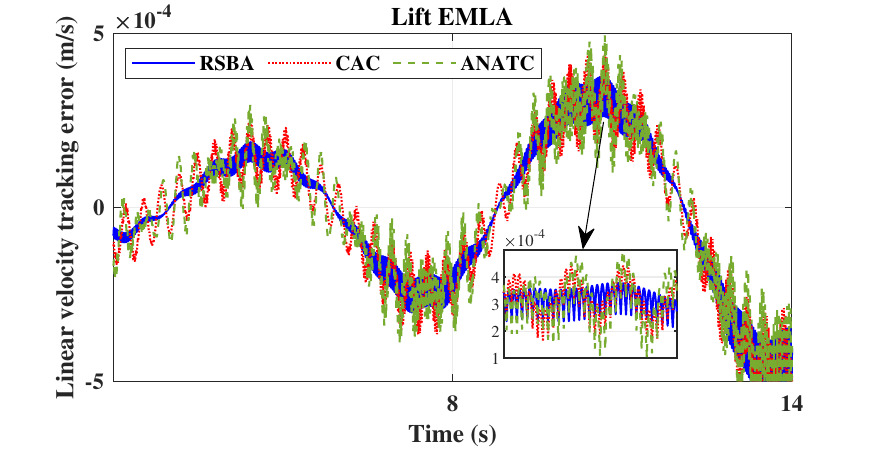}}
    \caption{Velocity tracking error in the lift piston of the studied 3-DoF HDRM}
    \label{v_e1}
\end{figure}

\begin{figure}[h!] 
    \centering
    \scalebox{0.9}
    {\includegraphics[trim={0.1cm 0.1cm 0.1cm
    0.1cm},clip,width=\columnwidth]{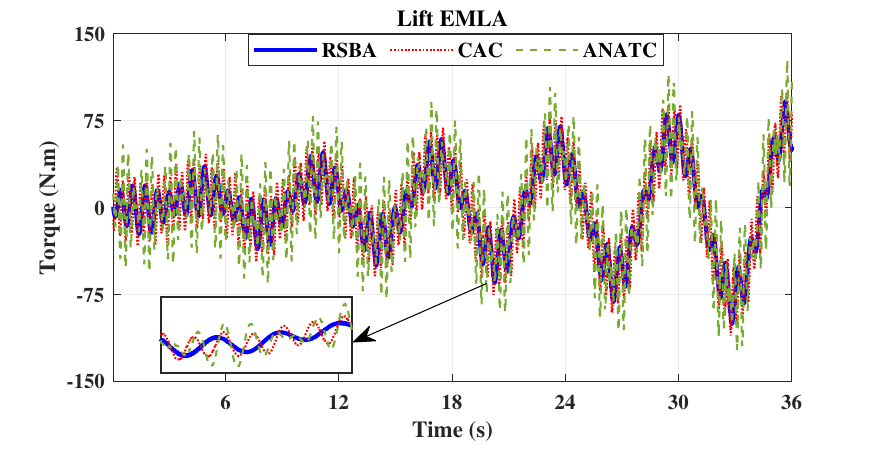}}
    \caption{Torque generated in the lift piston of the studied 3-DoF HDRM}
    \label{t1}
\end{figure}

In the same conditions, the three control strategies were implemented across all three EMLA scenarios to track the control tasks. Figs. (\ref{p_t1})-(\ref{t1}) depict the results pertaining to the lift piston. Figs. (\ref{p_t1}) and (\ref{p_e1}) demonstrate that the three controllers effectively perform linear position tracking for the first EMLA with high accuracy ($10^{-5}$). In addition, rapid convergence (less than $0.1$ sec) was observed among them.

\begin{figure}[h!] 
    \centering
    \scalebox{0.9}
    {\includegraphics[trim={0.1cm 0.1cm 0.1cm
    0.1cm},clip,width=\columnwidth]{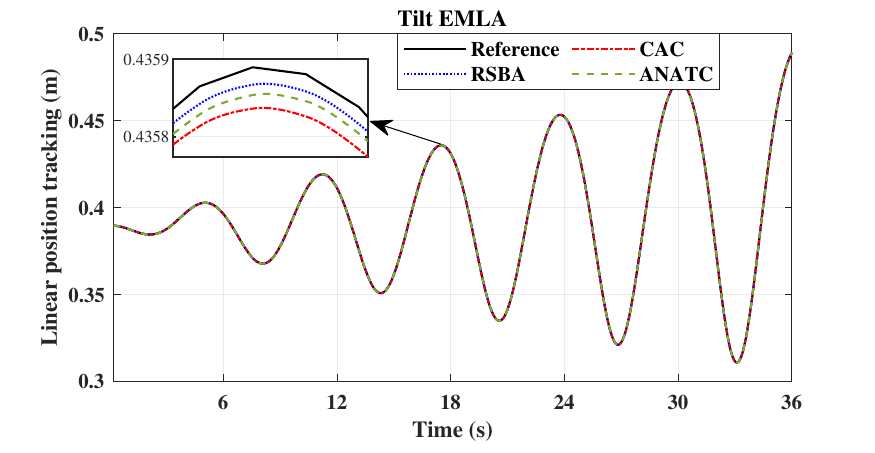}}
    \caption{Position tracking in the tilt piston of the studied 3-DoF HDRM}
    \label{p_t2}
\end{figure}

\begin{figure}[h!] 
    \centering
    \scalebox{0.9}
    {\includegraphics[trim={0.1cm 0.1cm 0.1cm
    0.1cm},clip,width=\columnwidth]{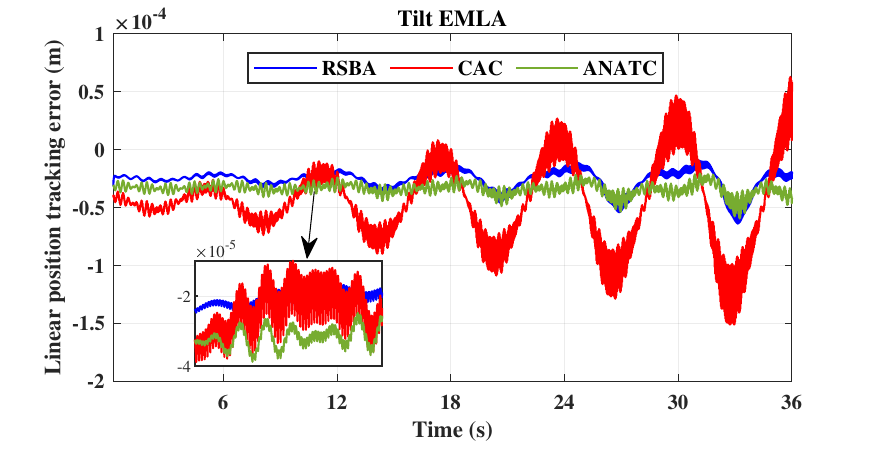}}
    \caption{Position tracking error in the tilt piston of the studied 3-DoF HDRM}
    \label{p_e2}
\end{figure}

\begin{figure}[h!] 
    \centering
    \scalebox{0.9}
    {\includegraphics[trim={0.1cm 0.1cm 0.1cm
    0.1cm},clip,width=\columnwidth]{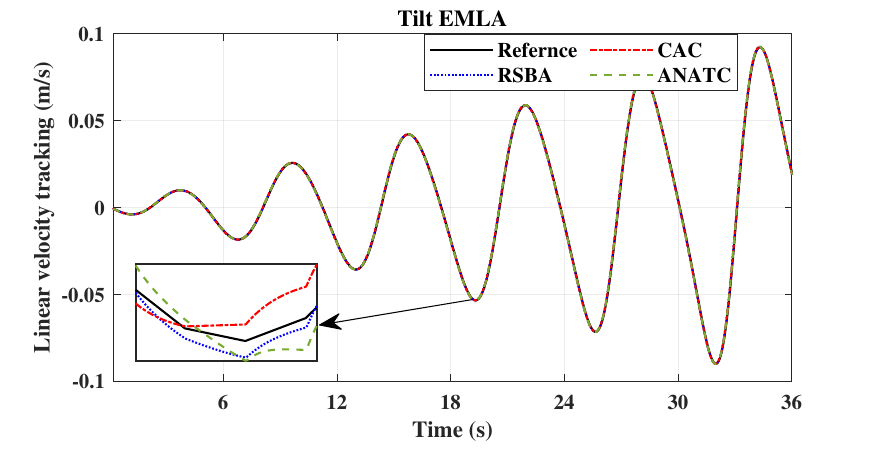}}
    \caption{Velocity tracking in the tilt piston of the studied 3-DoF HDRM}
    \label{v_t2}
\end{figure}

\begin{figure}[h!] 
    \centering
    \scalebox{0.9}
    {\includegraphics[trim={0.1cm 0.1cm 0.1cm
    0.1cm},clip,width=\columnwidth]{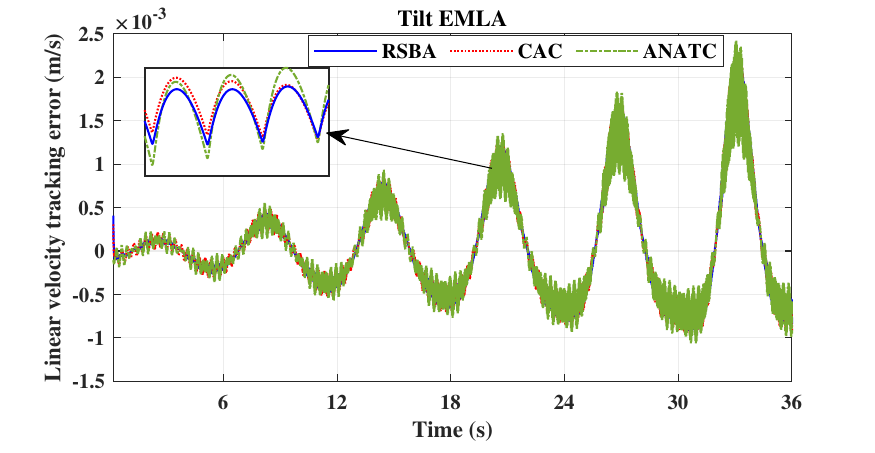}}
    \caption{Velocity tracking error in the tilt piston of the studied 3-DoF HDRM}
    \label{v_e2}
\end{figure}

\begin{figure}[h!] 
    \centering
    \scalebox{0.9}
    {\includegraphics[trim={0.1cm 0.1cm 0.1cm
    0.1cm},clip,width=\columnwidth]{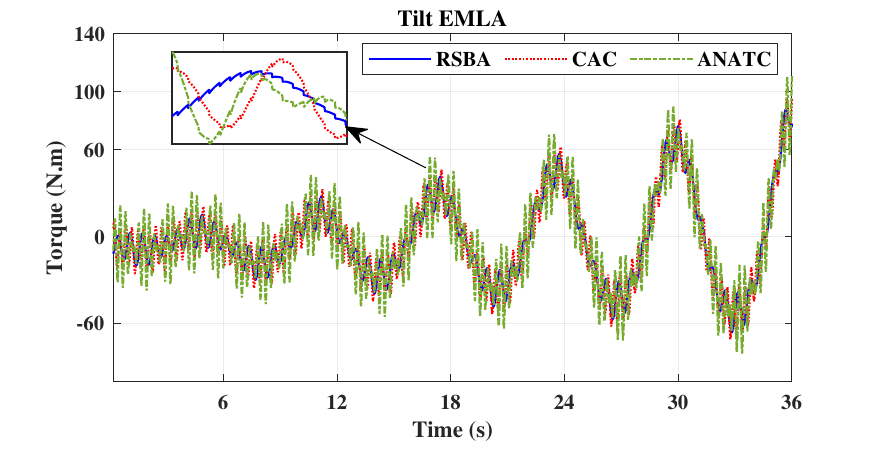}}
    \caption{Torque generated in the tilt piston of the studied 3-DoF HDRM}
    \label{t2}
\end{figure}

Similar results are observed for linear velocity tracking in Figs. (\ref{v_t1}) and (\ref{v_e1}). However, the RSBA control outcomes exhibited smoother and better accuracy in performance compared to the others. Furthermore, the tracking errors depicted in Figs. (\ref{p_e1}) and (\ref{v_e1}) indicate that while ANATC and CAC are competitive, they lag behind RSBA control. Similar to velocities, the torque exerted by RSBA for the lift EMLA was smoother and required less effort, as shown in Fig. (\ref{t1}). The results concerning the performances of the controllers for the second piston (tilt EMLA) are presented in Figs. (\ref{p_t2})-(\ref{t2}).

Similarly, Figs. (\ref{p_t2})-(\ref{v_e2}) demonstrate performance in tracking position and velocity for all three control strategies. An interesting observation is that in both lift and tilt EMLAs, RSBA and ANATC exhibited better accuracy in position tracking than CAC, whereas RSBA and CAC showed a better velocity tracking performance. Furthermore, RSBA and ANATC showed similar results in position tracking for the tilt EMLA (Fig. \ref{p_e2}), although the ANATC error is $0.000001$ less than that of RSBA. 
Fig. (\ref{t2}) illustrates the torque efforts related to the tilt EMLA. With the exception of the RSBA control result, which is smoother, CAC occasionally outperformed ANATC in torque during certain periods. 

\begin{figure}[h!] 
    \centering
    \scalebox{0.9}
    {\includegraphics[trim={0.1cm 0.1cm 0.1cm
    0.1cm},clip,width=\columnwidth]{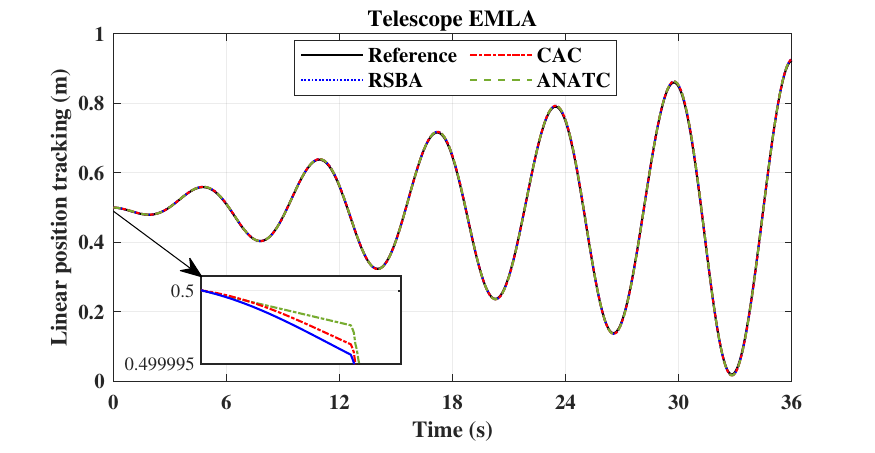}}
    \caption{Position tracking in the telescope piston of the studied 3-DoF HDRM}
    \label{p_t3}
\end{figure}

\begin{figure}[h!] 
    \centering
    \scalebox{0.9}
    {\includegraphics[trim={0.1cm 0.1cm 0.1cm
    0.1cm},clip,width=\columnwidth]{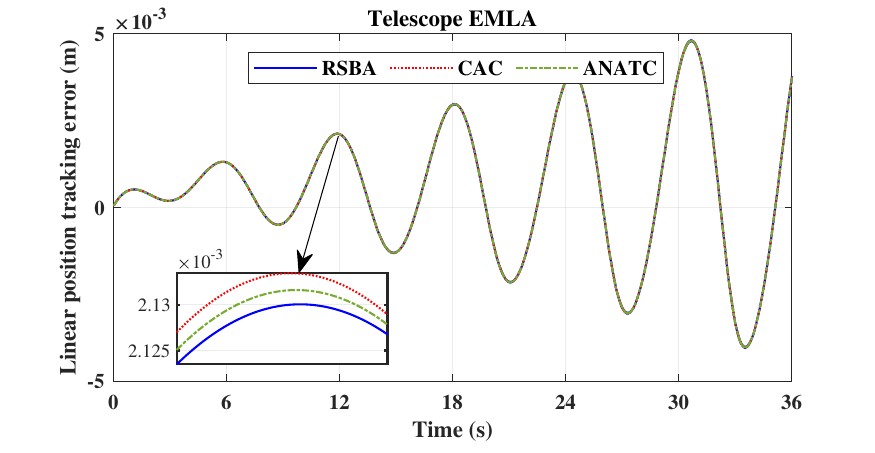}}
    \caption{Position tracking error in the telescope piston of the studied 3-DoF HDRM}
    \label{p_e3}
\end{figure}

\begin{figure}[h!] 
    \centering
    \scalebox{0.9}
    {\includegraphics[trim={0.1cm 0.1cm 0.1cm
    0.1cm},clip,width=\columnwidth]{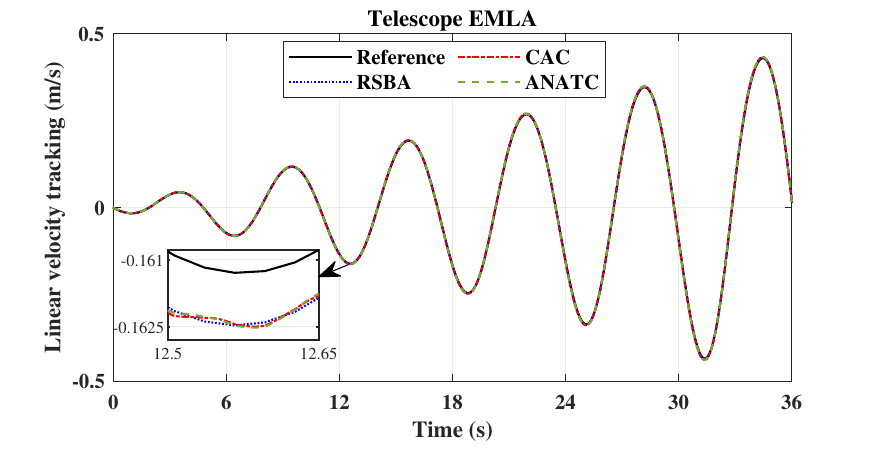}}
    \caption{Velocity tracking in the telescope piston of the studied 3-DoF HDRM}
    \label{v_t3}
\end{figure}

\begin{figure}[h!] 
    \centering
    \scalebox{0.9}
    {\includegraphics[trim={0.1cm 0.1cm 0.1cm
    0.1cm},clip,width=\columnwidth]{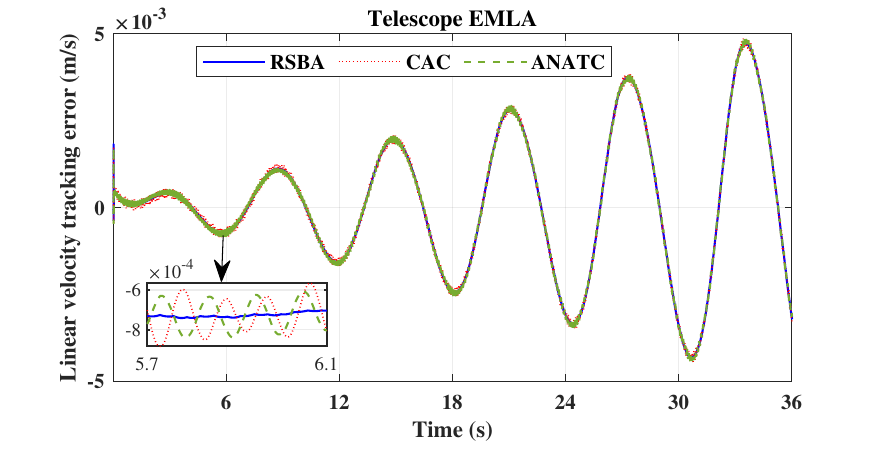}}
    \caption{Velocity tracking error in the telescope piston of the studied 3-DoF HDRM}
    \label{v_e3}
\end{figure}

\begin{figure}[h!] 
    \centering
    \scalebox{0.9}
    {\includegraphics[trim={0.1cm 0.1cm 0.1cm
    0.1cm},clip,width=\columnwidth]{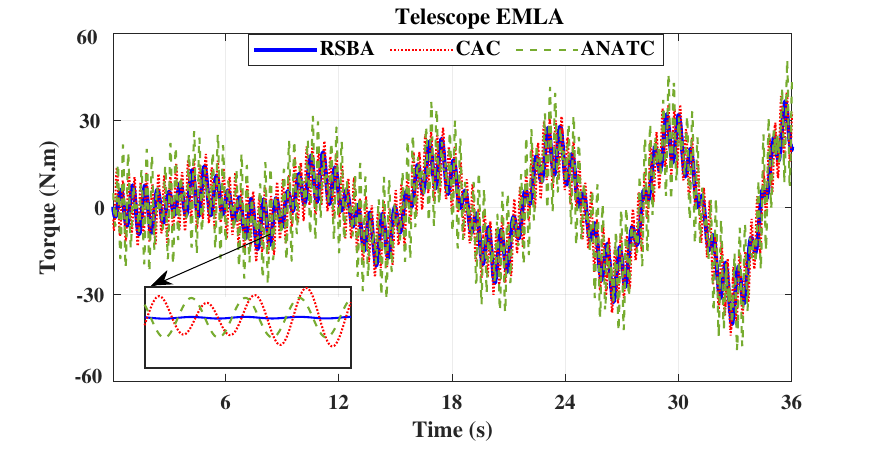}}
    \caption{Torque generated in the telescope piston of the studied 3-DoF HDRM}
    \label{t_3}
\end{figure}

Similarly, Figs (\ref{p_t3})-(\ref{t_3}) depict the results of controllers implemented on the last EMLA (telescope). 
In the close-scale Fig. (\ref{p_t3}), the speed convergence of the real position states to the reference is illustrated. It can be observed that the position raised by RSBA converged with a steeper slope than the others, followed by CAC and ANATC, although all the methods demonstrated high-speed convergence. Interestingly, the accuracies of both position and velocity tracking for all three methods are competitive (Fig. \ref{p_e3}). However, the error of position tracking by employing all three control strategies in the last EMLA increased compared with the previous EMLAs, in contrast with velocity results. 

\begin{table}[h!]
  \captionsetup{position=top}
  \caption{Control performance of each EMLA implemented by the RSBA control, CAC \cite{liu2023command}, and ANATC \cite{zhang2023adaptive} in tracking the linear motion reference}
  \centering
  \scriptsize
  \begin{tabular}{ccccc}
    \toprule
    \toprule
    \textcolor{black}{\textbf{HDRM}} &\textcolor{black}{\textbf{Convergence}} & \textcolor{black}{\textbf{RSBA}} &
    \textcolor{black}{\textbf{CAC \cite{liu2023command}}} &
    \textcolor{black}{\textbf{ANATC \cite{zhang2023adaptive}}}  \\
        \textcolor{black}{\textbf{joint}} &\textcolor{black}{\textbf{criteria}} & \textcolor{black}{\textbf{approach}} &
    \textcolor{black}{\textbf{approach}} &
    \textcolor{black}{\textbf{approach}} \\
    \midrule
    \multirow{4}{*}{\textbf{\textcolor{black}{Lift}}} & \textcolor{black}{{Pos. error (m)}} & \textcolor{darkergreen}{$0.00006$} & \textcolor{black}{$0.00009$} & \textcolor{black}{$0.00007$}  \\
 & \textcolor{black}{{Vel. error (m/s)}} & \textcolor{darkergreen}{$0.00038$} & \textcolor{black}{$0.00042$} & \textcolor{black}{$0.00049$}  \\
 & \textcolor{black}{{T. effort (N.m)}} & \textcolor{darkergreen}{$93.75$} & \textcolor{black}{$104.22$} & \textcolor{black}{$112.53$}  \\
 & \textcolor{black}{{Con. speed (s)}} & \textcolor{darkergreen}{$0.083$} & \textcolor{black}{$0.091$} & \textcolor{black}{$0.095$}  \\
\hdashline
        \multirow{4}{*}{\textbf{\textcolor{black}{Tilt}}} &\textcolor{black}{{Pos. error}} {(m)} & \textcolor{black}{$0.000065$} & \textcolor{black}{$0.000112$} & \textcolor{darkergreen}{$0.000064$}  \\
   \ & \textcolor{black}{{Vel. error}} {(m/s)} & \textcolor{darkergreen}{$0.00208$} & \textcolor{black}{$0.00221$} & \textcolor{black}{$0.00230$}  \\
       & \textcolor{black}{{T. effort}} {(N.m)} & \textcolor{darkergreen}{$102$} & \textcolor{black}{$111$} & \textcolor{black}{$120$}  \\
                   & \textcolor{black}{{Con. speed}} {(s)} & \textcolor{darkergreen}{$0.085$} & \textcolor{black}{$0.094$} & \textcolor{black}{$0.095$}  \\
   \hdashline
        \multirow{4}{*}{\textbf{\textcolor{black}{Telescope}}} &\textcolor{black}{{Pos. error}} {(m)} & \textcolor{darkergreen}{$0.00213$} & \textcolor{black}{$0.00214$} & \textcolor{black}{$0.00215$}  \\
    & \textcolor{black}{{Vel. error}} {(m/s)} & \textcolor{darkergreen}{$0.0047$} & \textcolor{black}{$0.0049$} & \textcolor{black}{$0.0048$}  \\
       & \textcolor{black}{{T. effort}} {(N.m)} & \textcolor{darkergreen}{$35$} & \textcolor{black}{$41$} & \textcolor{black}{$51$}  \\
                   & \textcolor{black}{{Con. speed}} {(s)} & \textcolor{darkergreen}{$0.086$} & \textcolor{black}{$0.095$} & \textcolor{black}{$0.098$} \\
    \midrule
            \multirow{3}{*}{\textbf{\textcolor{black}{Normalized}}} &\textcolor{black}{{Pos. error}} {(m)} & \textcolor{darkergreen}{$1.000$} & \textcolor{black}{$1.040$} &  \textcolor{black}{$1.010$} \\
    & \textcolor{black}{{Vel. error}} {(m/s)} & \textcolor{darkergreen}{$1.000$} & \textcolor{black}{$1.090$} & \textcolor{black}{$1.100$}  \\
       \multirow{1}{*}{\textbf{\textcolor{black}{average}}}& \textcolor{black}{{T. effort}} {(N.m)} & \textcolor{darkergreen}{$1.000$} & \textcolor{black}{$1.110$} & \textcolor{black}{$1.230$}  \\
                   & \textcolor{black}{{Con. speed}} {(s)} & \textcolor{darkergreen}{$1.000$} & \textcolor{black}{$1.100$} & \textcolor{black}{$1.130$} \\
    \bottomrule
    \bottomrule
  \end{tabular}
  
  \label{VI}
  \begin{tablenotes} 
\item[-]- T. effort assigns the torque amplitudes generated.
\item[-]- Pos. error assigns the position error.
\item[-]- Vel. error assigns the velocity error.
\item[-]- Con. speed assigns the convergence speed.
\end{tablenotes}
\end{table}

\begin{figure}[h!] 
    \centering
    \scalebox{1}
    {\includegraphics[trim={0.1cm 0.1cm 0.1cm
    0.1cm},clip,width=\columnwidth]{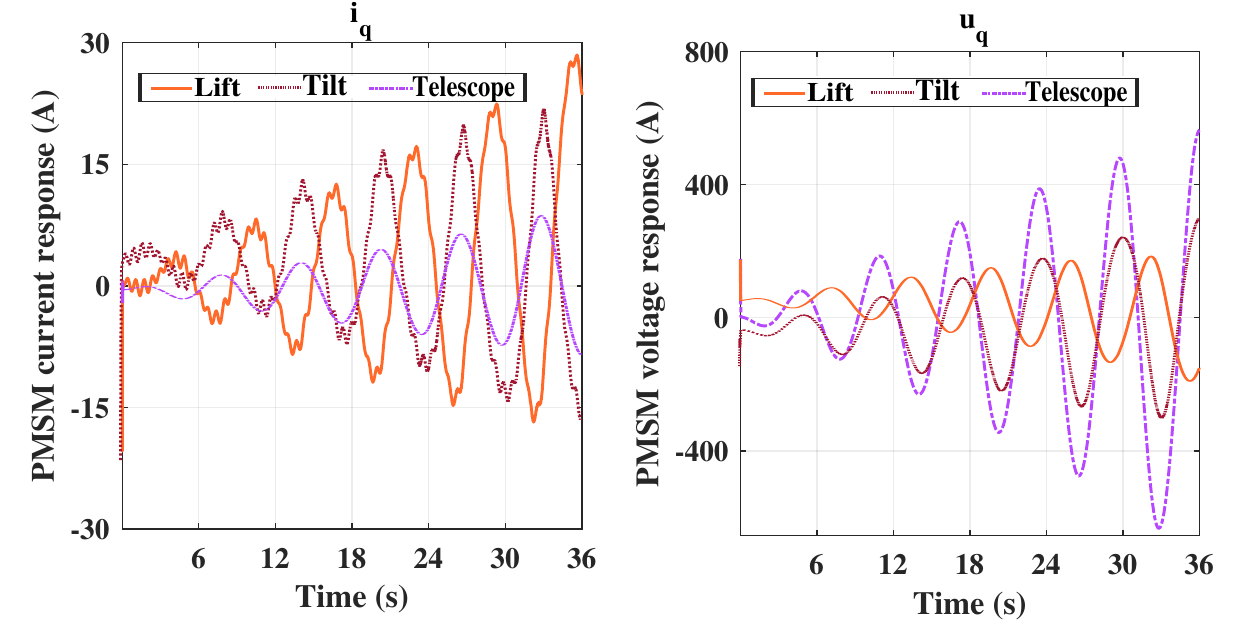}}
    \caption{The q-axis current and voltage responses in the lift, tilt, and telescope pistons of the studied 3-DoF HDRM by implementing RSBA control}
    \label{pmsm123}
\end{figure}

\begin{figure*}[h!] 
    \centering
    \includegraphics[trim={0.1cm 0.1cm 0.1cm 0.1cm},clip,width=13.5cm]{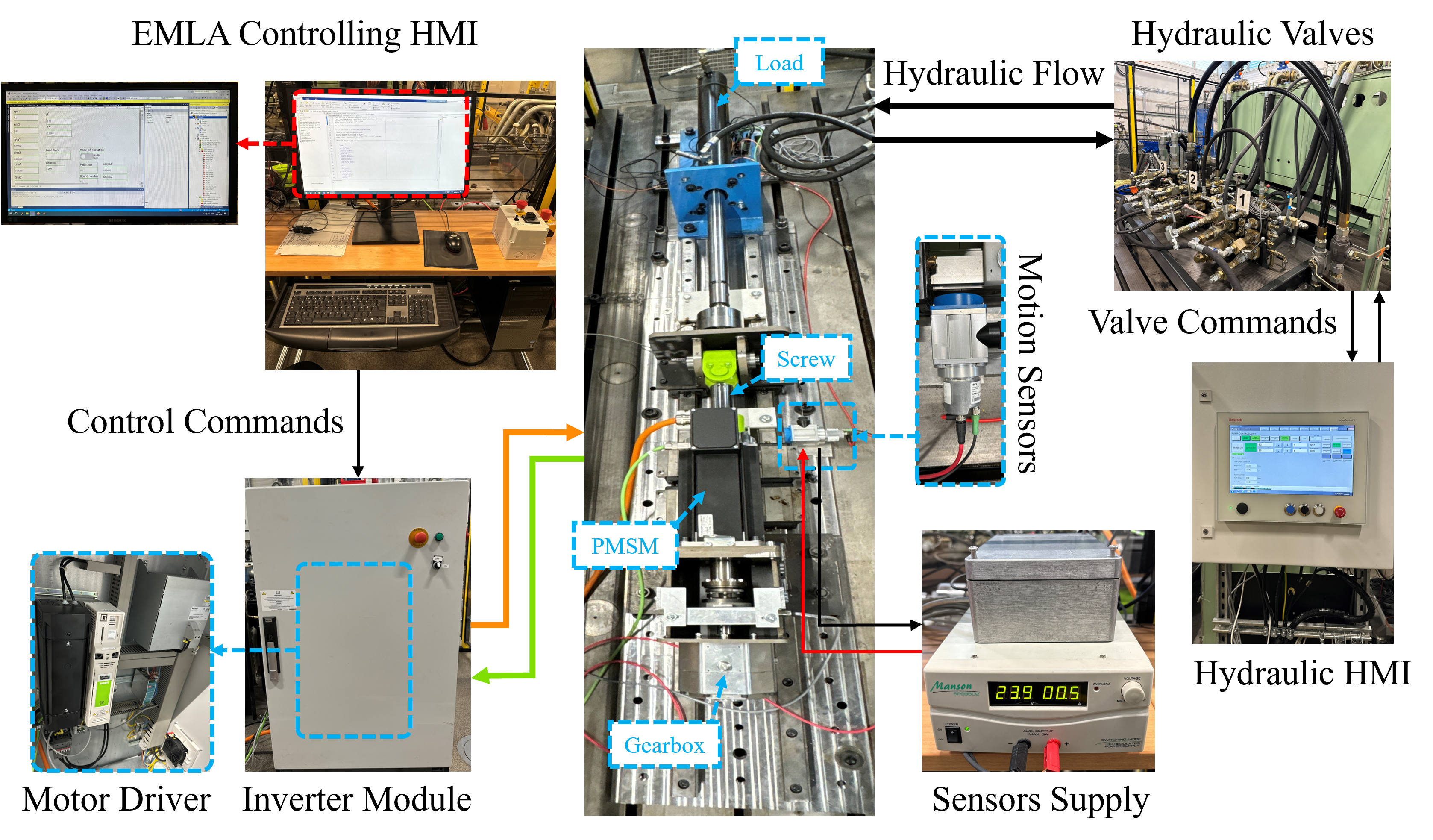}
    \caption{The setup of the PMSM-powered EMLA prototype aligned with a linear load generated by an EHA.}
    \label{e_xxxx}
\end{figure*}
According to Fig. (\ref{t_3}), the highest torque effort for this scenario is associated with ANATC, followed by CAC and RSBA control. 
The performances of all three control strategies implemented in the three EMLAs are summarized in Table \ref{VI}. Except for one instance involving the position outcomes of ANATC in lift and tilt EMLAs, we observe an increase in errors from one EMLA to the next for all three algorithms. We can witness a similar pattern in the convergence speed of the real states to the references, such that the subsequent EMLA had an approximately quicker convergence compared with the next one. In the table, to simplify comparisons, the average values of all three control performances based on the RSBA approach were normalized, we set the RSBA control results as our reference point ($=1$) and calculated the other approach results as ratios relative to the RSBA control results. We can assert that considering the payload and dynamics of the HDRM, the RSBA results slightly outperformed the other methods for three studied PMSM-powered EMLAs in simulations.
As previously mentioned, to maximize torque per ampere, we set the d-axis reference current $i^*_d$ to zero. When the d-axis controller operates effectively, then $i_d$ and $u_d$ will be minimal. Hence, besides control performance, the q-axis current and voltage responses generated by RSBA control in lift, tilt, and telescope EMLAs are presented in Fig. (\ref{pmsm123}).

\section{Experimental results}
\label{experiment}
The experiments are conducted on a PMSM-powered EMLA mechanism, which serves as a prototype for actuating one of the joints in an upcoming 3-DoF fully electrified HDRM. According to specifications and safety standards, EMLA can sustain and move a load of up to approximately $75$ kN. Fig. \ref{e_xxxx} illustrates the experimental setup for testing the control algorithms on the EMLA mechanism. To generate an adjustable external load, we employed an EHA, which is mechanically coupled to the load side of the EMLA, thereby enabling the generation of a pushing linear force on the actuator under study. The amplitude of the EHA's linear force can be controlled via the hydraulic HMI, which regulates the valves of the hydraulic piston. The RSBA, CAC, and ANATC controlling algorithm were separately uploaded to the Unidrive controller (model: M700-064 00350 A) through the EMLA controlling HMI to manage and monitor the operation of the servo motor of the EMLA. The mentioned controller sends the voltage commands to an inverter module in order to supply the servo motor. The adopted servo motor in the EMLA mechanism for the experiment is a three-phase $380$/$480$ VAC $11.6$ kW Nidec PMSM. The characteristics of the EMLA components, including the electric motor, gearbox, and ball screw, are listed in Table VII. Both control signals and communications between the controller, inverter module, and other components were managed via an EtherCAT network. This setup allowed for immediate control and surveillance, ensuring the motion control was precise and complied with the established safety and performance standards. Control systems operated with a sampling rate of $1,000$ Hz, and the torque of the controller was measured directly using a 16-bit message. We considered $C=[1,0]$, implying that the observer utilizes the linear position information from the position sensor as the output of the setup. Here, the reference position trajectory $x_{1d}$ for the experimental PMSM-powered EMLA was designed based on quantic polynomials in Chapter 13 of Ref. \cite{jazar2010theory} and was defined as a repetitive path from $0.05$ m to $0.30$ m, with two stops during forward movement and three stops during backward movement. The control parameters used for the RSBA control were as $\beta_{1} = 15$ $\beta_{2} = 1500$, $\beta_{3} = 5$ $\beta_{4} = 1$, $\zeta_{1, 2, 3, 4} = 100$, $\delta_{1, 2, 3, 4} = 100$, $\sigma_{1, 2, 3, 4} = 0.001$, $H = 20 \cos(y) ^4 +  20\sin(y) ^4$, $\ell=1, \quad m = 200 e^{-0.001t}$, and $\bm{p} = [1.4078\hspace{0.2cm}-0.1975\hspace{0.1cm};\hspace{0.1cm} -0.1975\hspace{0.2cm}4.4535]$; for CAC were as $k_1 = 2$, $k_2 = 150$, $k_{3, 5} = 0.01$, $k_4 = 0.9$, $\varpi_{1,2,4} = 1.8$, $\varpi_{3,5} = 0.02$, $m_{1, 3} = 1$, and $m_2 = 12$; and for ANATC, they were as $k_1 = 0.1$, $k_2 = 0.55$, $g = 0.1$, $\lambda_1 = 0.8$, $\lambda_2 = 0.75$, $\ell_{1,2} = 0.25$, and $b = 2$.

\begin{table}[h!]
  \captionsetup{position=top}
  \caption{PMSM and gearbox parameters of the experimental EMLA}
  \centering
  \small
\begin{tabular}{ccc}
\toprule
\toprule
Parameter & Value & Unit \\
\midrule PM Magnetic Flux & 0.134 & Wb \\
 Number of Pole Pairs & 4 & - \\
 Rated Power & 11.6 & kW \\
Rated Current & 23.1 & A \\
Peak Current & 48.2 & A \\
 Rated Torque & 37 & N.m \\
 Peak Torque & 77 & N.m \\
 Rated Speed & 3000 & rpm \\
 Maximum Speed & 3877 & rpm\\
 Phase Resistance & 0.08 & $\Omega$ \\
 Phase Inductance & 2.42 & mH\\
 Gear Ratio & 7.7 & - \\
Screw Lead & 16 & mm \\
 Screw Diameter & 63 & mm \\
 Screw Lead Angle & 4.55 & Degree $^{\circ}$ \\
\bottomrule
\bottomrule
\end{tabular}
\end{table}
To introduce a variety of tasks, we divided the tests into two parts: 1) upper-moderate velocity and gradually increasing load ranging from $7$ kN to $75$ kN, and 2) the maximum velocity and high load condition of $75$ kN.

\subsection{Experiment 1: Upper-Moderate Velocity and  Gradually Increasing Load}
\label{exp1}
\begin{figure}[h!] 
    \centering
    \scalebox{0.9}
    {\includegraphics[trim={0.1cm 0.1cm 0.1cm
    0.1cm},clip,width=\columnwidth]{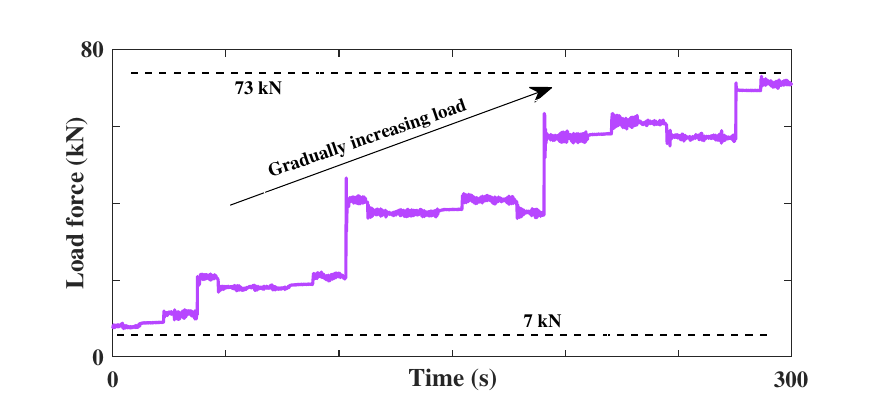}}
    \caption{Experiment 1, the load force applied to the experimental PMSM-powered EMLA.}
    \label{force1}
\end{figure}

The first experiment assumed that the EMLA would track the desired linear position at an upper-moderate velocity of $0.026$ m/s, under a gradually increasing load ranging from $7$ kN to $75$ kN. In this case, a $20$-kN load was added approximately $40$ s after beginning tracking. Subsequently, a $40$-kN load was added at around $100$ s, a $60$-kN load at about $200$ s, and, finally, the last load of $70$ kN was applied at $280$ s. Fig. \ref{force1} illustrates the pushing load force applied by the EHA to the EMLA during the experiment.

\begin{figure}[h!] 
    \centering
    \scalebox{0.9}
    {\includegraphics[trim={0.1cm 0.1cm 0.1cm
    0.1cm},clip,width=\columnwidth]{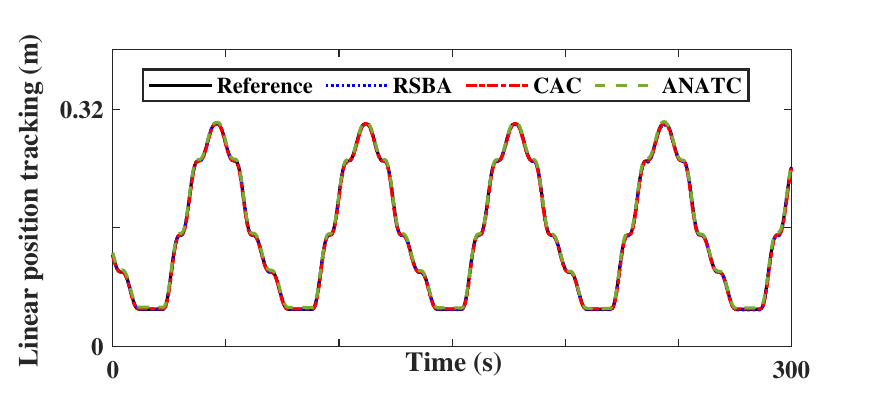}}
    \caption{Experiment 1, position tracking in the experimental PMSM-powered EMLA}
    \label{track1}
\end{figure}

It clearly shows that the load is not smoothly constant and frequently changes, with a $5$-kN amplitude, due to the vibration from the EHA affecting the EMLA. This variability in the load could be beneficial, as the frequent changes may more closely simulate the realistic load conditions encountered in off-road environments, where a multi-DoF PMSM-powered EMLA-actuated HDRM is intended to operate. We investigated the performances of the same control strategies that were compared in Section \ref{c-simulation}. Hence, Fig. \ref{track1} illustrates the EMLA's tracking of the desired linear position, which RSBA, CAC, and ANATC controlled in the same condition.

\begin{figure}[h!] 
    \centering
    \scalebox{0.9}
    {\includegraphics[trim={0.1cm 0.1cm 0.1cm
    0.1cm},clip,width=\columnwidth]{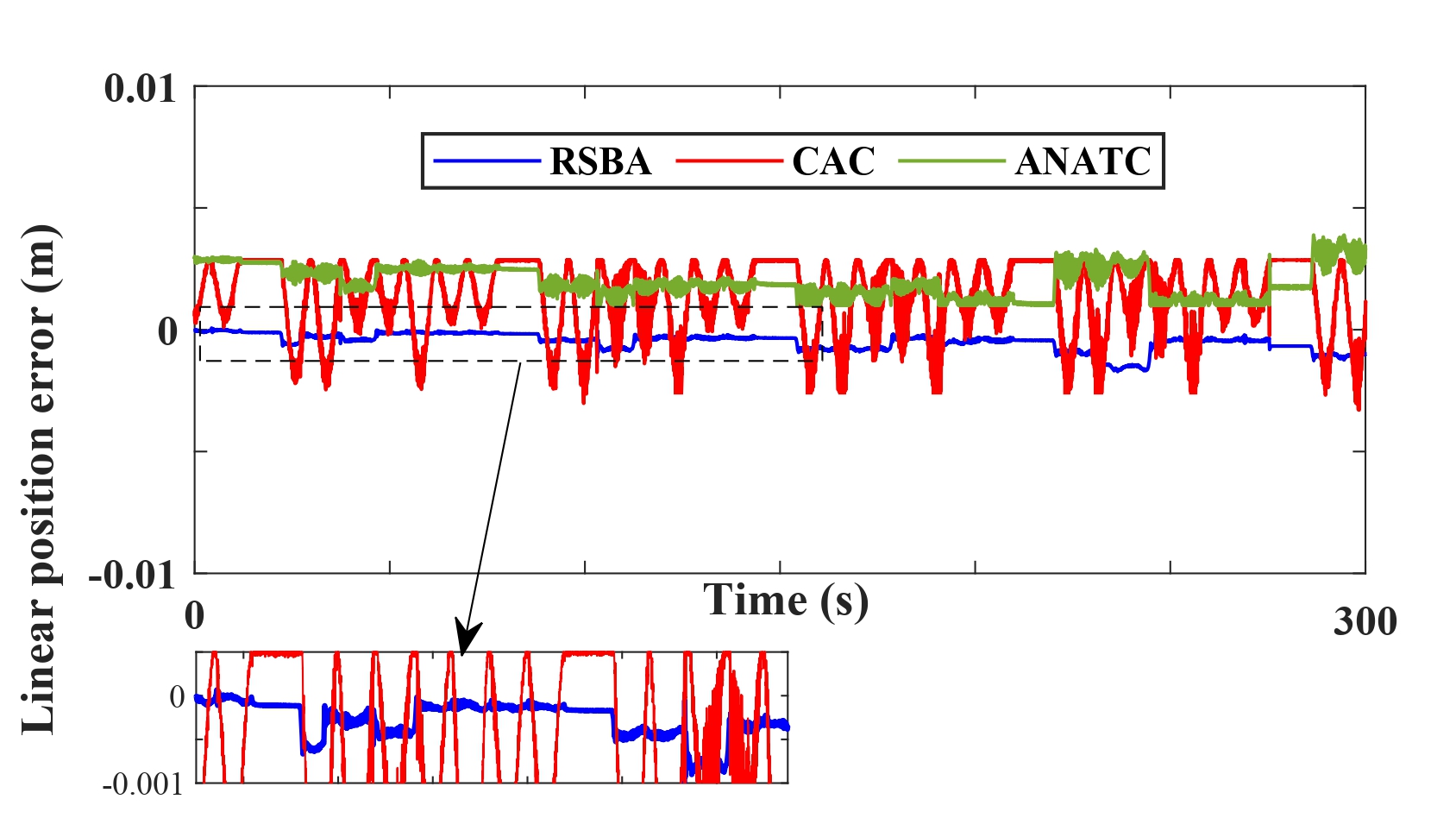}}
    \caption{Experiment 1, position tracking error in the experimental PMSM-powered EMLA ($x_{1d}-x_1$)}
    \label{err1}
\end{figure}

All three control algorithms could manage the experimental EMLA, though with significant differences, displayed in Fig. \ref{err1} as the tracking errors of each control algorithm.
This figure presents the linear position tracking errors: $0.0008$ m for RSBA control, $0.004$ m for CAC, and $0.0027$ m for ANATC. Interestingly, the values indicate that the tracking accuracy increased during actual tests compared to the simulation results. Although the position tracking error of RSBA in the experiment was approximately $10$ times its simulation result, it remains acceptable at $0.8$ mm. However, the errors in CAC and ANATC were about $100$ times their respective simulation results, approximately $4$ and $2.5$ mm. Their tracking errors in the simulation were acceptable; however, the observed position errors of CAC and ANATC in the experiment may pose challenges, especially when the number of PMSM-powered EMLA prototypes implemented in an integrated HDRM system increases, affecting the error of the manipulator's end effector in the task space. Thus, the noted accuracies confirm that the robustness and control performance of the RSBA framework surpassed those of the other two control algorithms.

\begin{figure}[h!] 
    \centering
    \scalebox{0.9}
    {\includegraphics[trim={0.1cm 0.1cm 0.1cm
    0.1cm},clip,width=\columnwidth]{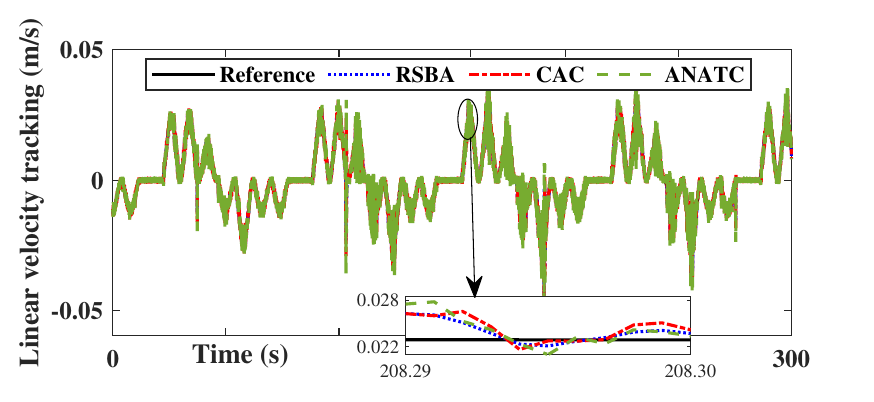}}
    \caption{Experiment 1, velocity tracking in the experimental PMSM-powered EMLA}
    \label{vel1}
\end{figure}

\begin{figure}[h!] 
    \centering
    \scalebox{0.9}
    {\includegraphics[trim={0.1cm 0.1cm 0.1cm
    0.1cm},clip,width=\columnwidth]{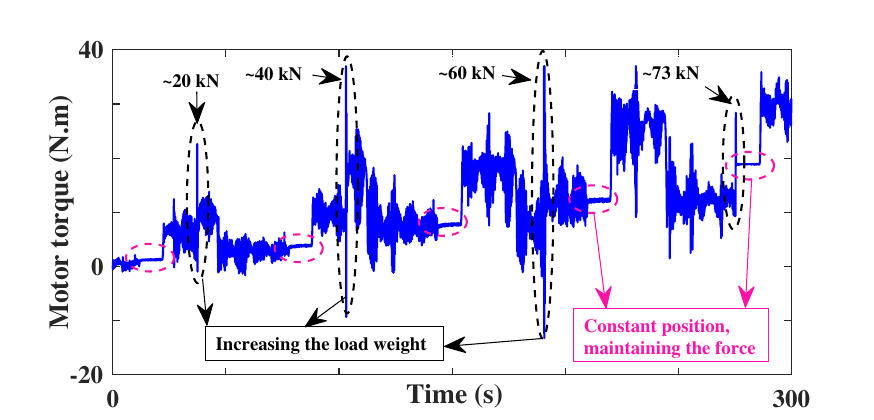}}
    \caption{Experiment 1, motor torque generated in the experimental PMSM-powered EMLA}
    \label{tor_e1}
\end{figure}

Fig. \ref{vel1} illustrates the tracking of linear velocity. The enlarged image of this figure highlights the frequent changes in the linear velocity data of control-applied EMLA every $0.01$ s, demonstrating the effects of high-frequency load fluctuations on the received velocity data; regardless, Fig. \ref{err1} confirmed the effectiveness of the proposed RSBA control framework, achieving smooth position control. Similar to the linear velocity, the motor torque generated by the experimental RSBA-applied PMSM-powered EMLA is depicted in Fig. \ref{tor_e1}. This figure highlights frequent fluctuations intended to compensate for fluctuating load forces. It confirms that the torque remains within the nominal value ($37$ Nm). Despite frequent changes in load force, affecting torque and velocity, Fig. \ref{track1} validated the quick response and robustness of the RSBA control, which ensured smooth positioning of the experimental EMLA.

\subsection{Experiment 2: The High Velocity and Load Condition}
\label{exp2}
\begin{figure}[h!] 
    \centering
    \scalebox{0.9}
    {\includegraphics[trim={0.1cm 0.1cm 0.1cm
    0.1cm},clip,width=\columnwidth]{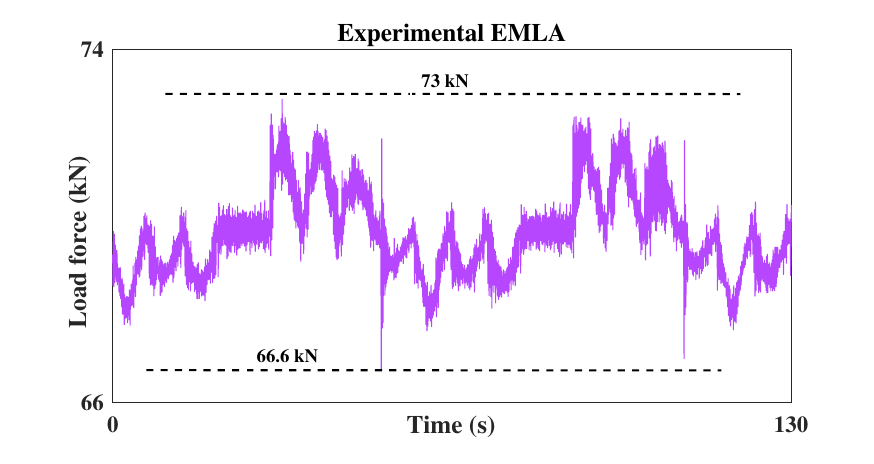}}
    \caption{Experiment 2, the load force applied to the experimental PMSM-powered EMLA.}
    \label{force2}
\end{figure}

The second experiment assumed that the EMLA would track the same desired linear position but at a nominally high velocity of $0.03$ m/s, under the heaviest load from the previous experiment, starting at $70$ kN. Fig. \ref{force2} illustrates the load force applied to the EMLA during the experiment. It clearly shows that the load is not constant and frequently changes, with a $5$-kN amplitude, due to the vibration from the EHA affecting the EMLA. The fluctuations in load could prove advantageous, as the regular alterations could better mimic the actual load conditions found in off-road settings where a multi-DoF PMSM-powered EMLA-actuated HDRM is designed to function.

\begin{figure}[h!] 
    \centering
    \scalebox{0.9}
    {\includegraphics[trim={0.1cm 0.1cm 0.1cm
    0.1cm},clip,width=\columnwidth]{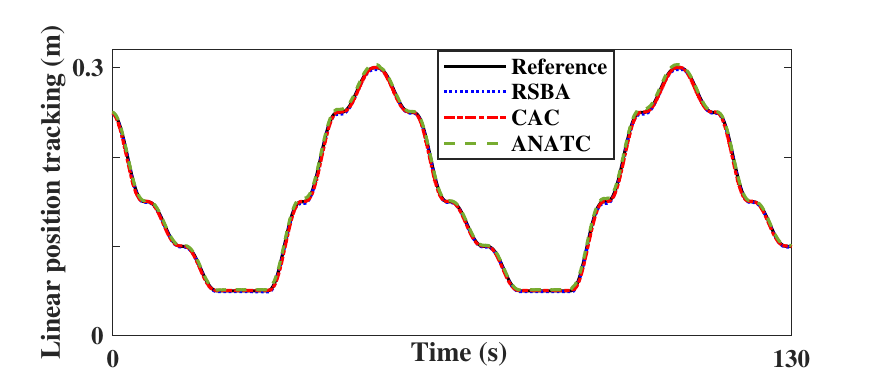}}
    \caption{Experiment 2, position tracking in the experimental PMSM-powered EMLA}
    \label{track2}
\end{figure}

\begin{figure}[h!] 
    \centering
    \scalebox{0.9}
    {\includegraphics[trim={0.1cm 0.1cm 0.1cm
    0.1cm},clip,width=\columnwidth]{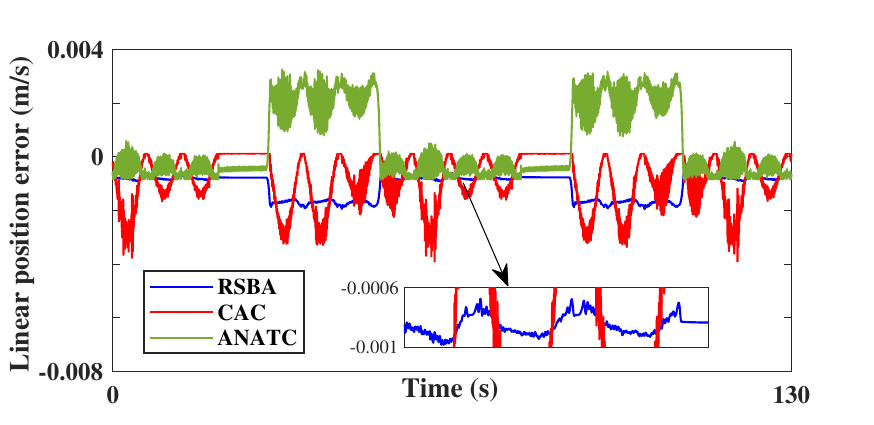}}
    \caption{Experiment 2, position tracking error in the experimental PMSM-powered EMLA ($x_{1d}-x_1$)}
    \label{err2}
\end{figure}

We investigated the performances of the same control strategies that were compared in Sections \ref{c-simulation} and \ref{exp1}. Fig. \ref{track2} illustrates the EMLA's tracking of the desired linear position, which RSBA, CAC, and ANATC controlled in the same experiment. Each of the three control algorithms was able to manage the experimental EMLA, albeit with notable differences in performance. These variations are illustrated in Fig. \ref{err2}, which shows the tracking errors for each control algorithm.

Fig. \ref{err2} presents the linear position tracking errors: $0.0008$ m for RSBA control, $0.004$ m for CAC, and $0.0025$ m for ANATC, which were close to the Experiment 1 results provided in Fig. \ref{err1}. Similarly, the performance of the three control algorithms deteriorated in the experiment compared to the simulation results (see Figs. \ref{p_e1}, \ref{p_e2}, and \ref{p_e3}). Because the error in the linear position tracking of RSBA control, CAC, and ANATC in Experiment 2 roughly matched that of Experiment 1, as visualized in Fig. \ref{err1}, it could validate the robustness of all three control algorithms in different conditions but with significantly different accuracy tracking. Like in Experiment 1, the position errors observed with CAC and ANATC in Experiment 2 could present difficulties, particularly as the number of PMSM-powered EMLA prototypes in an integrated HDRM system grows. This increase may impact the accuracy of the manipulator’s end effector within the task space. Thus, the noted accuracies confirm that the robustness and control performance of the RSBA framework surpassed those of the other two control algorithms.

\begin{figure}[h!] 
    \centering
    \scalebox{0.9}
    {\includegraphics[trim={0.1cm 0.1cm 0.1cm
    0.1cm},clip,width=\columnwidth]{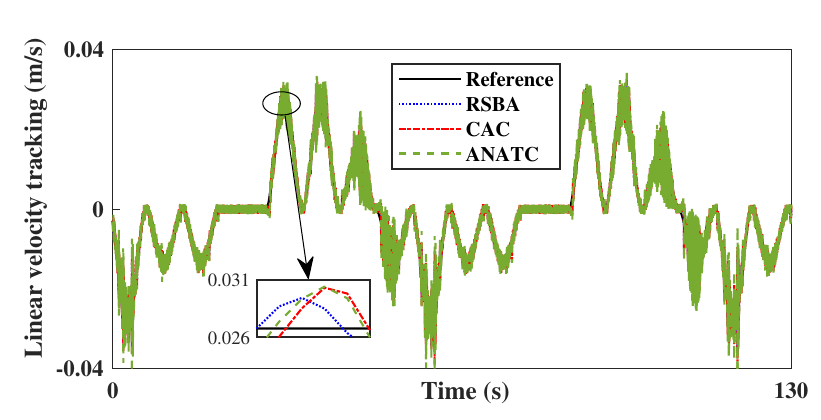}}
    \caption{Experiment 2, velocity tracking in the experimental PMSM-powered EMLA}
    \label{vel2}
\end{figure}

Fig. \ref{vel2} illustrates the tracking of linear velocity, and the magnified view of this figure emphasizes the frequent changes in the linear velocity data of the control-applied EMLA every $0.005$ s, showcasing the impact of high-frequency load fluctuations on the velocity data received. Nevertheless, Fig. \ref{track2} validates the efficacy of the proposed RSBA control framework, which successfully achieves smooth position control.

\begin{figure}[h!] 
    \centering
    \scalebox{0.9}
    {\includegraphics[trim={0.1cm 0.1cm 0.1cm
    0.1cm},clip,width=\columnwidth]{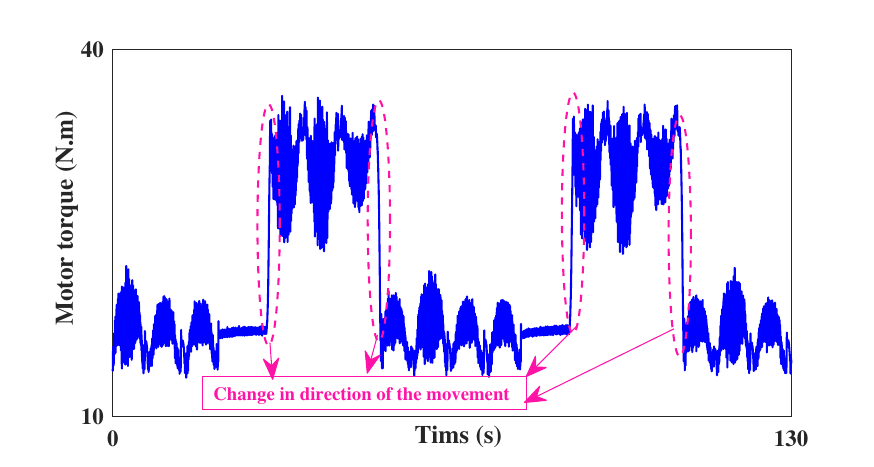}}
    \caption{Experiment 2, motor torque generated in the experimental PMSM-powered EMLA}
    \label{tor_e2}
\end{figure}

\begin{figure}[h!] 
    \centering
    \scalebox{0.9}
    {\includegraphics[trim={0.1cm 0.1cm 0.1cm
    0.1cm},clip,width=\columnwidth]{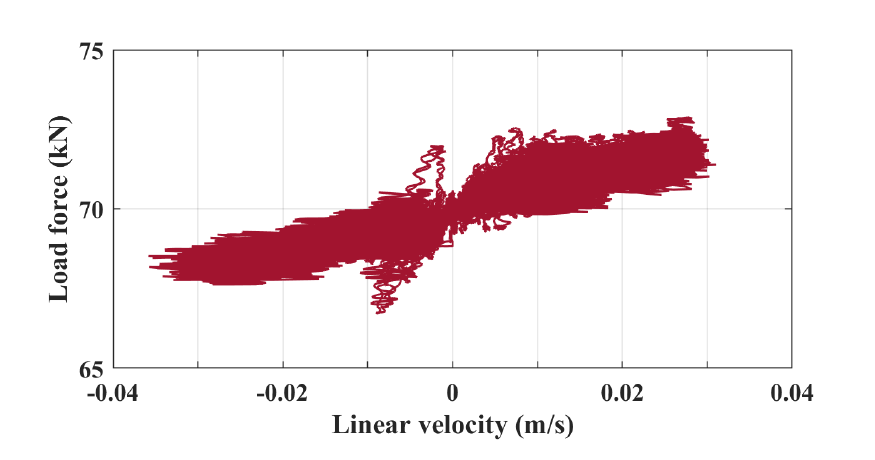}}
    \caption{Experiment 2, load force and linear velocity of the EMLA.}
    \label{for_vel_e2}
\end{figure}

\begin{figure}[h!] 
    \centering
    \scalebox{0.9}
    {\includegraphics[trim={0.1cm 0.1cm 0.1cm
    0.1cm},clip,width=\columnwidth]{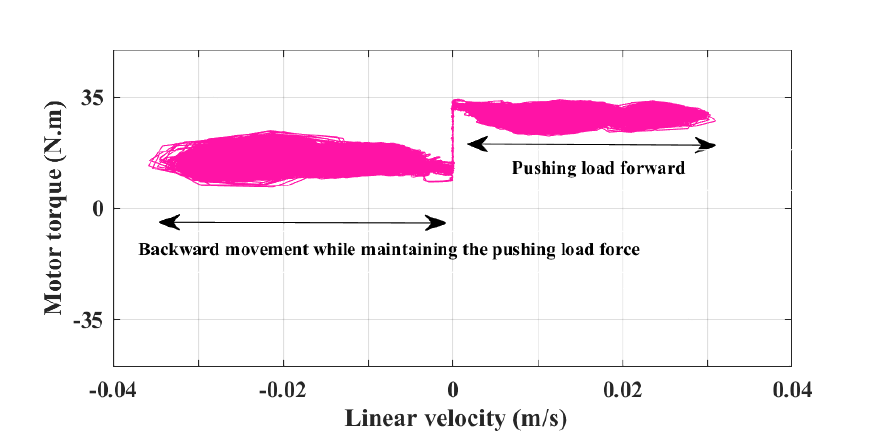}}
    \caption{Experiment 2, motor torque and linear velocity of the EMLA.}
    \label{tor_vel_e2}
\end{figure}

Similar to the linear velocity, the motor torque generated by the experimental RSBA-applied PMSM-powered EMLA is depicted in Fig. \ref{tor_e2}. This figure also highlights frequent variations intended to compensate for fluctuating load forces. It confirms that the torque remains within the nominal value (37 Nm). Figs. \ref{for_vel_e2} and \ref{tor_vel_e2} illustrate the variability in the load force and in the generated motor torque in relation to linear velocity for the experimental RSBA-applied PMSM-powered EMLA. Comparing these two figures highlighted could validate the torque effect-based robustness of RSBA control in managing the frequently changing heavy load. The immediate jump in the generated motor torque, when the velocity direction changed, validated the rapid responsiveness of the RSBA control, even as the load force at this point frequently varied.

The performances of all three control strategies implemented in two experiments are summarized in Table \ref{VIII}. As observed, all performance criteria for RSBA control—including tracking accuracy, torque effort, and convergence time—significantly surpassed those of the two other control algorithms, CAC and ANATC. In the table, the average values of two experiments on the PMSM-powered EMLA based on the RSBA approach were normalized, so we set the RSBA control results as our reference point and calculated the other approach results as ratios relative to the RSBA control results. The performance of CAC and ANATC, particularly in position tracking, after implementation of the experimental prototype, indicated that these algorithms might face challenges in multi-DoF PMSM-powered EMLA-actuated HDRMs, as the number of PMSM-powered EMLA prototypes in an integrated HDRM system increases, which adversely affects the error in the manipulator's end effector in the task space. Thus, the reported performances of all three algorithms confirm that the robustness and control performance of the RSBA framework surpasses those of the other two control algorithms. This could validate the superiority of the RSBA control in terms of accuracy and robustness, making it more beneficial for the application under study.

\begin{table}[h!]
  \captionsetup{position=top}
  \caption{Control performance of the experimental PMSM-powered EMLA implemented by the RSBA control, CAC \cite{liu2023command}, and ANATC \cite{zhang2023adaptive} in tracking the linear motion reference}
  \centering
  \scriptsize
  \begin{tabular}{ccccc}
    \toprule
    \toprule
    \textcolor{black}{\textbf{Case}} &\textcolor{black}{\textbf{Convergence}} & \textcolor{black}{\textbf{RSBA}} &
    \textcolor{black}{\textbf{CAC \cite{liu2023command}}} &
    \textcolor{black}{\textbf{ANATC \cite{zhang2023adaptive}}}  \\
        \textcolor{black}{\textbf{study}} &\textcolor{black}{\textbf{criteria}} & \textcolor{black}{\textbf{approach}} &
    \textcolor{black}{\textbf{approach}} &
    \textcolor{black}{\textbf{approach}} \\
    \midrule
    \multirow{3}{*}{\textbf{\textcolor{black}{Experiment}}} & \textcolor{black}{{Pos. error (m)}} & \textcolor{darkergreen}{$0.0008$} & \textcolor{black}{$0.004$} & \textcolor{black}{$0.0027$}  \\
 & \textcolor{black}{{Vel. error (m/s)}} & \textcolor{darkergreen}{$0.002$} & \textcolor{black}{$0.004$} & \textcolor{black}{$0.006$}  \\
\multirow{1}{*}{\textbf{\textcolor{black}{1}}} & \textcolor{black}{{T. effort (N.m)}} & \textcolor{darkergreen}{$35.0$} & \textcolor{black}{$39.5$} & \textcolor{black}{$40.0$}  \\
 & \textcolor{black}{{Con. speed (s)}} & \textcolor{darkergreen}{$0.12$} & \textcolor{black}{$0.22$} & \textcolor{black}{$0.28$}  \\
 \hdashline
     \multirow{3}{*}{\textbf{\textcolor{black}{Experiment}}} & \textcolor{black}{{Pos. error (m)}} & \textcolor{darkergreen}{$0.0008$} & \textcolor{black}{$0.004$} & \textcolor{black}{$0.0025$}  \\
 & \textcolor{black}{{Vel. error (m/s)}} & \textcolor{darkergreen}{$0.0025$} & \textcolor{black}{$0.0037$} & \textcolor{black}{$0.0032$}  \\
\multirow{1}{*}{\textbf{\textcolor{black}{2}}} & \textcolor{black}{{T. effort (N.m)}} & \textcolor{darkergreen}{$35.00$} & \textcolor{black}{$39.5$} & \textcolor{black}{$40.0$}  \\
 & \textcolor{black}{{Con. speed (s)}} & \textcolor{darkergreen}{$0.14$} & \textcolor{black}{$0.25$} & \textcolor{black}{$0.32$}  \\
 \midrule
            \multirow{3}{*}{\textbf{\textcolor{black}{Normalized}}} &\textcolor{black}{{Pos. error}} {(m)} & \textcolor{darkergreen}{$1.00$} & \textcolor{black}{$5.00$} &  \textcolor{black}{$3.25$} \\
    & \textcolor{black}{{Vel. error}} {(m/s)} & \textcolor{darkergreen}{$1.00$} & \textcolor{black}{$1.71$} & \textcolor{black}{$2.04$}  \\
       \multirow{1}{*}{\textbf{\textcolor{black}{average}}}& \textcolor{black}{{T. effort}} {(N.m)} & \textcolor{darkergreen}{$1.00$} & \textcolor{black}{$1.13$} & \textcolor{black}{$1.14$}  \\
                   & \textcolor{black}{{Con. speed}} {(s)} & \textcolor{darkergreen}{$1.00$} & \textcolor{black}{$1.81$} & \textcolor{black}{$2.31$} \\
    \bottomrule
    \bottomrule
  \end{tabular}
  
  \label{VIII}
  \begin{tablenotes} 
\item[-]- T. effort assigns the maximum of the torque amplitude.
\item[-]- Pos. error assigns the position error.
\item[-]- Vel. error assigns the velocity error.
\item[-]- Con. speed assigns the convergence speed.
\end{tablenotes}
\end{table}

\section {Conclusion}
This paper has introduced the RSBA control strategy as a solution to the control challenges inherent in the electrified $n_a$-DoF HDRM actuated by the synergy between a zero-emission EMLA and PMSM. The key findings of this study are as follows:
\begin{itemize}
    \item We developed a comprehensive, dynamic model for PMSM-powered EMLAs to capture mechanism motion intricacies, and we determined joint reference trajectories using B-spline curves for precise control task definition, thus providing a foundational modeling framework for designing effective control strategies.
    \item The measurement errors of linear motion states at the load side of the EMLAs have been compensated by employing an adaptive state observer algorithm.
    \item The proposed robust control strategy has mitigated non-triangular uncertainties, as well as torque and voltage disturbances affecting the control system of the EMLA-driven HDRM.
    \item We developed a modular configuration for each interacting subsystem of the $n_a$-DoF EMLA-driven HDRM, with the end goal of managing interactions between various components of the mechanism and ensuring exponential stability for the entire system.
\end{itemize}

These outcomes demonstrate the effectiveness of the proposed RSBA control strategy for the complex, fully electrified, EMLA-driven HDRM mechanism in achieving precise motion control. It is hoped that the outcomes of this paper may serve future research and contribute to unlocking the complete potential of HDRM electrification for greener and more efficient technological solutions, promoting a sustainable future.

\section*{Appendix A\\The proposed adaptive observer Stability}
Taken from the general solution for state space representation, we can solve \eqref{55} as follows \cite{heydari2024robust}:
\begin{equation}
\begin{aligned}
\label{709}
V_0 \leq&  V_0\left(t_0\right) e^{-\left\{\phi_{0}\left({t-t_0}\right)\right\}}+{\tilde\mu} \int_{t_0}^t e^{\left\{-\phi_{0}(t-T)\right\}}\hspace{0.2cm} dT \\
&+ \frac{1}{2} \int_{t_0}^t e^{-\phi_{0}(t-T)} m^2(T) \hspace{0.2cm} dT
\end{aligned}
\end{equation}
Because $-\phi^{-1}_0(e^{-\phi_0(t-t_0)})$ is always negative, we can interpret \eqref{709}, as follows:
\begin{equation}
\begin{aligned}
\label{800}
V_0 \leq&  V_0\left(t_0\right) e^{-\left\{\phi_{0}\left({t-t_0}\right)\right\}}+ \hspace{0.1cm}{\tilde{\mu}} \hspace{0.1cm} {\phi_{0}}^{-1}\\
&+ \frac{1}{2} \int_{t_0}^t e^{\left\{-\phi_{0}(t-T)\right\}} m^2(T) \hspace{0.2cm} dT 
\end{aligned}
\end{equation}
Based on \eqref{44}, and defining $p_{min} \in \mathbb{R}^{+}$ as the minimum eigenvalue of matrix $\bm{p}$, we can say:
\begin{equation}
\begin{aligned}
\label{801}
\hspace{-0.3cm}\|\bm{x_{eo}}\|^2 \leq & \frac{1}{p_{min}} V_0\left(t_0\right) e^{-\left\{\phi_{0}\left({t-t_0}\right)\right\}}+ \frac{1}{p_{min}} \hspace{0.1cm}{\tilde{\mu}} \hspace{0.1cm} {\phi_{0}}^{-1}\\
&+\frac{1}{2 p_{min}} \int_{t_0}^t e^{\left\{-\phi_{0}(t-T)\right\}} m^2\hspace{0.2cm}dT
\end{aligned}
\end{equation}
$\phi_{0}$ is a positive constant dependent on designable control gains that are freely chosen to satisfy the following condition:
\begin{equation}
\begin{aligned}
\label{802}
\frac{1}{2 p_{min}{\phi_{0}}}<1
\end{aligned}
\end{equation}
To continue the stability proof, we define a continuous operator, $Z(\cdot)$, as follows:
\begin{equation}
\begin{aligned}
\label{803}
&Z(\iota)=\frac{(2 p_{min})^{-1}}{\phi_{0}-\iota}>0, \hspace{0.2cm}\iota \in [0,\phi_{0})
\end{aligned}
\end{equation}
It is evident that by increasing $\iota$, $Z(\iota)$ increases, meaning that
$Z(\iota)\geq Z(0)=  \frac{1}{2 p_{min}{ \phi_{0}}}$. As we know $\phi_0 = \min [1, \underline{m} \ell] \leq 1$ from \eqref{55},
by designing $\mathbf{p}$ of which $p_{min}$ is sufficiently large such that $2 p_{min} \phi_{0} > 1$, we can easily find a small positive value $\bar{\iota} \in \iota$ that satisfies the following condition:
\begin{equation}
\begin{aligned}
\label{804}
0 < \bar{Z}=Z(\bar{\iota})=\frac{(2p_{min})^{-1}}{\phi_{0}-\bar{\iota}} < 1
\end{aligned}
\end{equation}
By multiplying $e^{\bar{\iota}(t-t_0)}$ to \eqref{801}, we reach:
\begin{equation}
\begin{aligned}
\label{805}
&{\|\bm{x_{eo}}\|^2}e^{\bar{\iota}(t-t_0)} \leq \\
  &\frac{1}{p_{min}} V_0(t_0) e^{{-(\phi_{0}-\bar{\iota})(t-t_0)}}+\frac{1}{p_{min}} \tilde{\mu} \phi_{0}^{-1}  e^{{\bar{\iota}(t-t_0)}} \\
 &+{(2p_{min})^{-1}} \int_{t_0}^t e^{{-\phi_{0}(t-T)+\bar{\iota}(t-t_0)}} m^2 \hspace{0.2cm} dT
\end{aligned}
\end{equation}
Because $0 \leq \bar{\iota} <\phi_{0}$, we can eliminate the decreasing element $e^{{-(\phi_{0}-\bar{\iota})(t-t_0)}}$ from the right-hand side of \eqref{805}:
\begin{equation}
\begin{aligned}
\label{806}
&{\|\bm{x_{eo}}\|^2}e^{\bar{\iota}(t-t_0)} \leq \\
&\frac{1}{p_{min}}V_0(t_0)+\frac{1}{p_{min}} \tilde{\mu} \phi_{0}^{-1} e^{{\bar{\iota}(t-t_0)}}\\
&+(2 p_{min})^{-1}  \int_{t_0}^t e^{-(\phi_{0}-\bar{\iota})(t-T)} m^2 e^{{\bar{\iota}(T-t_0)}}\hspace{0.1cm} dT
\end{aligned}
\end{equation}
Using this approach, we can describe functions $E_0$ and $E_{00}$, which are continuous and non-declining:
\begin{equation}
\begin{aligned}
\label{807}
&E_0 = \sup _{e \in(t-t_0)}  [\|\mathbf{x_{eo}}\|^2 e^{{\bar{\iota}(e-t_0)})}]\\
&E_{00}= \sup _{e \in(t-t_0)} [(m^2)  e^{\bar{\iota}(e-t_0)}]\\
\end{aligned}
\end{equation}
Next, by considering \eqref{806} and \eqref{807}, and conducting some straightforward mathematical manipulations while removing the decreasing term, we obtain:
\begin{equation}
\begin{aligned}
\label{808}
 {\|\bm{x_{eo}}\|^2}e^{\bar{\iota}(t-t_0)} \leq& \frac{1}{p_{min}}V_0(t_0)+\frac{1}{p_{min}} \tilde{\mu}\phi_{0}^{-1} e^{{\bar{\iota}(t-t_0)}}\\ 
 &+  \frac{(2 p_{min})^{-1}}{\phi_{0}-\bar{\iota}} E_{00}
\end{aligned}
\end{equation}
As $E_{00}$ does not exhibit a declining pattern, the left side of \eqref{808} will not exhibit a reduction. Therefore, with respect to the definition of $E_0$ in \eqref{807}, we can state that:
\begin{equation}
\begin{aligned}
\label{809}
E_0 \leq & \frac{1}{p_{min}} V_0(t_0)+ \frac{(2 p_{min})^{-1} }{\phi_{0}-\bar{\iota}} E_{00}+\frac{1}{p_{min}} \tilde{\mu} \phi_{0}^{-1} e^{\bar{\iota}(t-t_0)}
\end{aligned}
\end{equation}
Defining $E=\max {(E_0,E_{00})}$, we can have:
\begin{equation}
\begin{aligned}
\label{901}
E_0 \leq& \frac{1}{p_{min}} V_0\left(t_0\right)+\bar{Z} E+\frac{1}{p_{min}} \tilde{\mu} \phi_{0}^{-1} e^{\bar{\iota}(t-t_0)}
\end{aligned}
\end{equation}
Similar to \eqref{804}, it is possible to ensure the existence of a sufficiently large $\iota^*$, where $\phi_{0}>\iota^*>\bar{\iota}$ and $\overset{*}{Z}=Z(\iota^*)$, to satisfy the following condition \cite{heydari2024robust}:
\begin{equation}
\begin{aligned}
\label{902}
\overset{*}{Z}>\bar{Z}, \hspace{0.2cm} 0<\overset{*}{Z}<1, \hspace{0.1cm}\implies \hspace{0.1cm} \bar{Z} E \leq \overset{*}{Z} E_0
\end{aligned}
\end{equation}
\eqref{902} is justified as we can make $\bar{Z}$ sufficiently small in \eqref{804}. When we incorporate \eqref{902} into \eqref{901}, we arrive at:
\begin{equation}
\begin{aligned}
\label{903}
E_0 \leq& \frac{1}{p_{min}} V_0(t_0)+\overset{*}{Z} E_0(t)+\frac{1}{p_{min}} \tilde{\mu} \phi_{0}^{-1} e^{\bar{\iota}(t-t_0)}
\end{aligned}
\end{equation}
Afterward, we obtain:

\begin{equation}
\begin{aligned}
\label{904}
E_0 \leq \frac{\frac{1}{p_{min}}V_0\left(t_0\right)+\frac{1}{p_{min}} \tilde{\mu} \phi_{0}^{-1} e^{\bar{\iota}(t-t_0)}} {1-\overset{*}{Z}}
\end{aligned}
\end{equation}
Concerning \eqref{807}, we obtain:
\begin{equation}
\begin{aligned}
\label{905}
\|\bm{x_{eo}}\|^2 \leq \frac{\frac{1}{p_{min}}V_0\left(t_0\right) e^{-\bar{\iota}(t-t_0)}+\frac{1}{p_{min}} \tilde{\mu}\phi_{0}^{-1}} {1-\overset{*}{Z}}
\end{aligned}
\end{equation}
It is significant that:
\begin{equation}
\begin{aligned}
\label{906}
\sup _{t \in\left[t_0, \infty\right]}(\frac{\frac{1}{p_{min}}V(t_0) e^{-\bar{\iota}(t-t_0)}} {1-\overset{*}{Z}})\leq \frac{\frac{1}{p_{min}}V_0(t_0)} {1-\overset{*}{Z}}
\end{aligned}
\end{equation}
Thus, based on Definition 1, it is obvious from \eqref{905} that along with the adaptive algorithms provided in Eqs. \eqref{23}, $\bm{x_{eo}}$ including the state estimation error in \eqref{20} reaches a defined region $g_{0}\left(\bar{\tau}_0\right)$ in uniformly exponential convergence, such that:
\begin{equation}
\begin{aligned}
\label{907}
g_{0}\left(\bar{\tau}_0\right):=\left\{\|\bm{x_{eo}}\| \leq \bar{\tau}_0 := \sqrt{\frac{\frac{1}{p_{min}} \tilde{\mu} \phi_{0}^{-1}}{1-\overset{*}{Z}}}\right\}
\end{aligned}
\end{equation}
$\tilde{\mu}$ depends on the
disturbance and non-triangular uncertainty bounds (see Eqs. \eqref{43} and \eqref{77}). Because $p_{min}$, resulting from $\bm{p}$, and $\phi_{0}$, relying on the designable control parameters (see Eqs. \eqref{77}, \eqref{55}, \eqref{64}, \eqref{68}, \eqref{72}), can be chosen, it is possible to decrease the radius of the ball \eqref{97} as much as will satisfy \eqref{82}, \eqref{84}, and \eqref{92}.

\section*{Appendix B\\An EMLA-Actuated Joint Stability}
Taken from the general solution for state space representation, we can solve \eqref{78} as follows \cite{heydari2024robust}:
\begin{equation}
\begin{aligned}
\label{79}
V \leq&  V\left(t_0\right) e^{-\left\{\phi_{total}\left({t-t_0}\right)\right\}}\\
&+{\bar\mu}_{total} \int_{t_0}^t e^{\left\{-\phi_{total}(t-T)\right\}}\hspace{0.2cm} dT \\
&+ \frac{1}{2}\bar{\mu}^{-1} \sum_{i=1}^{4} \int_{t_0}^t e^{-\phi_{total}(t-T)} M_i^2(T) \hspace{0.2cm} dT
\end{aligned}
\end{equation}
Because $e^{-\phi_{total}(t-t_0)}$ is always decreasing, we can interpret \eqref{79} as follows:
\begin{equation}
\begin{aligned}
\label{80}
V \leq&  V\left(t_0\right) e^{-\left\{\phi_{total}\left({t-t_0}\right)\right\}}+ \hspace{0.1cm}{\bar{\mu}}_{total} \hspace{0.1cm} {\phi_{total}}^{-1}\\
&+ \frac{1}{2} \bar{\mu}^{-1} \sum_{i=1}^{4} \int_{t_0}^t e^{\left\{-\phi_{total}(t-T)\right\}} M_i^2(T) \hspace{0.2cm} dT 
\end{aligned}
\end{equation}
Based on \eqref{75}, and defining $\lambda_{min} \in \mathbb{R}$ as the minimum eigenvalue of matrix $\bm{\lambda}$, we can say:
\begin{equation}
\begin{aligned}
\label{81}
&\|\mathbf{P}\|^2 \leq\\
&\frac{2}{\lambda_{min}} V\left(t_0\right) e^{-\left\{\phi_{total}\left({t-t_0}\right)\right\}}+ \frac{2}{\lambda_{min}} \hspace{0.1cm}{\bar{\mu}}_{total} \hspace{0.1cm} {\phi_{total}}^{-1}\\
&+\frac{1}{\lambda_{min}} \bar{\mu}^{-1} \sum_{i=1}^{4} \int_{t_0}^t e^{\left\{-\phi_{total}(t-T)\right\}} M_i^2\hspace{0.2cm}dT
\end{aligned}
\end{equation}
$\bar{\mu}$ can be any positive constant, and $\phi_{total}$ is a positive constant dependent on designable control gains that are freely chosen to satisfy the following condition:
\begin{equation}
\begin{aligned}
\label{82}
\frac{1}{\lambda_{min}{\bar{\mu} \phi_{total}}}<1
\end{aligned}
\end{equation}
To continue the stability proof, we define a continuous operator, $Z(\cdot)$, as follows:
\begin{equation}
\begin{aligned}
\label{83}
&Z(\iota)= \frac{\lambda_{min}^{-1} \bar{\mu}^{-1}}{\phi_{total}-\iota}>0, \hspace{0.2cm}\iota \in [0,\phi_{total})
\end{aligned}
\end{equation}
It is evident that by increasing $\iota$, $Z(\iota)$ increases, meaning that
$Z(\iota)\geq Z(0)=  \frac{1}{\lambda_{min}{\bar{\mu} \phi_{total}}}$.
By making $\phi_{total}$ large enough, we can find a small positive value $\bar{\iota} \in \iota$ that satisfies the following condition:
\begin{equation}
\begin{aligned}
\label{84}
0 < \bar{Z}=Z(\bar{\iota})=\frac{(\lambda_{min}\bar{\mu})^{-1}}{\phi_{total}-\bar{\iota}} < 1
\end{aligned}
\end{equation}
By multiplying $e^{\bar{\iota}(t-t_0)}$ to \eqref{81}, we reach:
\begin{equation}
\begin{aligned}
\label{85}
&{\|\mathbf{P}\|^2}e^{\bar{\iota}(t-t_0)} \leq \frac{2}{\lambda_{min}} V(t_0) e^{{-(\phi_{total}-\bar{\iota})(t-t_0)}}\\
&+\frac{2}{\lambda_{min}} \bar{\mu}_{total} \phi_{total}^{-1} e^{{\bar{\iota}(t-t_0)}} \\
 &+{\lambda_{min}^{-1}\bar\mu^{-1}}\sum_{i=1}^{4} \int_{t_0}^t e^{{-\phi_{total}(t-T)+\bar{\iota}(t-t_0)}} M_i^2 \hspace{0.2cm} dT
\end{aligned}
\end{equation}
Because $0 \leq \bar{\iota} <\phi_{total}$, we can eliminate the decreasing element $e^{{-(\phi_{total}-\bar{\iota})(t-t_0)}}$ from the right-hand side of \eqref{85}:
\begin{equation}
\begin{aligned}
\label{86}
& {\|\mathbf{P}\|^2}e^{\bar{\iota}(t-t_0)} \leq \\
&\frac{2}{\lambda_{min}}V(t_0)+\frac{2}{\lambda_{min}} \bar{\mu}_{total} \phi_{total}^{-1} e^{{\bar{\iota}(t-t_0)}}\\
&+\lambda_{min}^{-1}\bar\mu^{-1} \sum_{i=1}^{4} \int_{t_0}^t e^{-(\phi_{total}-\bar{\iota})(t-T)} M_i^2 e^{{\bar{\iota}(T-t_0)}}\hspace{0.1cm} dT
\end{aligned}
\end{equation}
Using this approach, we can describe functions $E_0$ and $E_i$, which are continuous and non-declining:
\begin{equation}
\begin{aligned}
\label{87}
&E_0 = \sup _{e \in(t-t_0)}  [\|\mathbf{P}\|^2 e^{{\bar{\iota}(e-t_0)})}]\\
&E_i= \sup _{e \in(t-t_0)} [\sum_{i=1}^{4}(M_i^2)  e^{\bar{\iota}(e-t_0)}]\\
\end{aligned}
\end{equation}
Next, by considering \eqref{86} and \eqref{87}, and conducting some straightforward mathematical manipulations while removing the decreasing term, we obtain:
\begin{equation}
\begin{aligned}
\label{88}
 {\|\mathbf{P}\|^2}e^{\bar{\iota}(t-t_0)} \leq& \frac{2}{\lambda_{min}}V(t_0)+  \frac{\lambda_{min}^{-1}\bar\mu^{-1}}{\phi_{total}-\bar{\iota}} E_i\\ 
 &+\frac{2}{\lambda_{min}} \bar{\mu}_{total} \phi_{total}^{-1} e^{{\bar{\iota}(t-t_0)}}
\end{aligned}
\end{equation}
As $E_i$ does not exhibit a declining pattern, the left side of \eqref{88} will not exhibit a reduction. Therefore, with respect to the definition of $E_0$ in \eqref{87}, we can state that:
\begin{equation}
\begin{aligned}
\label{89}
E_0 \leq & \frac{2}{\lambda_{min}} V(t_0)+\frac{2}{\lambda_{min}} \bar{\mu}_{total} \phi_{total}^{-1} e^{\bar{\iota}(t-t_0)}\\
&+ \frac{\lambda_{min}^{-1} \bar\mu^{-1}}{\phi_{total}-\bar{\iota}} E_i
\end{aligned}
\end{equation}
Defining $E=\max {(E_0,E_1,...,E_4)}$, we can have:

\begin{equation}
\begin{aligned}
\label{91}
E_0 \leq& \frac{2}{\lambda_{min}} V\left(t_0\right)+\frac{2}{\lambda_{min}} \bar{\mu}_{total} \phi_{total}^{-1} e^{\bar{\iota}(t-t_0)}\\
&+\bar{Z} E
\end{aligned}
\end{equation}
By making $\phi_{total}$ large enough, which relies on control gains, and making $\bar{\iota}$ small enough, it becomes possible to ensure the existence of a sufficiently large $\iota^*$, where $\phi_{total}>\iota^*>\bar{\iota}$ and $\overset{*}{Z}=Z(\iota^*)$, to satisfy the following condition \cite{heydari2024robust}:
\begin{equation}
\begin{aligned}
\label{92}
\overset{*}{Z}>\bar{Z}, \hspace{0.2cm} 0<\overset{*}{Z}<1, \hspace{0.1cm}\implies \hspace{0.1cm} \bar{Z} E \leq \overset{*}{Z} E_0
\end{aligned}
\end{equation}
\eqref{92} is justified, as we can make $\bar{Z}$ sufficiently small in \eqref{84}. When we incorporate \eqref{92} into \eqref{91}, we arrive at:
\begin{equation}
\begin{aligned}
\label{93}
E_0 \leq& \frac{2}{\lambda_{min}} V(t_0)+\frac{2}{\lambda_{min}} \bar{\mu}_{total} \phi_{total}^{-1} e^{\bar{\iota}(t-t_0)}\\
&+\overset{*}{Z} E_0(t)
\end{aligned}
\end{equation}
Afterward, we obtain:
\begin{equation}
\begin{aligned}
\label{94}
E_0 \leq \frac{\frac{2}{\lambda_{min}}V\left(t_0\right)+\frac{2}{\lambda_{min}} \bar{\mu}_{total} \phi_{total}^{-1} e^{\bar{\iota}(t-t_0)}} {1-\overset{*}{Z}}
\end{aligned}
\end{equation}
Concerning \eqref{87}, we obtain:
\begin{equation}
\begin{aligned}
\label{95}
\|\mathbf{P}\|^2 \leq \frac{\frac{2}{\lambda_{min}}V\left(t_0\right) e^{-\bar{\iota}(t-t_0)}+\frac{2}{\lambda_{min}} \bar{\mu}_{total} \phi_{total}^{-1}} {1-\overset{*}{Z}}
\end{aligned}
\end{equation}
It is significant that
\begin{equation}
\begin{aligned}
\label{96}
\sup _{t \in\left[t_0, \infty\right]}(\frac{\frac{2}{\lambda_{min}}V(t_0) e^{-\bar{\iota}(t-t_0)}} {1-\overset{*}{Z}})\leq \frac{\frac{2}{\lambda_{min}}V(t_0)} {1-\overset{*}{Z}}
\end{aligned}
\end{equation}
Thus, based on Definition 1, it is obvious from \eqref{95} that along with the adaptive algorithms provided in Equations \eqref{23} and \eqref{32} and the control input \eqref{40}, $\mathbf{P}$, including the state estimation error in \eqref{27} and the tracking error in \eqref{30}, reaches a defined region $g_{total}\left(\bar{\tau}_0\right)$ in uniformly exponential convergence, such that:
\begin{equation}
\begin{aligned}
\label{97}
g_{total}\left(\bar{\tau}_0\right):=\left\{\|\mathbf{P}\| \leq \bar{\tau}_0 := \sqrt{\frac{\frac{2}{\lambda_{min}} \bar{\mu}_{total} \phi_{total}^{-1}}{1-\overset{*}{Z}}}\right\}
\end{aligned}
\end{equation}
$\bar{\mu}_{total}$ depends on the
disturbance and non-triangular uncertainty bounds (see Eqs. \eqref{43} and \eqref{77}). Because $\lambda_{min}$, resulting from $\bm{p}$, and $\phi_{total}$, relying on the designable control parameters (see Eqs. \eqref{77}, \eqref{55}, \eqref{64}, \eqref{68}, \eqref{72}), can be chosen, it is possible to decrease the radius of the ball \eqref{97} as much as will satisfy \eqref{82}, \eqref{84}, and \eqref{92}.

\section*{Appendix C\\$n_a$ DoF EMLA-Actuated Manipulator Stability}
If we denote each Lyapunov function of the \(k\)th joint, actuated by EMLA and provided in \eqref{72}, as \(\bar{V}_k\), where \(k\) represents the joint number ranging from 1 to \(n_a\), we can then obtain the following Lyapunov function for the entire HDRM system:
\begin{equation}
\begin{aligned}
\label{108}
&V_{total} =\bar{V}_1+\bar{V}_2+...+\bar{V}_{n_a}
\end{aligned}
\end{equation}
After the derivative of \eqref{108}, and calling the equations \eqref{78}, we obtain:
\begin{equation}
\begin{aligned}
\label{109}
\dot{V}_{total} \leq& - \bar{\phi} \bar{V}_{total} + \frac{1}{2} \sum_{k=1}^{n_a} {\chi}^{-1} \bar{M}_k^2+ \kappa
\end{aligned}
\end{equation}
where:
\begin{equation}
\begin{aligned}
\label{110}
&\bar{\phi}=\min(\phi_{{total}_1},...,\phi_{{total}_{n_a}}), \hspace{0.2cm}  \bar{M}^2_k= M^2_{i_k}\\
&{\chi}^{-1}=\max(\bar{\mu}_1^{-1},...,\bar{\mu}_{n_a}^{-1})\\
&\kappa=\bar{\mu}_{{total}_1}+...+\bar{\mu}_{{total}_{n_a}}
\end{aligned}
\end{equation}
where \(\phi_{{total}_k}\) represents \(\phi_{total}\), \(\bar{\mu}_k\) denotes \(\bar{\mu}\), \(M^2_{{i_k}}\) is \(M^2_1 + \ldots + M^2_4\), and \(\bar{\mu}_{{total}_k}\) signifies \(\bar{\mu}_{total}\) of the \(k\)th EMLA-actuated joint; refer to \eqref{77}.
Generally speaking, we can transform the equations provided in \eqref{108} as follows:
\begin{equation}
\begin{aligned}
\label{111}
V_{total}&=\\
&\frac{1}{2} \mathbf{P_{total}^{\top}} \bm{\lambda_{total}} \mathbf{P_{total}} + \frac{1}{2}\bm{\tilde{\theta}_{total}^{\top}} \bm{\Delta_{total}^{-1}} \bm{\tilde{\theta}_{total}}
\end{aligned}
\end{equation}
where:
\begin{equation}
\begin{aligned}
\label{112}
&&&\mathbf{P}_{total} = \begin{bmatrix}
    \bm{P_{{total}_1}} \\
    \vdots \\
    \bm{P_{{total}_{n_a}}} \\
\end{bmatrix}, \bm{\tilde{\theta}_{total}} = \begin{bmatrix}
    \bm{\tilde{\theta}_{1}} \\
    \vdots\\
    \bm{\tilde{\theta}_{n_a}} \\
\end{bmatrix}\\
&&&\bm{\lambda_{total}}=\begin{bmatrix}
    \bm{\lambda_{1}} & 0 & 0 & \ldots  & 0 \\
    0 & \bm{\lambda_{2}} & 0 & \ldots & 0 \\
\vdots & \vdots & \vdots & \vdots & \vdots \\
    0 & \ldots & 0 & 0 & \bm{\lambda_{n_a}} \\
    \end{bmatrix}\\
    &&&
\bm{\Delta_{total}^{-1}} = \begin{bmatrix}
    \bm{\Delta_{1}^{-1}} & 0 & 0 & \ldots  & 0 \\
    0 & \bm{\Delta_{2}^{-1}} & 0 & \ldots & 0 \\
    \vdots & \vdots & \vdots & \vdots & \vdots \\
    0 & \ldots & 0 & 0 & \bm{\Delta_{n_a}^{-1}} \\
\end{bmatrix}\\
\end{aligned}
\end{equation}
where \(\bm{P_{{total}_k}}\) represents $\bm{P}$, \(\bm{\lambda_k}\) denotes \(\bm{\lambda}\), \(\bm{\tilde{\theta}_k}\) corresponds to \(\bm{\tilde{\theta}}\), and \(\bm{\Delta_k}\) signifies \(\bm{\Delta}\) of the \(k\)th EMLA-actuated joint; refer to \eqref{78}.
Note that $\mathbf{P_{total}}:\mathbb{R}^{6} \rightarrow \mathbb{R}^{6 n_a}$, $\bm{\lambda_{total}}:\mathbb{R}^{6 \times 6} \rightarrow \mathbb{R}^{6n_a \times 6n_a}$, $\bm{\tilde\theta_{total}}:\mathbb{R}^{5} \rightarrow \mathbb{R}^{5na}$, and $\bm{\Delta_{total}}:\mathbb{R}^{5 \times 5} \rightarrow \mathbb{R}^{5n_a \times 5n_a}$.
Therefore, we can conclude the demonstration of Theorem 2 for the whole of the $n_a$ DoF EMLA-actuated HDRM with similar steps provided in \eqref{75} and Appendix A.

\bibliographystyle{IEEEtran}
\bibliography{ob_REVISED.bib}

\begin{IEEEbiography}[{\includegraphics[width=1in,height=1.25in,clip,keepaspectratio]{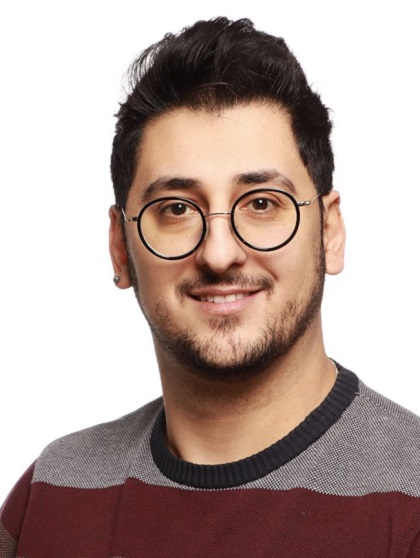}}]{Mehdi Heydari Shahna} earned a B.Sc. in electrical engineering from Razi University, Kermanshah, Iran, in 2015 and an M.Sc. in control engineering at Shahid Beheshti University, Tehran, Iran, in 2018. Since December 2022, he has been pursuing his doctoral degree in automation technology and mechanical engineering at Tampere University, Tampere, Finland. His research interests encompass robust control, nonlinear control of robotic systems, control of heavy-duty manipulators, fault-tolerant algorithms, and stability.
\end{IEEEbiography}

\begin{IEEEbiography}[{\includegraphics[width=1in,height=1.25in,clip,keepaspectratio]{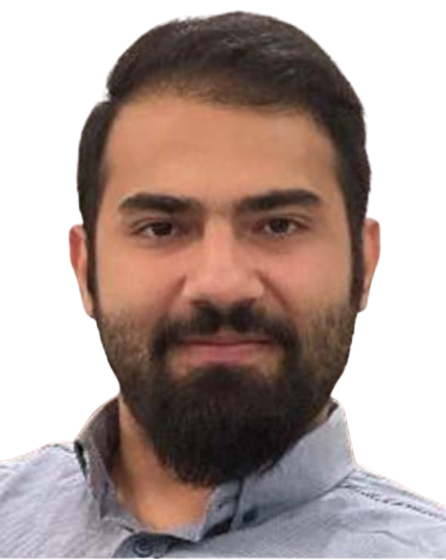}}]{Mohammad Bahari} earned a B.Sc. in electrical engineering power from Semnan University, Semnan, Iran, in 2015, followed by the completion of his M.Sc.  in electrical engineering power electronics and electric machines from Sharif University of Technology, Tehran, Iran, in 2019. Presently, he is engaged as a doctoral researcher at Tampere University, Tampere, Finland, focusing on design and control of an all-electric robotic e-boom. His research interests include multidisciplinary design optimization of electromechanical actuators, propulsion systems of electric vehicles, and high-precision electromagnetic sensors.  
\end{IEEEbiography}

\begin{IEEEbiography}[{\includegraphics[width=1in,height=1.25in,clip,keepaspectratio]{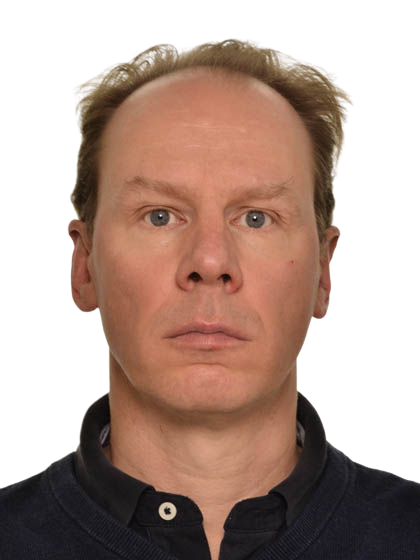}}]{Jouni Mattila}
received an M.Sc. and Ph.D. in automation engineering from Tampere University of Technology, Tampere, Finland, in 1995 and 2000, respectively. He is currently a professor of machine automation with the Unit of Automation Technology and Mechanical Engineering at Tampere University. His research interests include machine automation, nonlinear-model-based control of robotic manipulators, and energy-efficient control of heavy-duty mobile manipulators.
\end{IEEEbiography}

\vfill

\end{document}